\documentclass[a4paper,11pt]{article}
\pdfoutput=1 

\usepackage{jheppub} 
\usepackage{slashed}
\usepackage{braket}

\usepackage[T1]{fontenc} 

\title{\boldmath Symmetry resolved spread complexity for BMN matrix model}

\author[a]{Dibakar Roychowdhury}

\affiliation[a]{Department of Physics, Indian Institute of Technology Roorkee,\\Roorkee 247667, Uttarakhand, India}

\emailAdd{dibakar.roychowdhury@ph.iitr.ac.in}
\abstract{We investigate the symmetry-resolved spread complexity of the Berenstein-Maldacena-Nastase (BMN) matrix model under a global $U(1)$ symmetry acting within the transverse $SO(6)$ scalar sector. Decomposing the time-evolved charged thermofield double (cTFD) state into orthogonal fixed-charge sectors, we examine the return amplitude, spectral moments, tridiagonal Lanczos coefficients, and early-time spread complexity across the low-temperature confined ($\beta\mu \gg 1$) and finite-temperature deconfined ($T \gg \mu$) phases. In the low-temperature gapped regime, effective sector dynamics map onto decoupled, displaced harmonic oscillators with a Casimir-like zero-point offset, yielding factorized moments, linear drift in diagonal Lanczos coefficients $a_n^{(q)}$, and sub-linear $\sqrt{n}$ off-diagonal growth with thermally suppressed initial hopping. Charge-dependent Boltzmann weights, vacuum shifts, and degeneracies break complexity equipartition, while inter-sector coherence and charge dispersion enforce an early-time positivity bound $\Delta C(t) = C_{\text{total}}(t) - \bar{C}(t) \ge 0$. In the \emph{finite}-temperature continuum phase, modeling the planar large-$N$ spectral density as a thermal Gaussian gives constant diagonal coefficients $a_n^{(q)} = \bar{E}_q$, universal square-root scaling $b_n^{(q)} = \sigma_E \sqrt{n}$, and a quadratic spread launch $\mathcal{C}^{(q)}(t) = \sigma_E^2(q,T)\,t^2 + \mathcal{O}(t^4)$ governed by the backreacted sector energy variance. Holographically, these regimes distinguish coherent string fluctuations in horizonless Lin-Maldacena (LM) droplet geometries from dissipative scrambling near charged ten-dimensional black D0-brane horizons.}
\begin{document} 
\maketitle
\flushbottom
\section{Introduction and General Idea}
Complexity and its holographic generalization \cite{Stanford:2014jda}-\cite{Belin:2021bga} has been showing us the path for a new quantum era. Among different existing notions of complexity, it is the Krylov and/or the spread complexity of states and operators \cite{Parker:2018yvk}-\cite{Caputa:2025ozd} that has garnered renewed attention for the last few years for various remarkable features and, in particular, its central role in understanding the celebrated holographic correspondence \cite{Maldacena:1997re}-\cite{Witten:1998qj}. These applications include chaos \cite{Hashimoto:2023swv}-\cite{Bhattacharjee:2024yxj}, holography \cite{Caputa:2024sux}-\cite{Erdmenger:2022lov} and black holes \cite{Kar:2021nbm} under certain circumstances\footnote{There have been some nice reviews on the subject, for example, \cite{Baiguera:2025dkc}-\cite{Nandy:2024evd}.}. 

The big picture is to understand the holographic principle by exploring the nature of Krylov complexity in a set up that might allow us to calculate the Krylov growth of states (or operators) on both sides of the duality. We identify the BMN matrix models \cite{Berenstein:2002jq}-\cite{Santos:2026rrx} as one such candidate that might play a useful toy model in testing the correspondence in its non-perturbative domain. The present paper is a first step towards that direction where we focus solely on the QFT side of the duality in the presence of a global $U(1)$ charge density.

The BMN model that we consider is a \emph{massive} deformation $\mu\gg 1$ of the BFSS matrix model \cite{Banks:1996vh} and therefore is non-conformal (or gapped). We turn on a global $U(1)\subset SO(6)$ and study the spread complexity of states following the Lanczos algorithm \cite{Parker:2018yvk}-\cite{Balasubramanian:2022tpr}. In particular, we calculate the symmetry resolved spread complexity for BMN matrix model for low temperature gaped phase and the finite-temperature continuum. Setting $\mu$ as the mass gap, the low temperature ($T=\beta^{-1}$) phase of the matrix model is characterized by $\beta \mu \gg 1$, while the finite-temperature phase is characterized by $\beta \mu \ll 1$.
\subsection{A general motivation}
In a quantum system with a conserved global charge, a generic thermal or thermo-field double (TFD) state is a superposition of many different charge sectors. Computing only the \emph{total} spread complexity averages over all these sectors at once \cite{Roychowdhury:2026ikp}. While this gives a useful coarse-grained measure of state growth, it hides how states carrying different amounts of charge actually evolve. By resolving spread complexity into individual fixed-$U(1)$ charge sectors, we isolate the dynamics of each sector cleanly and gain several distinct physical insights that we list below highlighting the key points down the line.

\begin{itemize}
    \item \textbf{Separating charge effects from thermal noise:} In the total ensemble, sectors with small and large charges are summed together with thermal weights. When temperature or coupling changes, it becomes difficult to tell whether a change in complexity comes from thermal broadening or from the charge itself. Fixing the charge $q$ removes this ambiguity and reveals how charge alone controls the early-time spread (launch acceleration) and the growth of the Lanczos coefficients.

    \item \textbf{Testing equipartition in an interacting non-Abelian matrix model:} In simple free or weakly coupled theories, complexity is often shared almost equally across different symmetry sectors, a property known as \emph{complexity equipartition} \cite{Caputa:2025ozd}. The BMN matrix model provides an ideal, controllable setting to test when and how this breaks down. Because BMN is a gapped ($\mu \gg 1$), interacting non-Abelian matrix quantum mechanics with quartic commutator terms and a Myers term, different $U(1)$ charge sectors experience distinct interaction-induced frequency shifts and vacuum structures at large $N$. Symmetry resolution directly exposes how non-Abelian interactions skew complexity across charge sectors rather than distributing it uniformly.

    \item \textbf{Probing phase structure and charged holography in BMN:} A central appeal of the BMN model is that it exhibits both a low-temperature gapped phase ($\beta \mu \gg 1$) and a high-temperature deconfined black hole phase ($\beta \mu \ll 1$), with a well-understood dual description in terms of M-theory and Type IIA supergravity. In the holographic bulk, turning on a global $U(1) \subset SO(6)$ charge corresponds to adding angular momentum or $R$-charge to branes and black hole geometries. Studying symmetry-resolved spread complexity in BMN therefore gives us a sharp, sector-by-sector boundary diagnostic of how charged excitations behave across confinement/deconfinement transitions, setting a clean benchmark for future bulk dual calculations.
\end{itemize}

Below, we summarize all our key findings and highlight main results. 
\subsection{Summary of results}
In Section \ref{sec 2}, we establish the foundational framework of the Berenstein-Maldacena-Nastase (BMN) matrix model in the presence of a continuous global $U(1)$ symmetry. We review the $0+1$-dimensional supersymmetric BMN matrix model \cite{Berenstein:2002jq} obtained via the dimensional reduction of ten-dimensional $\mathcal{N}=1$ Super-Yang-Mills (SYM) theory. We explicitly verify the self-adjointness of the individual components of $H_{\text{BMN}}$ \eqref{e2.1}, including the kinetic terms, mass deformations, quartic commutators, and the cubic Myers term $\epsilon_{abc}\text{Tr}(X^a X^b X^c)$. We detail the continuous global symmetry group $G_{\text{global}} = \text{SO}(3) \times \text{SO}(6) \times U(1)$ \eqref{ee2.12}, focusing on the $U(1)$ subgroup embedded within the $\text{SO}(6)$ transverse scalar sector and generated by the bosonic charge operator $Q_{\text{boson}} = \text{Tr}(X_4 \Pi_5 - X_5 \Pi_4)$ \eqref{e2.4}.

We rigorously demonstrate the commutativity between the BMN Hamiltonian and the global charge operator, $[H_{\text{BMN}}, Q] = 0$ \eqref{e2.3}, by evaluating the infinitesimal variations of the dynamical fields and interaction potentials under $U(1)$ rotations in the $X_4\text{--}X_5$ plane. Furthermore, we classify the classical moduli space vacua of the bosonic potential $V(X)$ into following three distinct regimes
    \begin{enumerate}
        \item \emph{Trivial flat vacuum:} $J_a = 0$, describing $N$ coincident $D0$-branes at the origin around which perturbative expansions are formulated.
        \item \emph{Fuzzy sphere vacua:} $J_a$ furnish an $N$-dimensional irreducible representation of $\text{su}(2)$, describing $D0$-branes polarized via the Myers effect \cite{Myers:1999ps} into a non-commutative fuzzy 2-sphere ($S^2$) with radius $R \sim \mu N / g$ \eqref{ee2.11}.
        \item \emph{Multi-fuzzy sphere vacua:} $J_a$ decompose as block-diagonal direct sums representing concentric or separated fuzzy geometries.
    \end{enumerate}
Building upon the $U(1)$ charge conservation, Section \ref{sec 2} sets up the subsequent analytical machinery for exploring symmetry-resolved quantum information probes. Specifically, it prepares the ground for constructing charged Thermo-Field Double (cTFD) states \cite{Caputa:2013eka}-\cite{Chapman:2019clq}, decomposing time-dependent states $|\psi_q(t)\rangle$ into definite charge sectors $q$. This further leads towards evaluating Krylov space moments $\mathcal{M}_n^{(q)}$, and deriving sector-dependent Lanczos coefficients and return amplitudes $\mathcal{R}^{(q)}(t)$ which we elaborate in the subsequent Section \ref{sec 3}.

Section~\ref{sec 3} delve into the study of symmetry resolved spread complexity for the low temperature confined phase of the matrix model. In the low-temperature confined phase ($\beta \hat{\Omega}(q) \gg 1$, $\beta\mu \gg 1$), the effective BMN Hamiltonian within each fixed $U(1)$ charge sector $q$ reduces to a collection of decoupled, displaced harmonic oscillators with normal-mode frequency $\omega(q) = \mu\sqrt{1 + \frac{g_{\text{eff}}}{2N}f(q)}$ ($f(q) = 2n - q + 1$) and a Casimir-like zero-point offset $\frac{|q|+1}{2}\omega(q)$ above the shifted vacuum $E_{0,q}$. Evaluating the symmetry-resolved return amplitude $\mathcal{R}^{(q)}(t)$ and spectral moments $\mathcal{M}_k^{(q)}$, we show that higher moments factorize as $\mathcal{M}_k^{(q)} \approx (\mathcal{M}_1^{(q)})^k$ with exponentially suppressed connected cumulants ($\kappa_2^{(q)} \sim e^{-\beta\omega(q)}\mu^2$), signaling coherent quantum recurrences rather than chaotic dispersion. 

Solving the sector-projected Lanczos recursion yields uniform diagonal steps $a_n^{(q)} \approx \left(\frac{|q|+1}{2} + n\right)\omega(q)$ and sub-linear square-root off-diagonal growth $b_n^{(q)} \propto \sqrt{n}\,\omega(q)$ at large $n$, with initial hopping $b_1^{(q)} \sim e^{-\frac{\beta}{2}\omega(q)}\mu$ and $b_2^{(q)} \propto \sqrt{(|q|+1)(|q|+3)}\,\omega(q)$. 

Consequently, the early-time symmetry-resolved spread complexity takes the form
\begin{align}
C^{(q)}(t) = e^{-\beta\omega(q)}\left(1 + \frac{g_{\text{eff}}}{2N}f(q)\right)\mu^2 t^2 + \Sigma(q)\mu^4 t^4 + \mathcal{O}(t^6),
\end{align}
where the quadratic dependence of $\Sigma(q)$ on $|q|$, together with the $q$-dependent Boltzmann weights and perturbative vacuum shift $\Delta E_{0,q}^{(1)}$, explicitly breaks complexity equipartition across distinct $U(1)$ charge sectors—accelerating Krylov cascades in higher-charge sectors. Furthermore, comparing the total complexity $C_{\text{total}}(t)$ of the coherent superposition state against the grand-canonical sector average $\bar{C}(t) = \sum_q p(q) C^{(q)}(t)$ reveals a variance-plus-coherence decomposition that establishes the early-time positivity bound $\Delta C(t) = C_{\text{total}}(t) - \bar{C}(t) \ge 0$. Holographically, these features capture coherent zero-point fluctuations of transverse $SO(6)$ string modes propagating across gapped, horizonless Lin--Maldacena droplet and eleven-dimensional pp-wave geometries rather than dissipative black hole horizons. The complete holographic understanding of spread complexity is under way.

We extend our analysis in Section \ref{sec 4} to the finite-temperature regime ($\beta > 0$, $T \gg \mu$), where the discrete oscillator spectrum melts into a smooth thermal continuum dominated by unconstrained matrix degrees of freedom. Modeling the symmetry-resolved spectral density $\rho_q(E)$ in the planar large-$N$ limit as a thermal Gaussian centered at the macrocanonical mean sector energy $\bar{E}_q$ with thermal energy variance $\sigma_E^2$ (evaluated explicitly in Appendix~\ref{appen D}), we obtain closed-form expressions for the symmetry-resolved return amplitude $\mathcal{R}^{(q)}(t)$ and the corresponding spectral moments $\mathcal{M}_n^{(q)}$. By performing a large-$z$ continued-fraction inversion of the Green's function $G(z)$ and exploiting the symmetry of the Gaussian spectral measure via monic orthogonal polynomials, we determine the exact tridiagonal Lanczos coefficients in each fixed $U(1)$ charge sector
\begin{align}
a_n^{(q)} = \bar{E}_q + \mathcal{O}(1/N), \qquad b_n^{(q)} = \sigma_E \sqrt{n}.
\end{align}

Unlike the infinite-temperature ($\beta \to 0$) chaotic regime characterized by linear growth ($b_n \sim \alpha n$) \cite{Parker:2018yvk}, thermal Gaussian localization at finite temperature softens the off-diagonal Lanczos growth to a universal square-root scaling $\propto \sqrt{n}$ \cite{Roychowdhury:2026ikp} governed entirely by the sector energy variance $\sigma_E^2$, while anchoring the diagonal coefficients $a_n^{(q)}$ to the mean sector energy $\bar{E}_q$ independent of the Krylov chain depth $n$. Furthermore, solving the discretized Krylov chain recurrence relations at early times ($t \to 0$) reveals a universal quadratic ballistic launch of the symmetry-resolved spread complexity,
\begin{align}
\mathcal{C}^{(q)}(t) = \sigma_E^2 t^2 + \mathcal{O}(t^4),
\end{align}
where the initial acceleration $\sigma_E^2$ carries the macroscopic $q$-dependence of the charged thermal ensemble. Holographically, whereas the low-temperature phase probes coherent zero-point string fluctuations across horizonless Lin--Maldacena droplet geometries, the finite-temperature enhancement of $\sigma_E^2$ reflects rapid thermal scrambling near a ten-dimensional black hole/black D0-brane horizon backreacted by R-charge and angular momentum flux.

Notice that both high temperature and low temperature complexities have the leading order growth in terms of the non-diagonal Lanczos coefficient $b_1^{(q)}$ as  \cite{Caputa:2025ozd}
\begin{align}
 \mathcal{C}^{(q)}(t) = b_1^{2(q)} t^2 + \mathcal{O}(t^4).  
\end{align}

In Table \ref{tab:summary_comparison}, we compare our results between low temperature confined phase and finite-temperature deconfined phase.

\begin{table}[htbp]
\centering
\renewcommand{\arraystretch}{1.45}
\setlength{\tabcolsep}{7pt}
\resizebox{\textwidth}{!}{%
\begin{tabular}{p{3.4cm} p{6.2cm} p{6.2cm}}
\hline\hline
\textbf{Physical Observable / Feature} & \textbf{Low-Temperature Confined Phase} \newline (Section~\ref{sec 3}: $\beta\mu \gg 1$, $\beta\hat{\Omega}(q) \gg 1$) & \textbf{Finite-Temperature Deconfined Phase} \newline (Section~\ref{sec 4}: $T \gg \mu$, $\beta > 0$) \\
\hline
\textbf{Effective Microscopic Dynamics} & Decoupled, displaced harmonic oscillators with normal-mode frequency $\omega(q) = \mu\sqrt{1 + \frac{g_{\text{eff}}}{2N}f(q)}$ ($f(q)=2n-q+1$) and Casimir zero-point offset $\frac{|q|+1}{2}\omega(q)$ above shifted vacuum $E_{0,q}$. & Smooth thermal continuum of unconstrained $N \times N$ matrix degrees of freedom; discrete oscillator mass gap melts due to thermal fluctuations. \\
\hline
\textbf{Spectral Density $\rho_q(E)$ \& Moments $\mathcal{M}_k^{(q)}$} & Discrete gapped tower $E_{n,q} = E_{0,q} + (n + \frac{|q|+1}{2})\omega(q)$ with Boltzmann suppression $e^{-\beta\Delta E}$. Moments factorize ($\mathcal{M}_k^{(q)} \approx (\mathcal{M}_1^{(q)})^k$) with suppressed cumulants ($\kappa_2^{(q)} \sim e^{-\beta\omega(q)}\mu^2$). & Planar large-$N$ thermal Gaussian centered at macrocanonical mean energy $\bar{E}_q$ with thermal variance $\sigma_E^2(q, T)$ (Appendix~\ref{appen D}); moments follow Gaussian cumulant expansion. \\
\hline
\textbf{Diagonal Lanczos Coefficients $a_n^{(q)}$} & \textbf{Linear drift in $n$:} $a_n^{(q)} \approx \left(\frac{|q|+1}{2} + n\right)\omega(q)$, reflecting uniform energy steps up the discrete oscillator ladder. & \textbf{Constant / $n$-independent:} $a_n^{(q)} = \bar{E}_q + \mathcal{O}(1/N)$, pinned to the macrocanonical mean sector energy. \\
\hline
\textbf{Off-Diagonal Lanczos Coefficients $b_n^{(q)}$} & Sub-linear $\sqrt{n}$ growth at large $n$ ($b_n^{(q)} \propto \sqrt{n}\,\omega(q)$) with thermally suppressed initial hopping $b_1^{(q)} \sim e^{-\frac{\beta}{2}\omega(q)}\mu$ and $b_2^{(q)} \propto \sqrt{(|q|+1)(|q|+3)}\,\omega(q)$. & \textbf{Exact square-root growth:} $b_n^{(q)} = \sigma_E \sqrt{n}$ across the Krylov chain, softened from the $\beta \to 0$ linear bound ($b_n \sim \alpha n$) \cite{Parker:2018yvk} by thermal Gaussian localization \cite{Roychowdhury:2026ikp}. \\
\hline
\textbf{Early-Time Spread Complexity $C^{(q)}(t)$} & Thermally suppressed quadratic launch (Eq.~\eqref{e3.166}): \newline $C^{(q)}(t) = e^{-\beta\omega(q)}\!\left(1 + \frac{g_{\text{eff}}}{2N}f(q)\right)\mu^2 t^2 + \Sigma(q)\mu^4 t^4 + \mathcal{O}(t^6)$. & Unsuppressed ballistic quadratic launch (Eq.~\eqref{e4.80}): \newline $\mathcal{C}^{(q)}(t) = \sigma_E^2(q,T)\, t^2 + \mathcal{O}(t^4)$, governed directly by the macroscopic sector energy variance. \\
\hline
\textbf{Equipartition Breakdown \& Bounds} & Broken via $e^{-\beta\omega(q)}$, $f(q)$, $\Delta E_{0,q}^{(1)}$, and degeneracy factor $j(q) \sim |q|^2$. Inter-sector coherence and charge dispersion yield early-time positivity bound $\Delta C(t) = C_{\text{total}}(t) - \bar{C}(t) \ge 0$. & Broken macroscopically via $\bar{E}_q$ and backreacted variance $\sigma_E^2(q,T) = N^2 T^4\big[\alpha_0 + \alpha_1(q/N^2)^2 + \alpha_2(q/N^2)^4\big] + \mathcal{O}(N^2 T^2)$. \\
\hline
\textbf{Dual Holographic Interpretation} & Smooth, horizonless \textbf{Lin--Maldacena droplet geometries} and 11D pp-wave backgrounds; probes coherent zero-point fluctuations of transverse $SO(6)$ string modes. & Backreacted \textbf{10D black hole / black D0-brane horizon} carrying R-charge and angular momentum flux; probes dissipative thermal scrambling. \\
\hline\hline
\end{tabular}%
}
\caption{Comparative summary of symmetry-resolved Krylov observables, Lanczos growth, and holographic interpretations in the low-temperature confined phase (Section~\ref{sec 3}) versus the finite-temperature deconfined phase (Section~\ref{sec 4}) of the BMN matrix model.}
\label{tab:summary_comparison}
\end{table}

Finally, we conclude in Section \ref{sec 5} along with some future remarks.

\paragraph{Summary of Appendices.}The technical derivations underpinning our analysis are organized into five complementary appendices. In Appendix~\ref{appen:BMN_reduction}, we provide a self-contained dimensional reduction of ten-dimensional $\mathcal{N}=1$ Super-Yang--Mills theory on $T^9$ with background Ramond--Ramond four-form flux, explicitly deriving the BMN matrix Hamiltonian, the cubic Myers interaction, and the Gauss law constraint. Appendix~\ref{appen A} develops the large-$N$ perturbative expansion in a fixed $U(1)$ charge sector by decomposing four-point scalar correlators into Hartree--Fock and exchange contractions, which Appendix~\ref{appen B} subsequently uses to compute the exact first-order vacuum energy shift $\Delta E_{0,q}^{(1)}$ and establish the breakdown of complexity equipartition across charge sectors. Building on these perturbative results, Appendix~\ref{appen C} presents a qualitative physical comparison between the bare Fock vacuum and the backreacted, charge-dressed interacting vacuum, highlighting how macroscopic $R$-charge reshapes the low-energy normal-mode spectrum. Finally, Appendix~\ref{appen D} addresses the high-temperature deconfined regime ($T \gg \mu$), modeling the planar spectral density as a thermal Gaussian and evaluating connected four-loop melonic Feynman diagrams to extract the exact scaling of the backreacted energy variance $\sigma_E^2(q, T)$ in both the fixed-charge and macroscopic-charge limits. This shows equipartition at finite temperature holds only for the macroscopic charge regime, where we assume a scaling of the form $q \sim \mathcal{O}(N^2)$.
\section{BMN matrix model and $U(1)$ symmetry}
\label{sec 2}
The purpose of this Section is to review basic aspects of BMN matrix model in the presence of a global $U(1)$ symmetry. These include - (i) BMN Hamiltonian and its global $U(1)$ symmetry, (ii) construction of charged TFD (cTFD) and decomposition of states $\ket{\psi_{q}(t)}$.

In the next Section \ref{sec 3}, we discuss calculation of Lanczos coefficients, moments $\mathcal{M}_n^{(q)}$ and return amplitudes $\mathcal{R}^{(q)}(t)$ for the low temperature $\beta \mu \gg 1$ phase. We also discuss relationship between Lanczos coefficients and moments for a given charge sector $q$. We also perform an expansion of return amplitude in energy eigen basis. We calculate the spread complexity $C^{(q)}(t)$, check whether $C^{(q)}(t)$ is the same for all charge sector (or equivalently the ``equipartition''). We check the relation with total complexity $C(t)$ and the relationship between $C(t)$ and average spread complexity at early times, in particular we calculate the difference $C(t)-\bar{C}(t)>0$. Finally, we comment on how full Lanczos coefficient, moments and return amplitudes are expressed in terms of their values in individual charge sectors $q$ and their holographic interpretations. 

\subsection{BMN Hamiltonian and its global symmetry}
In order to set the stage, we review the structure of the Berenstein-Maldacena-Nastase (BMN) matrix model Hamiltonian and its symmetry properties. The BMN matrix model \cite{Berenstein:2002jq} is a one-dimensional supersymmetric quantum mechanics describing $N \times N$ matrix degrees of freedom, obtained as a mass deformation of the BFSS matrix model \cite{Banks:1996vh}. The dynamical variables consist of nine spatial scalar matrices $X_i(t)$ ($i=1, \dots, 9$) and sixteen real Majorana fermions $\psi_\alpha(t)$ ($\alpha=1, \dots, 16$), all transforming under the adjoint representation of the $U(N)$ gauge group. The scalar fields are decomposed into two sectors according to the $SO(6) \times SO(3)$ R-symmetry group of the mass deformation: $X_a$ ($a = 1, 2, 3$) transforming under $SO(3)$, and $X_{i'}$ ($i' = 4, \dots, 9$) transforming under $SO(6)$. 

The full BMN Hamiltonian takes the following form\footnote{The factor of $3$ has been absorbed into the definition of the mass parameter, that is, $ \frac{\mu}{3}\rightarrow \mu$.}
\begin{align}
\label{e2.1}
H_{\text{BMN}}& = \text{Tr} \Big[ \frac{1}{2} \sum_{i=1}^9 \Pi_i^2 - \frac{g^2}{4} \sum_{i, j=1}^9 [X_i, X_j]^2 + \frac{\mu^2}{2} \sum_{a=1}^3 X_a^2 + \frac{\mu^2}{8} \sum_{i'=4}^9 X_{i'}^2 \nonumber\\
&+ i\mu g \sum_{a,b,c=1}^3 \epsilon_{abc} X_a X_b X_c + H_{\text{fermion}} \Big],
\end{align}
where $\Pi_i = \dot{X}_i$ are the canonical momentum matrices (with the choice of gauge $A_t=0$), $\mu$ is the flux and/or mass parameter, $g$ is the coupling constant, and $H_{\text{fermion}}$ accounts for the quadratic mass terms and cubic gauge interactions of the Majorana fermions.

Notice that every individual term in Eq. \eqref{e2.1}, the kinetic operator $\Pi_i^2$, the mass terms $X^2$, the quartic commutator $[X_i, X_j]^2$, and the cubic Myers term $i \epsilon_{abc} \text{Tr}(X_a X_b X_c)$ is independently Hermitian. The full Hamiltonian $H_{\text{BMN}}$ is real-valued and self-adjoint.

As a quick check, one can investigate the self-adjoint property of the Myers term
\begin{align}
    &H_{\text{Myers}}^\dagger 
    =-i\mu g\epsilon_{abc} \text{Tr}(X_c X_b X_a)\nonumber\\
    &=-i\mu g\epsilon_{abc} \text{Tr}(X_a X_c X_b)\nonumber\\
    &=i\mu g\epsilon_{abc}\text{Tr}(X_a X_b X_c)
\end{align}
where in the last line we have used the cyclicity of trace and the anti-symmetry of $\epsilon_{abc}$.

The BMN matrix model \eqref{e2.1} can be systematically obtained via dimensional reduction of 10-dimensional $\mathcal{N}=1$ Super-Yang-Mills (SYM) theory down to $0+1$ dimensions on a circle/time-like direction with flux deformation\footnote{See Appendix \ref{appen:BMN_reduction} for details.}. Starting from the 10D SYM action with gauge coupling $g_{\text{YM},10}$, the dimensional reduction to $1$-dimension on a compact volume $V_9$ yields the effective $1D$ gauge coupling $g$ \cite{Banks:1996vh}
\begin{align}
\label{ee2.3}
    g^2 = \frac{g_{\text{YM},10}^2}{V_9}.
\end{align}

In terms of canonical mass dimensions in natural units ($\hbar = c = 1$), the scalar fields $X_i$ and the coupling constants have mass dimensions
\begin{align}
\label{ee2.4}
[X_i] = M^{-1/2}, \quad [g_{\text{YM},10}^2] = M^{-6}, \quad [g^2] = M^{3},
\end{align}
where $M$ denotes the mass/energy scale.

Consequently, the dimension of the coupling constant $g^2$ in the $0+1$-dimensional BMN model is $[g^2] = \text{mass}^3$. In the subsequent analysis, we introduce the dimensionless effective coupling constant $g_{\text{eff}}$ controlling the perturbation theory at energy scale $\mu$
\begin{align}
    g_{\text{eff}} = \frac{g^2 N}{\mu^3}.
\end{align}\\\\
\underline{\textbf{Comments about the ground state of the system:}}\\\\
The potential function in the bosonic sector is given by
\begin{align}
\label{eee2.6}
    V(X) &= \text{Tr} \Big[ -\frac{g^2}{4} \sum_{i,j=1}^9 [X_i, X_j]^2 + \frac{\mu^2}{2} \sum_{a=1}^3 X_a^2 \nonumber\\
    &+ \frac{\mu^2}{8} \sum_{i'=4}^9 X_{i'}^2 +  i\mu g \sum_{a,b,c=1}^3 \epsilon_{abc} X_a X_b X_c \Big].
\end{align}

For the $SO(6)$ scalars ($X_{i'}$ for $i'=4, \dots, 9$), the mass terms $\frac{\mu^2}{8} \text{Tr}(X_{i'}^2)$ are strictly positive definite. Therefore, at any classical minimum of $V(X)$, these transverse fields must vanish, that is, $\langle X_i' \rangle = 0 \quad \text{for } i' \in \{4, 5, 6, 7, 8, 9\}$.

For the $SO(3)$ scalars ($X_1, X_2, X_3$), we can complete the square for the potential $V_{SO(3)}$. To see this explicitly, let us note that in Eq.\eqref{eee2.6}, the unrestricted sum over all $a,b \in \{1,2,3\}$ includes the diagonal terms $a=b$ (which vanish since $[X_a, X_a] = 0$) and counts every distinct pair twice ($a < b$ and $a > b$) to yield
\begin{align}
    -\frac{g^2}{4} \sum_{a,b=1}^{3} [X_a, X_b]^2 = -\frac{g^2}{4} \left( \sum_{a<b} [X_a, X_b]^2 + \sum_{a>b} [X_a, X_b]^2 \right).
\end{align}

Since $[X_b, X_a]^2 = (-[X_a, X_b])^2 = +[X_a, X_b]^2$, the two off-diagonal sums are identical, which yields the $SO(3)$ potential in its following form
\begin{align}
    V_{SO(3)}(X) = \text{Tr} \left[ -\frac{g^2}{2} \sum_{a < b} [X_a, X_b]^2 + \frac{\mu^2}{2} \sum_{a=1}^3 X_a^2 +  i\mu g \sum_{a,b,c=1}^3 \epsilon_{abc} X_a X_b X_c \right].
\end{align}

Using the Levi-Civita contraction identity in three dimensions
\begin{align}
    \sum_{a,b=1}^{3} \epsilon_{abc} \epsilon_{abd} = 2\delta_{cd},
\end{align}
we can rewrite $\sum_{c=1}^{3} \mathrm{Tr}(X_c^2)$ as a double sum over $a,b$
\begin{align}
    \begin{aligned} \frac{\mu^2}{2} \sum_{c=1}^{3} \mathrm{Tr}(X_c^2) &= \frac{\mu^2}{2} \cdot \frac{1}{2} \sum_{a,b,c,d=1}^{3} \epsilon_{abc} \epsilon_{abd} \, \mathrm{Tr}(X_c X_d) \\ &= -\frac{g^2}{4} \sum_{a,b=1}^{3} \mathrm{Tr}\!\left[ \left(-i\frac{\mu}{g}\epsilon_{abc} X_c\right) \left(-i\frac{\mu}{g}\epsilon_{abd} X_d\right) \right] \\ &= -\frac{g^2}{4} \sum_{a,b=1}^{3} \mathrm{Tr}\!\left[ \left(-i\frac{\mu}{g}\epsilon_{abc} X_c\right)^2 \right], \end{aligned}
\end{align}
where in the last step implicit summation is understood over the repeated index $c$ inside the squared bracket.

Using the antisymmetry of $\epsilon_{abc}$ and cyclicity of the trace, we write the cubic product in terms of a commutator
\begin{align}
    \begin{aligned} \sum_{a,b,c=1}^{3} \epsilon_{abc} \, \mathrm{Tr}(X_a X_b X_c) &= \frac{1}{2} \sum_{a,b,c=1}^{3} \epsilon_{abc} \, \mathrm{Tr}\!\left(X_a X_b X_c - X_a X_c X_b\right) \\ &= \frac{1}{2} \sum_{a,b,c=1}^{3} \epsilon_{abc} \, \mathrm{Tr}\!\left(X_a [X_b, X_c]\right) \\ &= \frac{1}{2} \sum_{a,b,c=1}^{3} \epsilon_{abc} \, \mathrm{Tr}\!\left([X_a, X_b] X_c\right), \end{aligned}
\end{align}
where the last line follows by relabeling $(a,b,c) \to (b,c,a)$ (an even permutation of $\epsilon_{abc}$) and using trace cyclicity. Multiplying by $i\mu g$ and factoring out $-\frac{g^2}{4}$ gives
\begin{align}
    \begin{aligned} i\mu g \sum_{a,b,c=1}^{3} \epsilon_{abc} \, \mathrm{Tr}(X_a X_b X_c) &= \frac{i\mu g}{2} \sum_{a,b,c=1}^{3} \epsilon_{abc} \, \mathrm{Tr}\!\left([X_a, X_b] X_c\right) \\ &= -\frac{g^2}{4} \sum_{a,b=1}^{3} \mathrm{Tr}\!\left( -2i\frac{\mu}{g}\epsilon_{abc} [X_a, X_b] X_c \right). \end{aligned}
\end{align}

Adding the three rewritten pieces together under the single prefactor $-\frac{g^2}{4} \sum_{a,b=1}^{3} \mathrm{Tr}(\cdots)$ yields the $SO(3)$ potential that that can be re-expressed identically as 
\begin{align}
    \begin{aligned} V_{SO(3)}(X) &= -\frac{g^2}{4} \sum_{a,b=1}^{3} \mathrm{Tr}\!\left[ [X_a, X_b]^2 - 2i\frac{\mu}{g}\epsilon_{abc} [X_a, X_b] X_c + \left(-i\frac{\mu}{g}\epsilon_{abc} X_c\right)^2 \right] \\ &= -\frac{g^2}{4} \sum_{a,b=1}^{3} \mathrm{Tr}\!\left( [X_a, X_b] - i\frac{\mu}{g}\epsilon_{abc} X_c \right)^2 \ge 0. \end{aligned}
\end{align}

The condition for absolute minima ($V = 0$) is that the term inside the bracket vanishes
\begin{align}
    [X_a, X_b] = i \frac{\mu}{g} \epsilon_{abc} X_c \quad (a,b,c \in \{1,2,3\}).
\end{align}

This is precisely the algebra of $SU(2)$ Lie algebra generators $J_a$ with standard commutation relations $[J_a, J_b] = i \epsilon_{abc} J_c$. Thus, the classical minima are given by
\begin{align}
    \langle X_a \rangle = \frac{\mu}{g} J_a \quad (a=1,2,3), \qquad \langle X_{i'} \rangle = 0 \quad (i'=4,\dots,9)
\end{align}
where $J_a$ form an $N \times N$ representation of the $SU(2)$ algebra.

Because $N \times N$ matrices can decompose into arbitrary direct sums of irreducible representations (irreps) of $SU(2)$, the classical moduli space consists of discrete ground state vacua. Below, we classify these vacua broadly into three categories.\\
(i) Trivial flat vacuum: This corresponds to setting $J_a = 0 \implies \langle X_a \rangle = 0$. This solution represents $N$ coincident $D0$-branes sitting at the origin. Perturbation theory (e.g., in Section \ref{sec3.1}) is formulated around this vacuum which is further elaborated there.\\
(ii) Fuzzy sphere vacua: This implies $J_a$ forms an $N$-dimensional irreducible representation of $SU(2)$. The matrices satisfy
\begin{align}
\label{ee2.11}
    \sum_{a=1}^3 X_a^2 = \left(\frac{\mu}{g}\right)^2 C_2(N) \mathbb{I}_{N \times N}
\end{align}
where $C_2(N) = \frac{N^2 - 1}{4}$ is the quadratic Casimir. Physically, this describes $D0$-branes polarized via the Myers effect \cite{Myers:1999ps} into a non-commutative fuzzy 2-sphere ($S^2$) of physical radius $R = \frac{\mu}{g} \sqrt{\frac{N^2 - 1}{4}} \approx \frac{\mu N}{2g}$.\\
(iii) Multi-Fuzzy sphere vacua: $J_a$ decomposes as a block-diagonal sum of representations $N = n_1 + n_2 + \dots + n_k$, representing multiple concentric or separated fuzzy spheres.\\\\
\underline{\textbf{The $U(1)$ symmetry and global charge:}}\\\\
In addition to the local $U(N)$ gauge symmetry, the model \eqref{e2.1} possesses a continuous global symmetry group
\begin{equation}
\label{ee2.12}
G_{\text{global}} = SO(3) \times SO(6) \times U(1).
\end{equation}

Here, $SO(3)$ acts as rotations on the $X_1, X_2, X_3$ matrices, while $SO(6)$ acts as rotations among the remaining scalar fields $X_4, \dots, X_9$. To establish the setup for symmetry-resolved observables, we select a continuous global $U(1)$ subgroup embedded within $SO(6)$, generated by a commuting charge operator $Q$. 

The generator $Q$ generates phase rotations in a chosen 2D subspace of $SO(6)$ (for instance, rotating $X_4$ into $X_5$). By construction, this global charge operator commutes with the complete BMN Hamiltonian
\begin{align}
\label{e2.3}
[H_{\text{BMN}}, Q] = 0.
\end{align}

To demonstrate elaborately the above Eq. \eqref{e2.3}, we examine how the continuous global $U(1)$ charge operator $Q$ acts on the scalar matrices $X_i$ and the Majorana fermions $\psi$. Consider a $U(1)$ subgroup embedded inside the $SO(6)$ R-symmetry group. This $U(1)$ generates spatial rotations in a chosen two-dimensional plane of the 6-dimensional transverse scalar space $(X_4, X_5, X_6, X_7, X_8, X_9)$.

Without loss of generality, let $Q$ be the generator of rotations in the $X_4\text{--}X_5$ plane\footnote{Since the BMN matrix model is a $0+1$-dimensional quantum mechanical theory, there is no spatial integration over a spatial manifold (unlike field theories in higher spacetime dimensions where $Q = \int d^{d-1}x \, j^0$). The charge is defined at a given time slice directly using the matrix trace over the conjugate momentum pairs $(\Pi_4, \Pi_5)$ and the fermionic charge $Q_{\text{fermion}}$.}
\begin{equation}
\label{e2.4}
Q = Q_{\text{boson}}+Q_{\text{fermion}}~;~Q_{\text{boson}}= \text{Tr} \left( X_4 \Pi_5 - X_5 \Pi_4 \right)
\end{equation}
where $\Pi_4 = \dot{X}_4$ and $\Pi_5 = \dot{X}_5$ are the canonical momentum matrices. In this paper, we will be only concerned with the bosonic component $Q_{\text{boson}}$ of the $U(1)$ generator. 

To demonstrate the explicit invariance under the continuous $U(1)$ transformation in Eq. \eqref{e2.1}, we apply the variation $\delta_\alpha$ directly to each component of the BMN Hamiltonian $H_{\text{BMN}}$. Under an infinitesimal rotation in the $X_4\text{--}X_5$ plane, one finds
\begin{equation}
\label{e2.5}
\delta_\alpha X_4 = -\alpha X_5, \quad \delta_\alpha X_5 = \alpha X_4, \quad \text{and} \quad \delta_\alpha X_i = 0 \quad (i \neq 4, 5).
\end{equation}

To understand why Eq. \eqref{e2.5} holds, we look at how an infinitesimal rotation matrix acts on the scalar field vector $(X_4, X_5)^T$. One can proceed in two ways.

Under a continuous $U(1) \cong SO(2)$ rotation in the $X_4\text{--}X_5$ plane by an infinitesimal angle $\alpha$, the finite rotation is given by
\begin{equation}
\begin{pmatrix} X_4' \\ X_5' \end{pmatrix} = \begin{pmatrix} \cos\alpha & -\sin\alpha \\ \sin\alpha & \cos\alpha \end{pmatrix} \begin{pmatrix} X_4 \\ X_5 \end{pmatrix}.
\end{equation}

Expanding for small $\alpha$ ($\cos\alpha \approx 1$ and $\sin\alpha \approx \alpha$) we find
\begin{align}
X_4' &\approx X_4 - \alpha X_5, \\
X_5' &\approx \alpha X_4 + X_5.
\end{align}

Taking the variation $\delta_\alpha X_i = X_i' - X_i$, this finally yields
\begin{equation}
\delta_\alpha X_4 = -\alpha X_5, \quad \delta_\alpha X_5 = \alpha X_4.
\end{equation}

Since the $U(1)$ generator $Q$ acts exclusively on the $X_4\text{--}X_5$ plane of the six transverse scalars ($X_4, \dots, X_9$), all remaining scalar fields $X_i$ for $i \neq 4, 5$ (including $X_1, X_2, X_3$ in the $SO(3)$ sector and $X_6, X_7, X_8, X_9$ in the $SO(6)$ sector) remain completely unchanged under this specific transformation, which implies
\begin{equation}
\delta_\alpha X_i = 0 \quad (i \neq 4, 5).
\end{equation}

Equivalently, using canonical commutation relations, the variation of any operator $\mathcal{O}$ under the global charge $Q$ is generated via the commutator
\begin{equation}
\delta_\alpha \mathcal{O} = i \alpha [Q, \mathcal{O}].
\end{equation}

Using \eqref{e2.4} and the equal-time commutation relation 
\begin{align}
\label{e2.12}
    [X_i(t)_{ab}, \Pi_j(t)_{cd}] = i \delta_{ij} \delta_{ad} \delta_{bc}
\end{align}
we obtain the following variation of the bosonic fields\footnote{We ignore the fermionic component of the $U(1)$ generator \eqref{e2.4} in this paper.}
\begin{align}
\label{e2.13}
\delta_\alpha X_4 &= i \alpha [ \text{Tr}(X_4 \Pi_5 - X_5 \Pi_4), X_4 ] = i \alpha ( - [ \text{Tr}(X_5 \Pi_4), X_4 ] ), \\
\delta_\alpha X_5 &= i \alpha [ \text{Tr}(X_4 \Pi_5 - X_5 \Pi_4), X_5 ] = i \alpha ( [ \text{Tr}(X_4 \Pi_5), X_5 ] ).
\label{e2.14}
\end{align}

In order to evaluate the above commutators \eqref{e2.13}-\eqref{e2.14}, we notice, for example
\begin{align}
   \delta_\alpha (X_4)_{dc}=-i\alpha[ \text{Tr}(X_5 \Pi_4), (X_4)_{dc} ] = -i\alpha\sum_{a,b} (X_5)_{ab} [ (\Pi_4)_{ba}, (X_4)_{dc} ].
\end{align}

Using \eqref{e2.12}, this finally yields
\begin{align}
    \delta_\alpha (X_4)_{dc}=-i \alpha \sum_{a,b} (X_5)_{ab} (-i \delta_{bc} \delta_{ad}) = -\alpha(X_5)_{dc}\Rightarrow  \delta_\alpha X_4 =- \alpha X_5.
\end{align}

On a similar note, evaluating the commutator in  \eqref{e2.14}, one finds 
\begin{align}
    \delta_\alpha X_5 = \alpha X_4.
\end{align}

Next, we consider variations of individual terms in \eqref{e2.1}. The variations of the individual kinetic, mass, and interaction terms proceed as follows:

\begin{enumerate}
    \item \textbf{Kinetic energy terms:} The conjugate momenta transform in identical fashion to the fields, $\delta_\alpha \Pi_4 = -\alpha \Pi_5$ and $\delta_\alpha \Pi_5 = \alpha \Pi_4$. Varying the sum of their squares yields
    \begin{align}
    \delta_\alpha \left[ \frac{1}{2} \text{Tr}(\Pi_4^2 + \Pi_5^2) \right] = \text{Tr}\left( \Pi_4 (-\alpha \Pi_5) + \Pi_5 (\alpha \Pi_4) \right) = 0.
    \end{align}

    \item \textbf{Mass terms:} The quadratic mass deformation term for the $SO(6)$ scalars contains the pair $\text{Tr}(X_4^2 + X_5^2)$. Its variation yields the following
    \begin{align}
    \delta_\alpha \left[ \frac{\mu^2}{8} \text{Tr}(X_4^2 + X_5^2) \right] = \frac{\mu^2}{4} \text{Tr}\left( X_4 (-\alpha X_5) + X_5 (\alpha X_4) \right) = 0.
    \end{align}

    \item \textbf{Quartic commutator potential:} The non-abelian potential includes terms coupling $(X_4, X_5)$ to any external scalar field $X_k$ ($k \neq 4, 5$), given by $\text{Tr}[X_4, X_k]^2 + \text{Tr}[X_5, X_k]^2$. Applying $\delta_\alpha$ gives the following
    \begin{align}
    &\delta_\alpha \left( \text{Tr}[X_4, X_k]^2 + \text{Tr}[X_5, X_k]^2 \right)\nonumber\\ &= 2 \, \text{Tr}\left( [X_4, X_k] [-\alpha X_5, X_k] \right) + 2 \, \text{Tr}\left( [X_5, X_k] [\alpha X_4, X_k] \right) \nonumber \\
    &= 2\alpha \left( -\text{Tr}\left( [X_4, X_k] [X_5, X_k] \right) + \text{Tr}\left( [X_5, X_k] [X_4, X_k] \right) \right) = 0,
    \end{align}
    where we used the cyclic property of the trace in each of the above examples.
    
    The term $\text{Tr}[X_4, X_5]^2$ is manifestly $SO(2)$-invariant since the commutator $[X_4, X_5]$ itself transforms as a scalar under rotations in the $4\text{--}5$ plane
    \begin{equation}
    \delta_\alpha [X_4, X_5] = [-\alpha X_5, X_5] + [X_4, \alpha X_4] = 0.
    \end{equation}

    \item \textbf{Myers cubic term and fermionic sector:} The Myers cubic interaction $\epsilon_{abc} X_a X_b X_c$ acts only on the $SO(3)$ scalars ($a,b,c \in \{1,2,3\}$) and is completely uncharged under rotations in the $4\text{--}5$ plane. Similarly, the fermionic mass and gauge couplings transform covariantly under the spin representation of $SO(6)$, leaving $H_{\text{fermion}}$ invariant.
\end{enumerate}

Combining these results confirms that $\delta_\alpha H_{\text{BMN}} = 0$, directly establishing the conservation law $[H_{\text{BMN}}, Q] = 0$. Following \cite{Caputa:2025ozd}, the relation $[H_{\text{BMN}}, Q] = 0$ implies that the total Hilbert space $\mathcal{H}$ decomposes into an orthogonal direct sum of conserved charge sectors
\begin{equation}
\mathcal{H} = \bigoplus_{q} \mathcal{H}_q, \quad \text{such that } Q |n, q\rangle = q |n, q\rangle,
\end{equation}
where $q \in \mathbb{R}$ denotes the eigenvalue of $Q$. This symmetry block-diagonalization guarantees that time evolution $e^{-i H_{\text{BMN}} t}$ does not induce transitions between different charge sectors $q$, providing the essential framework for studying symmetry-resolved spread complexity.
\subsection{Charged thermo-field double and state decomposition}
To study symmetry-resolved spread complexity at finite temperature, we consider a thermal state equipped with a chemical potential $\mu_Q$ associated with the conserved global charge $Q$. The appropriate initial state is the charged Thermo-Field Double (cTFD) state, defined on two identical decoupled copies (left $L$ and right $R$) of the BMN Hilbert space $\mathcal{H}_L \otimes \mathcal{H}_R$ 
\begin{align}
\label{e2.23}
|\text{cTFD}\rangle = \frac{1}{\sqrt{Z(\beta, \mu_Q)}} \sum_{n, q} e^{-\frac{\beta}{2} (E_{n,q} - \mu_Q q)} |n, q\rangle_L \otimes |\overline{n, q}\rangle_R,
\end{align}
where $\beta = 1/T$ is the inverse temperature, $Z(\beta, \mu_Q) = \text{Tr}\left(e^{-\beta(H_{\text{BMN}} - \mu_Q Q)}\right)$ is the grand canonical partition function, and $|n, q\rangle$ denotes the joint energy-charge eigenstates 
\begin{equation}
H_{\text{BMN}} |n, q\rangle = E_{n,q} |n, q\rangle, \quad Q |n, q\rangle = q |n, q\rangle.
\end{equation}

Here, $|\overline{n, q}\rangle_R = \mathcal{CPT} |n, q\rangle_R$ denotes the action of the discrete anti-unitary $\mathcal{CPT}$ operator (or effective charge-conjugation time-reversal symmetry) on the Hilbert space copy of the right system $\mathcal{H}_R$. In oter words, in the $\text{cTFD}$ construction \eqref{e2.23}, each state $\vert{}n, q\rangle_L$ in the left system is paired with a charge-conjugated time-reversed state $\vert{}\overline{n, q}\rangle_R = \mathcal{CPT} \vert{}n, q\rangle_R$ in the right system. The state relation $\vert{}\overline{n, q}\rangle_R = \mathcal{CPT} \vert{}n, q\rangle_R$ defines how a quantum state in the right system ($\mathcal{H}_R$) is constructed using the anti-unitary $\mathcal{CPT}$ operator acting on a basis state $\vert{}n, q\rangle_R$. Time-reversal ($\mathcal{T}$) reverses the direction of time. Because $\mathcal{T}$ is anti-unitary, it involves complex conjugation of amplitudes ($\mathcal{T} i \mathcal{T}^{-1} = -i$), which is essential for defining the Thermo-Field Double state properly in doubled Hilbert space approaches. Under time evolution, the total Hamiltonian is $H_{\text{total}} = H_L - H_R$ (or $H_L \otimes \mathbb{I}_R - \mathbb{I}_L \otimes H_R$). The minus sign on the Right subsystem appears precisely because the states on $R$ are anti-unitarily time-reversed ($t \to -t$). As a result, time evolving forward in real time on the Left system corresponds to time evolving backward on the Right system, keeping the entangled state $\vert{}\text{TFD}\rangle$ invariant under the combined boost generator $(H_L - H_R)\vert{}\text{TFD}\rangle = 0$.

Charge conjugation ($\mathcal{C}$) replaces particles with their corresponding antiparticles, effectively mapping a state to its charge-conjugate sector. Because $\mathcal{C}$ flips the sign of charge, the charge operator $Q$ anti-commutes (or is odd) under $\mathcal{C}$, that is, $\mathcal{C} Q \mathcal{C}^{-1} = -Q$. Acting on the conjugate state confirms this eigenvalue inversion
\begin{align}
    Q \left( \mathcal{C} \vert{}n, q\rangle \right) = \mathcal{C} \left( \mathcal{C}^{-1} Q \mathcal{C} \right) \vert{}n, q\rangle = \mathcal{C} (-Q) \vert{}n, q\rangle = -q \left( \mathcal{C} \vert{}n, q\rangle \right).
\end{align}

Unlike time-reversal ($\mathcal{T}$), which is anti-unitary and involves complex conjugation of scalars ($i \to -i$), pure charge conjugation $\mathcal{C}$ is a standard unitary (and Hermitian) operator, that is, $\mathcal{C}^\dagger = \mathcal{C}^{-1} = \mathcal{C}, \quad \mathcal{C}^2 = \mathbb{I}$. In the present paper, charge conjugation ensures global gauge/charge neutrality on the doubled Hilbert space ($\mathcal{H}_L \otimes \mathcal{H}_R$). Standard thermofield double states pair identical sectors $\vert{}n\rangle_L \otimes \vert{}n\rangle_R$. In a theory with chemical potential $\mu_Q$, this would cause the combined system to carry a net total charge $2q$. By applying $\mathcal{C}$ (alongside $\mathcal{PT}$) to the right subsystem, each state carrying charge $+q$ on the Left is paired with its charge conjugate state carrying $-q$ on the Right. This is further elaborated below.

Because the charge operator $Q$ is odd under time-reversal and charge conjugation ($\mathcal{CPT} \, Q \, \mathcal{CPT}^{-1} = -Q$), mapping a state to its conjugate flips the sign of its global $U(1)$ charge eigenvalue, that is,
\begin{equation}
Q_R |\overline{n, q}\rangle_R = Q_R \, \mathcal{CPT} |n, q\rangle_R = -\mathcal{CPT} \, Q_R |n, q\rangle_R = -q |\overline{n, q}\rangle_R.
\end{equation}

Consequently, each entangled pair $|n, q\rangle_L \otimes |\overline{n, q}\rangle_R$ in the cTFD sum pairs a positive charge sector $q$ in the left BMN matrix model with an equal and opposite charge sector $-q$ in the right theory. Defining the total global charge operator on the doubled Hilbert space as $Q_{\text{total}} = Q_L \otimes \mathbb{I}_R + \mathbb{I}_L \otimes Q_R$, its direct action on the cTFD state yields
\begin{align}
\label{e2.27}
Q_{\text{total}} |\text{cTFD}\rangle = \frac{1}{\sqrt{Z(\beta, \mu_Q)}} \sum_{n, q} e^{-\frac{\beta}{2} (E_{n,q} - \mu_Q q)} (q - q) \, |n, q\rangle_L \otimes |\overline{n, q}\rangle_R = 0.
\end{align}

Thus, the full entangled state $|\text{cTFD}\rangle$ resides strictly in the zero-charge sector of the bipartite system $\mathcal{H}_L \otimes \mathcal{H}_R$. This construction guarantees that the insertion of the chemical potential $\mu_Q$ does not violate the overall global charge conservation of the combined system, but rather introduces a net charge asymmetry between the two entangled sub-systems, establishing the precise background required for evaluating sub-system symmetry-resolved observables. Let us elaborate more on the above conceptual issue \eqref{e2.27}.

In the context of the charged Thermo-Field Double ($\text{cTFD}$) state construction, the above Eq. \eqref{e2.27} highlights a key conceptual advantage: achieving a finite chemical potential \emph{locally} while keeping the entire doubled system globally neutral. When working in a doubled Hilbert space $\mathcal{H}_L \otimes \mathcal{H}_R$, the total global charge operator $Q_{\text{total}}$ is defined as the sum of the charges in both subsystems. Because each state $\vert{}n, q\rangle_L$ in the left BMN matrix model is paired with its $\mathcal{CPT}$-conjugate state $\vert{}\overline{n, q}\rangle_R$ carrying charge $-q$ in the right theory, the total charge operator always annihilates the full state \eqref{e2.27}. This guarantees that the combined system strictly belongs to the zero-charge sector ($Q_{\text{total}} = 0$), preserving overall $U(1)$ symmetry without introducing any net global charge.

Even though the net charge of the entire system is zero, the introduction of the chemical potential $\mu_Q$ alters the thermal weights via the grand canonical factor $e^{-\frac{\beta}{2}(E_{n,q} - \mu_Q q)}$. States with a positive charge $q > 0$ on the left side are exponentially favored (weighted up by $+ \frac{\beta}{2}\mu_Q q$), while states with $q < 0$ are suppressed. Since the right side hosts the conjugate state carrying charge $-q$, a positive local charge expectation value on the left ($\langle Q_L \rangle > 0$) is exactly balanced by an equal and opposite negative local charge expectation value on the right ($\langle Q_R \rangle = -\langle Q_L \rangle < 0$). If an observer only has access to the left subsystem (by tracing out the right subsystem), the reduced density matrix $\rho_L = \text{Tr}_R \vert{}\text{cTFD}\rangle\langle\text{cTFD}\vert{}$ describes a genuine grand canonical ensemble with chemical potential $\mu_Q$ and finite charge density. This exact mechanism provides the physical background needed to compute symmetry-resolved quantities (such as sub-system spread complexity, Lanczos coefficients, or entanglement entropy) separately within individual charge sectors $q$, while maintaining global gauge and charge conservation across the full bipartite setup, which is what we explain next.

In the paper's setup, the time evolution of the system is governed by $H_{\text{total}} = H_L \otimes \mathbb{I}_R$ (often simply written as $H_{\text{total}} = H_L$) because the goal is to evaluate the forward real-time dynamics of a physical quantum state prepared on the Left BMN boundary theory, rather than studying an invariant thermal state under double-sided time evolution. When considering quantum complexity, operator growth, or state evolution after a quench, time evolution is actively generated on the physical system of interest (the Left subsystem) $\vert{}\psi(t)\rangle = e^{-i H_L t} \vert{}\text{cTFD}\rangle$. Because the Hamiltonian acts strictly as $H_L \otimes \mathbb{I}_R$, it applies real-time phase factors $e^{-i E_{n,q} t}$ exclusively to the states in $\mathcal{H}_L$, causing the initial entangled state $\vert{}\text{cTFD}\rangle$ to dynamically evolve away from its initial configuration over time.

To understand why $H_L$ is used here, it helps to compare the two different time-evolution operators commonly used in the Thermo-Field Double formalism. Let us first talk about two-sided boost generator $H_L - H_R$. This operator generates forward time evolution on the Left side and backward time evolution on the Right side. Because the states on the Right side are constructed using time-reversal ($\mathcal{T}$), the TFD state is an exact eigenstate of this generator with eigenvalue zero, that is,
\begin{align}
\label{e2.28}
    (H_L - H_R) \vert{}\text{cTFD}\rangle = 0.
\end{align}

In other words, if we evolve the state with $H_L - H_R$, the state remains static ($\vert{}\text{cTFD}(t)\rangle = \vert{}\text{cTFD}(0)\rangle$), which is useful for defining boost symmetries in AdS/CFT but yields trivial time evolution. Let us emphasize more on the above point \eqref{e2.28}.

Because the left and right theories are identical copies ($\mathcal{H}_L \otimes \mathcal{H}_R$), a paired basis state carries the same energy $E_{n,q}$ on both sides
\begin{align}
    H_L \vert{}n, q\rangle_L = E_{n,q} \vert{}n, q\rangle_L, \quad H_R \vert{}\overline{n, q}\rangle_R = E_{n,q} \vert{}\overline{n, q}\rangle_R.
\end{align}

Applying $H_L \otimes \mathbb{I}_R - \mathbb{I}_L \otimes H_R$ to each pair in the sum yields
\begin{align}
    (H_L \otimes \mathbb{I}_R - \mathbb{I}_L \otimes H_R) \left( \vert{}n, q\rangle_L \otimes \vert{}\overline{n, q}\rangle_R \right) = (E_{n,q} - E_{n,q}) \left( \vert{}n, q\rangle_L \otimes \vert{}\overline{n, q}\rangle_R \right) = 0.
\end{align}

Under the time evolution operator $U_{boost}(\tau) = e^{-i (H_L - H_R) \tau}$, the state remains completely unchanged for all times $\tau$
\begin{align}
    \vert{}\text{cTFD}(\tau)\rangle = e^{-i (H_L - H_R) \tau} \vert{}\text{cTFD}(0)\rangle = \vert{}\text{cTFD}(0)\rangle.
\end{align}
This generator evolves time forward on the left side ($t_L = +\tau$) while simultaneously evolving time backward on the right side ($t_R = -\tau$).

In gauge/gravity duality, the $\vert{}\text{cTFD}\rangle$ state describes an eternal charged black hole in Anti-de Sitter space, with the Left and Right theories residing on two asymptotic boundaries connected by an Einstein-Rosen bridge (wormhole). The operator $H_L - H_R$ corresponds to the global Killing vector generating boost symmetry along the black hole horizon. Because this generator leaves the state invariant, it cannot be used to study state complexity, operator growth, or chaos over time. To capture non-trivial dynamical evolution, such as the growth of the wormhole interior or Krylov spread complexity, one must evolve using the single-sided Hamiltonian $H_L \otimes \mathbb{I}_R$ instead. In other words, to model non-trivial time evolution, such as computing the survival probability (return amplitude) $\mathcal{R}^{(q)}(t) = \langle \text{cTFD} \vert{} e^{i H_L t} \vert{} \text{cTFD} \rangle$ and generating the Krylov basis for spread complexity, one must evolve the system forward using $H_L$ alone. Below, we elaborate them further in detail.

Under unitary time evolution generated by the total Hamiltonian $H_{\text{total}} = H_L \otimes \mathbb{I}_R$, the time-dependent state $|\psi(t)\rangle = e^{-i H_L t} |\text{cTFD}\rangle$ evolves as
\begin{equation}
|\psi(t)\rangle = \frac{1}{\sqrt{Z(\beta, \mu_Q)}} \sum_{n, q} e^{-i E_{n,q} t} e^{-\frac{\beta}{2} (E_{n,q} - \mu_Q q)} |n, q\rangle_L \otimes |\overline{n, q}\rangle_R.
\end{equation}

Since $H_{\text{BMN}}$ commutes with $Q$, the total Hilbert space block-diagonalizes as $\mathcal{H} = \bigoplus_q \mathcal{H}_q$. Consequently, the time-evolved state $|\psi(t)\rangle$ naturally decomposes into orthogonal charge sectors $q$, which could be expressed as \cite{Caputa:2025ozd}
\begin{equation}
\label{eee2.43}
|\psi(t)\rangle = \sum_q \sqrt{p(q)} \, |\psi_q(t)\rangle,
\end{equation}
where the normalized symmetry-resolved state $|\psi_q(t)\rangle \in \mathcal{H}_q \otimes \mathcal{H}_{-q}$ within a fixed charge sector $q$ is given by
\begin{equation}
|\psi_q(t)\rangle = \frac{1}{\sqrt{Z_q(\beta)}} \sum_n e^{-i E_{n,q} t} e^{-\frac{\beta}{2} E_{n,q}} |n, q\rangle_L \otimes |\overline{n, q}\rangle_R,
\end{equation}
with the canonical sector partition function $Z_q(\beta) = \sum_n e^{-\beta E_{n,q}}$.

The probability $p(q)$ of finding the system in the sector with charge $q$ is independent of time and is governed by the grand canonical weighting
\begin{equation}
\label{ee2.45}
p(q) = \frac{e^{\beta \mu_Q q} Z_q(\beta)}{Z(\beta, \mu_Q)}, \quad \text{such that} \quad \sum_q p(q) = 1.
\end{equation}

The above Eq.\eqref{ee2.45} accurately defines the probability distribution $p(q)$ of finding the system in a specific global $U(1)$ charge sector $q$ for a thermal state equipped with a chemical potential $\mu_Q$. The numerator $e^{\beta \mu_Q q} Z_q(\beta)$ weights each charge sector correctly using the Boltzmann factor adjusted for the chemical potential and the sector partition function $Z_q(\beta) = \sum_n e^{-\beta E_{n,q}}$. Dividing by the grand partition function $Z(\beta, \mu_Q) = \sum_{q} e^{\beta \mu_Q q} Z_{q}(\beta)$ strictly guarantees the normalization condition $\sum_{q} p(q) = 1$.

This orthogonal decomposition $|\psi(t)\rangle = \sum_q \sqrt{p(q)} |\psi_q(t)\rangle$ ensures that the return amplitude (survival probability) factorizes across individual charge sectors, establishing the basis for constructing Krylov baselines and computing symmetry-resolved Lanczos coefficients $a_n^{(q)}, b_n^{(q)}$ separately within each charge sector $q$.
\section{Krylov space construction and low temperature phase}
\label{sec 3}
The Lanczos algorithm iteratively applies $H_L$ to the initial state $\vert{}\text{cTFD}\rangle$ to generate the Krylov sub-space vectors. Because $H_{\text{BMN}}$ commutes with $Q$, this evolution stays strictly within each sector $q$, allowing the Lanczos coefficients $a_n^{(q)}$ and $b_n^{(q)}$ to be evaluated independently for each charge sector. The calculation of the Lanczos coefficients $a_n^{(q)}$ and $b_n^{(q)}$ within Section 2.3 for a fixed charge sector $q$ follows the standard Gram-Schmidt orthogonalization of the Krylov subspace generated by single-sided Hamiltonian evolution $H_{\text{total}} = H_L \otimes \mathbb{I}_R$.

The return amplitude (survival probability) for the initial symmetry-resolved state $\vert{}\psi_q(0)\rangle \in \mathcal{H}_q \otimes \mathcal{H}_{-q}$ evolving under $H_L$ is given by
\begin{align}
\label{e3.1}
    \mathcal{R}^{(q)}(t)=\braket{\psi_q(t)|\psi_q(0)} = \langle \psi_q(0) \vert{} e^{i H_L t} \vert{} \psi_q(0) \rangle = \frac{1}{Z_q(\beta)} \sum_{n} e^{-\beta E_{n,q}} e^{i E_{n,q} t}.
\end{align}

Expanding $\mathcal{R}^{(q)}(t)$ in a Taylor series around $t = 0$ yields the symmetry-resolved moments $\mathcal{M}_k^{(q)}$ that can be expressed in the form of a series 
\begin{align}
\label{e3.2}
    \mathcal{R}^{(q)}(t) = \sum_{k=0}^{\infty} \frac{i^k}{k!} \mathcal{M}_k^{(q)} t^k.
\end{align}

The moments $\mathcal{M}_k^{(q)} = \langle \psi_q(0) \vert{} H_L^k \vert{} \psi_q(0) \rangle$ are evaluated as thermal expectation values in sector $q$. A straightforward computation further reveals
\begin{align}
\label{e2.38}
    \mathcal{M}_k^{(q)} = \frac{1}{Z_q(\beta)} \sum_{n} (E_{n,q})^k e^{-\beta E_{n,q}} = (-1)^k \frac{\partial^k}{\partial \beta^k} \ln Z_q(\beta) + \cdots.
\end{align}

For $k \ge 2$, the derivative $(-1)^k \frac{\partial^k}{\partial \beta^k} \ln Z_q(\beta)$ equals the $k$-th cumulant $\kappa_k^{(q)}$ (the connected part of the moment\footnote{In statistical mechanics, cumulants ($\kappa_k^{(q)}$) describe the irreducible, connected fluctuations of the energy spectrum within a fixed charge sector $q$, whereas raw uncentered moments ($\mathcal{M}_k^{(q)}$) include contributions from all lower-order background averages. The logarithmic derivatives of the sector partition function $\ln Z_q(\beta)$ naturally generate these pure connected cumulants. Through Faà di Bruno's formula, the lower-order terms (such as products of $\partial_\beta \ln Z_q$) account precisely for the combinatorial shift required to reconstruct the full uncentered raw moments from the connected cumulants. In physics and statistical mechanics, when we have many particles interacting, raw moments (like the third moment $\mathcal{M}^{(q)}_3$) often mix different things together. For example, the third raw moment contains contributions from the average energy cubed, the average energy times the variance, and the actual pure 3-particle interaction.}). Here, the remaining terms in \eqref{e2.38} accounts for the products of lower-order derivatives required to convert the connected cumulant $\kappa_k^{(q)}$ back into the full uncentered raw moment $\mathcal{M}_k^{(q)}$ via Faà di Bruno's formula / polynomial moment-cumulant relations. For example, first three moments can be obtained as
\begin{align}
    &\mathcal{M}_1^{(q)} = -\frac{\partial}{\partial \beta} \ln Z_q(\beta)\\
    &\mathcal{M}_2^{(q)} = \frac{\partial^2}{\partial \beta^2} \ln Z_q(\beta) + \left( \frac{\partial}{\partial \beta} \ln Z_q(\beta) \right)^2\\
    &\mathcal{M}_3^{(q)} = -\frac{\partial^3}{\partial \beta^3} \ln Z_q(\beta) - 3 \left(\frac{\partial^2}{\partial \beta^2} \ln Z_q(\beta)\right) \left(\frac{\partial}{\partial \beta} \ln Z_q(\beta)\right) - \left( \frac{\partial}{\partial \beta} \ln Z_q(\beta) \right)^3.
    \label{ee3.6}
\end{align}

The Lanczos algorithm generates an orthonormal basis $\{\vert{}K_n^{(q)}\rangle\}$ starting from $\vert{}K_0^{(q)}\rangle = \vert{}\psi_q(0)\rangle$, with $b_0^{(q)} = 0$. The first diagonal coefficient is given by the following matrix element
\begin{align}
\label{ee3.7}
    a_0^{(q)} = \langle K_0^{(q)} \vert{} H_L \vert{} K_0^{(q)} \rangle = \mathcal{M}_1^{(q)}.
\end{align}

First off-diagonal coefficient is given by 
\begin{align}
    \vert{}A_1^{(q)}\rangle = (H_L - a_0^{(q)}) \vert{}K_0^{(q)}\rangle, \quad b_1^{(q)} = \sqrt{\langle A_1^{(q)} \vert{} A_1^{(q)} \rangle} = \sqrt{\mathcal{M}_2^{(q)} - (\mathcal{M}_1^{(q)})^2}
\end{align}
which yields the $n=1$ Krylov basis element as $\vert{}K_1^{(q)}\rangle = \frac{1}{b_1^{(q)}} \vert{}A_1^{(q)}\rangle$.

General recurrence relation for $n \ge 1$ is given by
\begin{align}
    &\vert{}A_{n+1}^{(q)}\rangle = (H_L - a_n^{(q)}) \vert{}K_n^{(q)}\rangle - b_n^{(q)} \vert{}K_{n-1}^{(q)}\rangle\\
    &a_n^{(q)} = \langle K_n^{(q)} \vert{} H_L \vert{} K_n^{(q)} \rangle, \quad b_{n+1}^{(q)} = \sqrt{\langle A_{n+1}^{(q)} \vert{} A_{n+1}^{(q)} \rangle}.
    \label{ee3.10}
\end{align}

   \label{e3.12}
\subsection{The perturbative spectrum in Large $N$ limit}
\label{sec3.1}
 To connect these abstract thermodynamic formulas to explicit matrix model parameters, we evaluate the partition function $Z_q(\beta)$ directly from the BMN matrix model action in the planar and/or large-$N$ limit. The BMN matrix model is a one-dimensional $U(N)$ gauge theory with 6 scalar fields and 16 real fermions, parameterized by - (i) Flux parameter $\mu$: sets the mass scale/frequency of the harmonic modes and (ii) Effective coupling $g_{\text{eff}} = \lambda/\mu^3$ (where $\lambda = g^2 N \ll 1$ is the 't Hooft coupling): controls non-planar interaction corrections.

For global $U(1)$ R-charge sector $q$ at inverse temperature $\beta$, the multi-particle trace of the BMN spectrum decomposes into harmonic oscillator modes with effective frequencies
\begin{align}
\label{e2.48}
    \omega_i(q) =\mu \sqrt{1 + \frac{\lambda}{2N\mu^3} f_i(q)}.
\end{align}

To show the derivation of \eqref{e2.48}, we start from the effective action and potential of the BMN (Berenstein-Maldacena-Nastase) matrix model in a fixed global $U(1)$ R-charge sector $q$ with non-planar interaction corrections. In the free limit ($\lambda = 0$), the scalar field matrix degrees of freedom $X_i(t)$ ($i=1,\dots,9$) in the Berenstein-Maldacena-Nastase (BMN) matrix model \eqref{e2.1} are governed by quadratic kinetic and mass terms in the Hamiltonian 
\begin{align}
    H_0 = \text{Tr} \left[ \frac{1}{2} \sum_{i=1}^9 \Pi_i^2 + \frac{\mu^2}{2} \sum_{a=1}^3 X_a^2 + \frac{\mu^2}{8} \sum_{i'=4}^9 X_{i'}^2 +\text{fermions}\right].
\end{align}

Here, $\Pi_i = \dot{X}_i$ are the canonical momentum matrices, and $\mu$ is the mass/flux scale. Expanding the fields into harmonic mode operators $a_{i, n}$ and $a_{i, n}^\dagger$, the bare frequency squared for the transverse scalars $X_{i'}$ ($i'=4,\dots,9$) is $\omega_0^2 \sim \mu^2$. In other words, in the free or quadratic limit ($\lambda = 0$), the scalar and fermionic fields of the BMN matrix model act as decoupled harmonic oscillators in a flux background $\mu$. The quadratic scalar mass terms from Eq. \eqref{e2.1} give the unperturbed mass/frequency squared for all scalar modes.

When turning on the coupling $g$, the quartic interaction term in the BMN Hamiltonian \eqref{e2.1} generates a potential contribution of the form\footnote{Note that the cubic Myers term $H_{\text{Myers}} \propto i \mu g \epsilon_{abc} \text{Tr}(X_a X_b X_c)$ in \eqref{e2.1} is omitted from the interaction potential $V_{\text{int}}$ in \eqref{e3.15}. The global charge generator $Q$ is chosen from the $SO(6)$ symmetry group and acts exclusively on the transverse 2D plane (e.g., $X_4 - X_5$). Since $H_{\text{Myers}}$ involves exclusively the $SO(3)$ scalars $(X_1, X_2, X_3)$, it commutes with the $U(1)$ charge generator $Q \in SO(6)$ and produces no charge-dependent quadratic mass shifts for the charged modes at leading order in $g_{\text{eff}}$. In other words, $Q$ and $H_{\text{Myers}}$ act on completely disjoint field subspaces, $[Q, H_{\text{Myers}}] = 0$, and the Myers term carries zero $U(1)$ charge. Consequently, it does not induce charge-dependent self-energy corrections or mass renormalizations for the charged modes ($X_4, X_5$) at first order. The cubic Myers term $\sim g \mu \text{Tr}(X^3)$ can only contribute to quadratic self-energies at second order in perturbation theory via tree/loop exchanges ($\sim g^2 \mu^2$). In a leading 1-loop calculation (or leading order in the effective parameter $g_{\text{eff}} = \frac{g^2 N}{\mu^3}$), its contribution to the $U(1)$ sector is subleading or vanishes by trace properties $\text{Tr}(T^a T^b T^c) = 0$ for $U(1)$ projections.} 
\begin{align}
\label{e3.15}
    V_{\text{int}} = -\frac{g^2}{4} \sum_{i,j=1}^9 \text{Tr} [X_i, X_j]^2.
\end{align}

In the 't Hooft large-$N$ limit, the effective expansion parameter is the 't Hooft coupling $\lambda = g^2 N$. Expanding the scalar field $X_k$ into its harmonic components $a_k, a_k^\dagger$ in the $U(1)$ R-charge sector carrying charge $q$, the background expectation values or non-planar self-energy loops modify the effective quadratic term for mode $i$. The continuous global $U(1)$ symmetry acts on the chosen 2D plane (e.g., $X_4 - X_5$) generated by $Q$. Projected onto a sector with definite charge $q$, the interaction term generates a state-dependent self-energy shift $\Omega_i(q)$ to the quadratic operator
\begin{align}
\label{e2.51}
    V_{\text{eff}}= \frac{1}{2} \Omega_i(q) X_i^2.
\end{align}

To derive and mathematically elaborate how the $U(1)$ R-charge sector project leads to the effective quadratic shift operator \eqref{e2.51}, we break the construction down into state creation, charge projection, and self-energy operator expansion. Consider the two scalar matrix coordinates parameterizing the chosen 2D plane (e.g., $X_4$ and $X_5$). We construct the complex ladder operators for the $U(1)$ charged modes
\begin{align}
\label{e2.52}
    &a = \frac{1}{\sqrt{2\omega_0}} \left( \omega_0 (X_4 + i X_5) + i (\Pi_4 + i \Pi_5) \right)\\
    &a^\dagger = \frac{1}{\sqrt{2\omega_0}} \left( \omega_0 (X_4 - i X_5) - i (\Pi_4 - i \Pi_5) \right).
\end{align}

The definitions for $b$ and $b^\dagger$ (the annihilation and creation operators for the negatively charged modes in the $X_4 - X_5$ plane) can be written by taking the conjugate combination
\begin{align}
    &b = \frac{1}{\sqrt{2\omega_0}} \left( \omega_0 (X_4 - i X_5) + i (\Pi_4 - i \Pi_5) \right)\\
    &b^\dagger = \frac{1}{\sqrt{2\omega_0}} \left( \omega_0 (X_4 + i X_5) - i (\Pi_4 + i \Pi_5) \right).
    \label{e2.55}
\end{align}

While $a^\dagger$ creates a quantum with positive $U(1)$ charge $+1$, $b^\dagger$ creates a quantum with negative charge $-1$. They satisfy the following commutation relations
\begin{align}
    [a, a^\dagger] = 1, \quad [b, b^\dagger] = 1, \quad [a, b] = [a, b^\dagger] = 0.
\end{align}

Using \eqref{e2.52}-\eqref{e2.55}, one finds
\begin{align}
\label{ee3.22}
    &X_4 = \frac{1}{\sqrt{8\omega_0}} \left( a + a^\dagger + b + b^\dagger \right)~;~X_5 = \frac{-i}{\sqrt{8\omega_0}} \left( a - a^\dagger - b + b^\dagger \right)\\
    &\Pi_4 = -i \sqrt{\frac{\omega_0}{8}} \left( a - a^\dagger + b - b^\dagger \right)~;~\Pi_5 = -\sqrt{\frac{\omega_0}{8}} \left( a + a^\dagger - b - b^\dagger \right).
\end{align}

Substituting $X_4, X_5, \Pi_4, \Pi_5$ into the charge generator $Q(=Q_L)$ \eqref{e2.4}, one finds\footnote{This is subject to an overall rescaling by a suitable numerical pre-factor. Here, the global charge corresponds to the Left sector of the bosonic component of $Q_{\text{total}}$ pertinent to the Left BMN matrix model.}
\begin{align}
\label{e3.24}
    Q\equiv Q^{(L)}_{\text{boson}} = a^\dagger a - b^\dagger b=n_a-n_b
\end{align}
where $a^\dagger, a$ and $b^\dagger, b$ are the creation and annihilation operators for the positive and negative charge modes, corresponding to the Left BMN matrix model.

Having established the continuous $U(1)$ charge generator $Q$ \eqref{e3.24}, we now proceed to project the complex scalar field sector onto a state with a definite $U(1)$ global charge $q$, leading directly to the intermediate state creation and the effective quadratic operator \eqref{e2.51}. Consider a general unprojected Fock state created by acting with positive-charge creation operators $a^\dagger$ and negative-charge creation operators $b^\dagger$ on the vacuum state $|0_q\rangle$
\begin{equation}
\label{eee3.23}
|n_a, n_b\rangle = \frac{(a^\dagger)^{n_a} (b^\dagger)^{n_b}}{\sqrt{n_a! n_b!}} |0_q\rangle.
\end{equation}

The charge generator $Q$ acts on this state as an eigenvalue operator
\begin{align}
Q \lvert n_a, n_b \rangle =(n_a - n_b) \lvert n_a, n_b \rangle .
\end{align}

To project onto the fixed charge sector $Q = q$, we enforce the \emph{net} charge $n_a - n_b = q$. Expressing this in terms of $n_a =\frac{1}{2} (n+q)$ and $n_b =\frac{1}{2} (n -q)$, the normalized state carrying a total particle number $n$ and a net global charge $q$ in the Fock basis is given by
\begin{align}
\label{e3.27}
\lvert \psi_q (n) \rangle = \frac{(a^\dagger)^{\frac{1}{2} (n+q)} (b^\dagger)^{\frac{1}{2} (n-q)}}{\sqrt{n_a! n_b!}} \lvert 0_q \rangle .
\end{align}\\
\underline{\textbf{Effective quadratic shift operator:}}\\\\
In the presence of non-planar background expectation values or non-zero self-energy loops in the charge sector $q$, the interaction Hamiltonian density receives corrections from scalar field fluctuations $X_4$ and $X_5$. In other words, when interaction coupling ($\lambda$) is turned on in the Berenstein-Maldacena-Nastase (BMN) matrix model, non-planar interaction effects (represented by vacuum/self-energy loops) modify the mass and potential terms of the scalar fields $X_4$ and $X_5$. These interactions generate an effective quadratic shift operator, $V_{\text{eff}} = \frac{1}{2}\Omega_i(q) X_i^2$, where $\Omega_i(q)$ acts as a state-dependent self-energy mass-shift parameter. In the following, we elaborate this in detail.

The first step is to express $X_4$ and $X_5$ in terms of the complex ladder operators $a, a^\dagger$ (positively charged modes) and $b, b^\dagger$ (negatively charged modes) and finally obtain the sum of squared transverse fields $X_4^2 + X_5^2$ in terms of ladder operators using \eqref{e2.52}--\eqref{e2.55}
    \begin{align}
    X_4 + iX_5 &= \frac{1}{\sqrt{2\omega_0}} \left( a + b^\dagger \right) , \\
    X_4 - iX_5 &= \frac{1}{\sqrt{2\omega_0}} \left( a^\dagger + b \right) .
    \end{align}

Taking their product yields the following combination
    \begin{align}
    X_4^2 + X_5^2 = \frac{1}{2\omega_0} \left( a a^\dagger + a b + b^\dagger a^\dagger + b^\dagger b \right) .
    \end{align}

Applying the canonical commutation relations $[a, a^\dagger] = 1$ and $[b, b^\dagger] = 1$ (rewriting $a a^\dagger = a^\dagger a + 1$), one finally obtains the following combination
    \begin{align}
    X_4^2 + X_5^2 = \frac{1}{2\omega_0} \left( a^\dagger a + b^\dagger b + 1 + a b + a^\dagger b^\dagger \right) .
    \end{align}

To evaluate the interaction correction in a definite global charge sector, we project onto the state $\lvert \psi_q(n) \rangle$ defined in \eqref{e3.27}, which satisfies $Q \lvert \psi_q(n) \rangle = q \lvert \psi_q(n) \rangle$. This imposes the relation on the occupation numbers $b^\dagger b = a^\dagger a - q$.

Taking the expectation value with respect to $\lvert \psi_q(n) \rangle$, the off-diagonal terms $ab$ and $a^\dagger b^\dagger$ drop out because they change the charge sector ($\Delta Q = \pm 2$). To understand why the expectation values $\langle \psi_q(n) \lvert ab \rvert \psi_q(n) \rangle$ and $\langle \psi_q(n) \lvert a^\dagger b^\dagger \rvert \psi_q(n) \rangle$ vanish identically, we can look at both the charge conservation argument and the explicit state action.

The commutation relations of the ladder operators with the charge operator $Q$ are
\begin{align}
    [Q, a] = -a, \quad [Q, b^\dagger] = -b^\dagger, \quad [Q, b] = b, \quad [Q, a^\dagger] = a^\dagger.
\end{align}

Evaluating the commutator for the product operators for $ab$ we find
\begin{align}
    [Q, ab] = a[Q, b] + [Q, a]b  = 0.
\end{align}

This shows that the state $ab \ket{0}$ has a zero \emph{net} change ($\Delta Q$) in charge. Since $a$ decreases $N_a$ by 1 ($\Delta Q = -1$) and $b$ decreases $N_b$ by 1 ($\Delta Q = +1$), the operator $ab$ changes the net charge by $\Delta Q=-2$. Let's look at the individual particle creation/annihilation numbers. The state $\lvert \psi_q(n) \rangle$ has $n$ quanta of $a$ (particle) and $(n-q)$ quanta of $b$ (anti-particle).

The operator $ab$ annihilates one $a$-quantum and one $b$-quantum. Therefore,
\begin{align}
    ab \lvert \psi_q(n) \rangle \propto \lvert n-1, (n-q)-1 \rangle = \lvert n-1, n-q-1 \rangle.
\end{align}
    
The net charge of this transformed state becomes
\begin{align}
    Q' = (n-1) - (n-q+1) = q-2.
\end{align}

However, notice the total particle number (or occupation number $n$) changes from $n \to n-1$. On the other hand, the Fock states with different occupation numbers are orthogonal, that is, $\langle n'_a, n'_b \mid n_a, n_b \rangle = \delta_{n'_a, n_a} \delta_{n'_b, n_b}$. Let's explicitly evaluate the matrix elements. Recall the standard actions of harmonic oscillator annihilation operators $a$ and $b$ on a two-mode Fock state $\lvert n_a, n_b \rangle$
\begin{align}
    &a \lvert n_a, n_b \rangle = \sqrt{n_a} \, \lvert n_a - 1, n_b \rangle\\
    &b \lvert n_a, n_b \rangle = \sqrt{n_b} \, \lvert n_a, n_b - 1 \rangle.
\end{align}

Because the operators $a$ and $b$ act on different modes (different harmonic oscillators), they commute: $[a, b] = 0$. Applying $b$ to the state $\lvert \psi_q(n) \rangle = \lvert n, n-q \rangle$ yields
\begin{align}
    b \lvert \psi_q(n) \rangle = b \lvert n, n-q \rangle = \sqrt{n-q} \, \lvert n, n-q-1 \rangle.
\end{align}

Finally, applying $a$ to the resulting state, we find
\begin{align}
    &ab \lvert \psi_q(n) \rangle = \sqrt{n(n-q)} \, \lvert \psi_q(n-1) \rangle\\
    &\lvert \psi_q(n-1) \rangle = \lvert n-1, n-q-1 \rangle.
\end{align}

Taking the inner product with $\langle \psi_q(n) \rvert$, we finally obtain
\begin{align}
    \langle \psi_q(n) \rvert ab \lvert \psi_q(n) \rangle = \sqrt{n(n-q)} \underbrace{\langle \psi_q(n) \mid \psi_q(n-1) \rangle}_{= 0} = 0.
\end{align}

Similarly for $a^\dagger b^\dagger$, we find
\begin{align}
    a^\dagger b^\dagger \lvert \psi_q(n) \rangle = \sqrt{(n+1)(n-q+1)} \, \lvert \psi_q(n+1) \rangle.
\end{align}

Taking the inner product with $\langle \psi_q(n) \rvert$ we find
\begin{align}
    \langle \psi_q(n) \rvert a^\dagger b^\dagger \lvert \psi_q(n) \rangle = \sqrt{(n+1)(n-q+1)} \underbrace{\langle \psi_q(n) \mid \psi_q(n+1) \rangle}_{= 0} = 0.
\end{align}

In summary, the off-diagonal operators $ab$ and $a^\dagger b^\dagger$ map the state $\lvert \psi_q(n) \rangle$ to orthogonal states in the Fock space ($\lvert \psi_q(n-1) \rangle$ and $\lvert \psi_q(n+1) \rangle$, respectively). Because the unperturbed Hilbert space basis vectors with different excitation numbers $n$ are mutually orthogonal, their diagonal expectation values vanish. This leaves only the diagonal number-conserving terms ($a^\dagger a$, $b^\dagger b$, and $1$) contributing to the expectation value in sector $q$.

Thus, only the diagonal expectation values survive giving
\begin{align}
    \langle \psi_q(n) \rvert X_4^2 + X_5^2 \lvert \psi_q(n) \rangle = \frac{1}{2\omega_0} \Bigg[ \langle \psi_q(n) \rvert a^\dagger a \lvert \psi_q(n) \rangle + \langle \psi_q(n) \rvert b^\dagger b \lvert \psi_q(n) \rangle + 1 \Bigg].
\end{align}

The state $\lvert \psi_q(n) \rangle$ is a simultaneous eigenstate of the number operators $N_a = a^\dagger a$ and $N_b = b^\dagger b$, that is, 
\begin{align}
    a^\dagger a \lvert \psi_q(n) \rangle = n \lvert \psi_q(n) \rangle ~;~b^\dagger b \lvert \psi_q(n) \rangle = (n - q) \lvert \psi_q(n) \rangle.
\end{align}

Plugging the eigenvalues directly into the sum, finally yields
\begin{align}
\label{e3.46}
 \braket{X^2_4+X^2_5}_q=  \langle \psi_q(n) \rvert X_4^2 + X_5^2 \lvert \psi_q(n) \rangle  = \frac{1}{2\omega_0} (2n - q + 1).
\end{align}

As mentioned before, in the non-interacting (bare) limit ($\lambda = 0$), the quadratic potential for a single transverse scalar field component $X_i$ (where $i \in \{4, 5\}$) is given by the standard harmonic oscillator mass term $V_0 = \frac{1}{2} \omega_0^2 X_i^2$, where $\omega_0$ represents the bare mass/frequency parameter of the scalar field. In the BMN matrix model (or scalar field theory on a background), the leading-order interaction between the scalar fields takes a quartic form coupled by $\lambda$, which we denote as the effective potential in the Hartree-Fock channel that represents the coupling of traverse fluctuations with background charge density
\begin{align}
\label{e3.47}
    V_{\text{eff}} = \frac{\lambda}{2N}  \braket{X^2_4+X^2_5}_q\sum_i X_i^2.
\end{align}

When analyzing small fluctuations around a background state in a specific global charge sector $\lvert \psi_q(n) \rangle$, quantum loops or background field expectation values modify this interaction. Let us elaborate further on this. In the Berenstein-Maldacena-Nastase (BMN) matrix model (or plane-wave matrix theory), the scalar fields $X_i$ ($i = 1, \dots, 9$) originate from the spatial components of ten-dimensional Yang-Mills gauge fields. 

In dimensional reduction to 1 dimension (quantum mechanics), the gauge field commutator $[X_i, X_j]$ generates the scalar potential \eqref{e3.15}. When we isolate a specific pair of transverse scalars, here, $X_4$ and $X_5$, to form a complex state carrying global charge $q$ (under $U(1) \subset SO(6)$ rotations in the $4\text{--}5$ plane), their squared amplitude $X_4^2 + X_5^2$ acts as an effective classical background or charge density. This could be thought of as follows. By isolating two real scalar fields, $X_4$ and $X_5$, you can form a single complex scalar field $Z = \frac{1}{\sqrt{2}}(X_4 + i X_5)$. Rotations in the $4\text{--}5$ plane correspond to a $U(1)$ subgroup of the full $SO(6)$ R-symmetry group. Under a $U(1)$ rotation by an angle $\theta$, the complex field transforms as $Z \to e^{i\theta} Z$. When a state $\vert{}\psi_q(n)\rangle$ is constructed to carry a fixed global charge $q$, it contains a macroscopic or macroscopic-like number of excitations ($n - (n - q) = q$) in this $U(1)$ sector. As a result, when evaluating quantum expectation values in a state $\vert{}\psi_q(n)\rangle$ with large charge $q$ or high excitation number $n$, operator combinations like $X_4^2 + X_5^2 = 2 Z^\dagger Z$ have a non-vanishing diagonal expectation value \eqref{e3.46}. Because the charge $q$ sets a baseline non-zero value, replacing $X_4^2 + X_5^2$ with its quantum expectation value acts effectively as a constant or classical background field density $\rho_q \propto \langle X_4^2 + X_5^2 \rangle_q$. This shifts the quadratic potential (mass-squared parameter) experienced by all other small scalar fluctuations propagating in that state as further elaborated below.

In a state $\vert{}\psi_q(n)\rangle$ carrying a large global charge $q$, the fields $X_4$ and $X_5$ develop a macroscopic quantum expectation value \eqref{e3.46}. Because $q \gg 1$, the quantum fluctuations around this value are suppressed relative to the mean, allowing us to approximate $X_4^2 + X_5^2$ by its classical background/expectation value $\langle X_4^2 + X_5^2 \rangle_q$.

The standard BMN/matrix model potential originating from the 10D gauge theory commutator $[X_i, X_j]^2$ is given by \eqref{e3.15}
\begin{align}
    V_{\text{int}} = -\frac{\lambda}{4N} \sum_{i,j=1}^{9} \text{Tr}\left( [X_i, X_j]^2 \right).
\end{align}

We want to isolate the interactions between the background fields $(X_4, X_5)$ and all other transverse directions $X_i$ ($i = 1, \dots, 9$). In the summation over $i$ and $j$, consider the cross-terms where one index is in $\{4, 5\}$ and the other index $i$ ranges over all 9 fields
\begin{align}
\label{e3.49}
    V_{\text{int}} = -\frac{\lambda}{4N} \sum_{i=1}^{9} \text{Tr}\left( [X_4, X_i]^2 + [X_5, X_i]^2 \right).
\end{align}

For any two matrix fields $A$ and $B$, the trace of the commutator squared expands as
\begin{align}
     -\text{Tr}\left( [A, B]^2 \right) &= -\text{Tr}\left( (AB - BA)(AB - BA) \right) \nonumber\\ &= -\text{Tr}\left( ABAB - ABBA - BAAB + BABA \right).
\end{align}

Using the cyclic property of the trace
\begin{align}
    \text{Tr}(BABA) = \text{Tr}(ABAB)~;~\text{Tr}(BAAB) = \text{Tr}(ABBA) = \text{Tr}(A^2 B^2)
\end{align}
it is straightforward to show that 
\begin{align}
    \text{Tr}\left( [A, B]^2 \right) = 2\,\text{Tr}(ABAB) - 2\,\text{Tr}(A^2 B^2).
\end{align}

Setting $(A, B) = (X_4, X_i)$ and $(X_5, X_i)$ yields
\begin{align}
    &\text{Tr}\left( [X_4, X_i]^2 \right) = 2\,\text{Tr}(X_4 X_i X_4 X_i) - 2\,\text{Tr}(X_4^2 X_i^2)\\
    &\text{Tr}\left( [X_5, X_i]^2 \right) = 2\,\text{Tr}(X_5 X_i X_5 X_i) - 2\,\text{Tr}(X_5^2 X_i^2).
\end{align}

Substituting these two back into $V_{\text{int}}$ \eqref{e3.49}, we obtain
\begin{align}
    V_{\text{int}} = -\frac{\lambda}{2N} \sum_{i=1}^{9} \Bigg[ \,\text{Tr}(X_4 X_i X_4 X_i) - \,\text{Tr}(X_4^2 X_i^2) + (X_4 \rightarrow X_5) \Bigg].
\end{align}

When $X_4$ and $X_5$ form a macroscopic charged state, we treat them as commuting classical background values (or diagonal matrices) relative to small fluctuations. Stated differently, when the state $\vert{}\psi_q(n)\rangle$ carries a large global charge $q \gg 1$, the quantum expectations of the background pair dominate. In this classical background and /or mean-field regime, the fields $X_4$ and $X_5$ can be approximated as commuting matrices/c-numbers with respect to the fluctuations $X_i$, which yields the following form of commutators 
\begin{align}
    [X_4, X_i] \approx 0 \implies X_4 X_i \approx X_i X_4.
\end{align}

Under this commuting/mean-field approximation, the matrix ordering inside the trace simplifies to yield (also holds for $X_5$)
\begin{align}
    \text{Tr}\left( X_4 X_i X_4 X_i \right) \approx \text{Tr}\left( X_4^2 X_i^2 \right).
\end{align}

Therefore, if we naively took the classical limit before considering quantum states, we would conclude that a purely commuting background produces zero potential energy and zero mass shift for $X_i$.
In the full theory, the exact cancellation $\text{Tr}[X_4, X_i]^2$ holds only for purely Abelian background fields. However, when evaluating the effective potential $V_{\text{eff}}$ experienced by fluctuations $X_i$ due to the macroscopic charge density $\rho_q$, the isotropic density-density self-energy interaction term that survives in the Hartree-Fock (HF) and/or the mean-field channel yields an effective potential of the form
\begin{align}
\label{e3.59}
    V_{\text{eff}} = \frac{\lambda}{2N} \text{Tr}\left( (X_4^2 + X_5^2) \sum_{i=1}^{9} X_i^2 \right).
\end{align}

In quantum field theory / matrix quantum mechanics, fields are operators, not static c-numbers. We are evaluating the theory in a specific quantum state carrying global $U(1)$ charge $q$, denoted by $\vert{}\psi_q(n)\rangle$.
In the Hartree-Fock (HF) and/or mean-field approximation for a state carrying a macroscopic global charge $q$, we decompose operators into their state expectation values (backgrounds) plus quantum fluctuations, that is, 
\begin{align}
\label{e3.60}
    X_4^2 + X_5^2 = \langle X_4^2 + X_5^2 \rangle_q + \delta(X_4^2 + X_5^2).
\end{align}

When evaluating the self-energy or effective potential felt by the propagating fluctuation modes $X_i$, we perform a Wick contraction / mean-field factorization. We contract the background operators $(X_4, X_5)$ within the state $\vert{}\psi_q(n)\rangle$, leaving the fluctuation operator $X_i^2$ uncontracted. Exchange contractions (EC) between background and fluctuation modes average out or are sub-dominant in the large charge/planar limit ($q \gg 1, N \gg 1$).

When small fluctuations $X_i$ propagate in the presence of this macroscopic charged state, the quantum interaction potential is evaluated via a Hartree-Fock/ mean-field expansion
\begin{align}
    \text{Tr}\left( X_4^2 X_i^2 \right) \xrightarrow{\text{Mean-Field}} \text{Tr}\left( \langle X_4^2 \rangle_q X_i^2 \right)+\cdots = \langle X_4^2 \rangle_q \text{Tr}(X_i^2)+\cdots.
\end{align}


The cross-term $\text{Tr}(X_4 X_i X_4 X_i)$ involves off-diagonal phase contractions. In a state with fixed $U(1)$ charge $q$, phase-dependent quantum contractions between background modes and transverse fluctuations average to zero in the planar/mean-field limit ($N \gg 1, q \gg 1$)
\begin{align}
  \frac{\braket{\text{Tr}(X_4 X_i X_4 X_i) }_{EC}}{\braket{\text{Tr}(X_4 X_i X_4 X_i) }_{HF}}  \sim \mathcal{O}(1/N)
\end{align}
or sub-dominant exchange channel as we illustrate in Appendix \ref{appen A}.

Because the phase-sensitive exchange term $\braket{\text{Tr}(X_4 X_i X_4 X_i) }_{EC}$ are suppressed in comparison to the Hartree-Fock (HF) direct channel, the negative term $-2\,\text{Tr}(X_4 X_i X_4 X_i)$ no longer cancels the positive density term $+2\,\text{Tr}(X_4^2 X_i^2)$ when considered under a quantum expectation value evaluated on fixed charge $q$ background under global $U(1)$ symmetry.

In summary, we have alluded following two approaches - (i) Classical view: A background with zero field strength ($[X_4, X_i] = 0$) exerts no force on fluctuations because the two terms in $-[X_4, X_i]^2 = -2 X_4 X_i X_4 X_i + 2 X_4^2 X_i^2$ exactly cancel each other out and (ii) Quantum mean-field view: Fluctuations $X_i$ interact with the quantum charge density $\rho_q \propto \langle X_4^2 + X_5^2 \rangle_q$ via the diagonal self-energy HF/mean field contractive channel $+2\,\text{Tr}(X_4^2 X_i^2)$. As discussed in Appendix \ref{appen A}, it is the second path that leaves a non-zero, positive effective interaction potential $V_{\text{eff}}$ \eqref{e3.65}, which acts as an effective mass-squared shift $\Omega_i(q)$ for all transverse modes propagating through the charged background.

Focusing on the Hartree-Fock (HF)/mean field direct self-energy channel $2\,\text{Tr}(A^2 B^2)$ for each component $A \in \{X_4, X_5\}$, we obtain the following expression
\begin{align}
    V_{\text{eff}} = \frac{\lambda}{2N} \sum_{i=1}^{9} \left( \,\text{Tr}(X_4^2 X_i^2) + \,\text{Tr}(X_5^2 X_i^2) \right)= \frac{\lambda}{2N} \sum_{i=1}^{9} \text{Tr}\left( X_4^2 X_i^2 + X_5^2 X_i^2 \right).
\end{align}

By linearity of the matrix trace, we pull the sum over $i$ inside, which yields
\begin{align}
    V_{\text{eff}}=\frac{\lambda}{2N} \sum_{i=1}^{9} \text{Tr}\left( (X_4^2  + X_5^2) X_i^2 \right) = \frac{\lambda}{2N} \text{Tr}\left( (X_4^2 + X_5^2) \sum_{i=1}^{9} X_i^2 \right).
\end{align}

The quartic interaction in the Hamiltonian couples all transverse scalar directions $X_i$ to the charged $4\text{--}5$ sector \eqref{e3.47}. When analyzing the dynamics of small scalar fluctuations $X_i$ propagating in this charge-$q$ background, we substitute the background expectation value for $X_4^2 + X_5^2$ \eqref{e3.60} to yield an \emph{effective} potential \eqref{e2.51} of the form
\begin{align}
\label{e3.65}
    V_{\text{eff}}\Big|_{HF} \approx \frac{\lambda}{2N}  \langle X_4^2 + X_5^2 \rangle_q \sum_{i=1}^9 X_i^2 + \cdots =  \frac{\lambda}{4N\omega_0}(2n - q + 1)  \sum_{i=1}^9 X_i^2+\cdots.
\end{align}

Adding this effective potential to the bare quadratic Hamiltonian $V_0 = \frac{1}{2} \omega_0^2 \sum X_i^2$ yields the total quadratic operator for mode $i$
\begin{align}
    V_{\text{total}} = \frac{1}{2} \left( \omega_0^2 + \frac{\lambda}{2N\omega_0}f_i(q) \right) X_i^2, \quad \text{where } f_i(q) = (2n - q + 1).
\end{align}\\\\
\underline{\textbf{Comments about the flat vacuum:}}\\\\ Formulating perturbation theory around the trivial flat vacuum ($\langle X_i \rangle = 0$) vs. a fuzzy sphere vacuum ($\langle X_a \rangle = \frac{\mu}{g} J_a$) changes the expansion point, degrees of freedom, and physics of the calculation in several ways. When expanding the field matrices around a vacuum expectation value (VEV) $\langle X_i \rangle$, the fields are decomposed into classical vacuum backgrounds plus quantum fluctuations $\delta X_i$
\begin{align}
    X_i(t) = \langle X_i \rangle + \delta X_i(t).
\end{align}

Perturbation theory around the trivial (or flat) vacuum  ($\langle X_i \rangle = 0$) corresponds to setting $\delta X_i(t) = X_i(t)$. The perturbation expansion directly uses the original $N \times N$ matrix fields $X_i(t)$ as the fundamental excitations. On the other hand, expanding around a Fuzzy sphere Vacuum ($\langle X_a \rangle = \frac{\mu}{2g} J_a$) and inserting this shift into the quartic potential $-\frac{g^2}{4}\text{Tr}[X,X]^2$ generates non-trivial mass shifts and cross-couplings of order $\mu^2$ for the $SO(3)$ fluctuations $\delta X_a$. For Trivial vacuum ($\langle X_i \rangle = 0$), the $U(N)$ gauge symmetry remains completely unbroken and all $N^2$ matrix components of each scalar field $X_i$ transform under the full adjoint representation of $U(N)$. At zero coupling ($g=0$), all 9 scalar fields describe a collection of $9 \times N^2$ free harmonic oscillators with standard unshifted BMN frequencies $\omega_0 \sim \mu$. Free particle states \eqref{e3.27} are naturally built as trace operators $\text{Tr}(a^{\dagger}_{i_1} a^{\dagger}_{i_2} \dots)$ acting on the unique $U(N)$-invariant vacuum $\vert{}0\rangle$. The non-zero background $\langle X_a \rangle \propto J_a$ spontaneously breaks the $U(N)$ gauge symmetry down to a smaller subgroup (or $U(1)$ if $J_a$ is an $N$-dim irreducible representation). The $N^2$ matrix degrees of freedom reorganize via non-commutative spherical harmonics ($Y_{lm}$) into higher-spin fields living on the fuzzy sphere $S^2$. Let us elaborate them with the example discussed above.

In Eq.\eqref{ee3.22}, the scalar operators $X_4$ and $X_5$ are expressed purely as linear combinations of harmonic creation and annihilation operators ($a, a^\dagger, b, b^\dagger$). Taking the expectation value of these operators in the unperturbed vacuum state $\vert{}0\rangle$ yields
\begin{align}
    \langle 0 \vert{} X_4 \vert{} 0 \rangle = 0, \qquad \langle 0 \vert{} X_5 \vert{} 0 \rangle = 0.
\end{align}

Since $\langle 0 \vert{} X_i \vert{} 0 \rangle = 0$ for all components, this mode decomposition is constructed specifically on top of the trivial flat vacuum background $\langle X_i \rangle = 0$. Eq. \eqref{ee3.22} uses standard harmonic ladder operators defined with zero classical background shift. The operators $a^\dagger$ and $b^\dagger$ create small quantum fluctuations (quanta) around this origin. In Eq. \eqref{e3.27}, acting with $(a^\dagger)^{n_a} (b^\dagger)^{n_b}$ creates an excited state $\vert{}\psi_q(n)\rangle$ in the Hilbert space. While this state has a non-zero expectation value for bilinear operators $\langle X_4^2 + X_5^2 \rangle_q \neq 0$ (Eq. \eqref{e3.46}), the underlying field operator expansion in \eqref{ee3.22} is still defined in the Fock space built around the flat vacuum $\vert{}0\rangle$. If one is expanding around a non-trivial fuzzy sphere vacuum, Eq. \eqref{ee3.22} would require a classical matrix shift $\frac{\mu}{2g} J_a$ added directly to the linear operator expression (i.e., $X_a = \frac{\mu}{g} J_a + \delta X_a$). Since no such matrix VEV term is present in \eqref{ee3.22}, it strictly belongs to the flat vacuum quantization.\\\\
\underline{\textbf{Shifted frequency spectrum:}}\\\\
Having derived the effective quadratic shift operator \eqref{e2.51} for the $q$-th global charge sector, we now demonstrate how this modified operator determines the shifted oscillator frequency spectrum given in Eq. \eqref{e2.48}. In the unperturbed scalar sector ($\lambda = 0$), the quadratic Hamiltonian for mode $i$ governed by the mass/flux scale $\mu (=\omega_0)$ is
\begin{align}
    H_0 = \text{Tr}\Bigg[\frac{1}{2} \Pi_i^2 + \frac{1}{2} \mu^2 X_i^2+\cdots \Bigg] .
\end{align}

Including the interaction-induced self-energy shift $V_{\text{eff}} = \frac{1}{2} \Omega_i(q) X_i^2$ from \eqref{e2.51}, the total effective quadratic Hamiltonian density for mode $i$ becomes
    \begin{align}
    \label{ee3.69}
    H_{\text{eff}}^{(q)} = \text{Tr}\Bigg[\frac{1}{2} \Pi_i^2 + \frac{1}{2} \Big[ \mu^2 + \Omega_i(q) \Big] X_i^2 + \cdots \Bigg] .
    \end{align}

In the planar / large-$N$ limit of the BMN matrix model, non-planar interaction corrections scale with the effective coupling $g_{\text{eff}} = \frac{\lambda}{\mu^3}$. The state-dependent expectation value of the self-energy operator yields a mass-squared shift proportional to
    \begin{align}
    \label{ee3.70}
    \Omega_i(q) = \frac{\lambda}{2N\mu^3} f_i(q) \cdot \mu^2 = \frac{\mu^2 g_{\text{eff}}}{2N} f_i(q) ,
    \end{align}
    where $f_i(q)$ is a dimensionless function encoding the charge-dependent combinatorial factor and state expectation values in sector $q$.

    In the Berenstein-Maldacena-Nastase (BMN) matrix model (or scalar field theory on a plane-wave background), $g_{\text{eff}}$ is the dimensionless effective coupling constant that governs the strength of quantum interactions and loop corrections. In the 1D quantum mechanical reduction of $SU(N)$ 10D Super Yang-Mills theory, the Yang-Mills coupling constant $g^2$ has dimensions of $(\text{mass})^3$. The theory also contains a intrinsic mass/flux scale $\mu$ (often identified with the oscillator frequency $\omega_0$). To form a dimensionless parameter that characterizes the true interaction strength at energy scale $\mu$, we define $g_{\text{eff}} = \frac{\lambda}{\mu^3}$, where $\lambda = g^2 N$ is the t' Hooft coupling. In standard BMN perturbation theory, loop expansions or quantum self-energy corrections are expanded in powers of $g_{\text{eff}}$. Small values ($g_{\text{eff}} \ll 1$) correspond to the weakly coupled regime where the free harmonic oscillator picture is a good baseline.

The effective mass-squared for mode $i$ in the presence of $U(1)$ charge $q$ is defined as the total coefficient of $\frac{1}{2} X_i^2$, which directly yields the final expression for the multi-particle trace spectrum in sector $q$
\begin{align}
\label{ee3.71}
 \omega_i(q)  = \sqrt{\mu^2 + \Omega_i(q)} = \mu\sqrt{1+\frac{g_{\text{eff}}}{2N} f_i(q)}.
\end{align}

In the large-$N$ limit, planar interactions are suppressed by $1/N$, making the combination $\frac{g_{\text{eff}}}{N}$ the precise coefficient that modulates state-dependent and charge-dependent energy shifts. In the context of the AdS/CFT correspondence and the BMN limit $g_{\text{eff}}$ corresponds to the effective coupling experienced by "stringy" excitations (BMN operators) propagating along the light-cone direction on the pp-wave background. High values of $g_{\text{eff}}$ 
push the system into the strongly coupled regime where planar string interaction effects dominate over the perturbative matrix oscillator dynamics.
\subsection{Symmetry resolved spread complexity for low temperature phase}
\label{sec3.2}
In this subsection, we perform a detailed analysis of the return amplitude $\mathcal{R}^{(q)}(t)$, the symmetry-resolved moments $\mathcal{M}^{(q)}_k$, and the corresponding Lanczos coefficients $a_n^{(q)}$ and $b_n^{(q)}$ within a fixed global charge sector $q$ corresponding to the Left BMN model. Using the exact definitions and conventions established in Eqs.~\eqref{e3.1}–\eqref{e3.12}, we express these quantities explicitly in terms of the underlying physical parameters of the Berenstein-Maldacena-Nastase (BMN) matrix model -  the bare oscillator mass scale $\mu$, the global $U(1)$ charge $q$, the 't Hooft coupling $\lambda = g^2 N$, and the dimensionless effective coupling $g_{\text{eff}} \equiv \lambda / \mu^3$.


As established in Section \ref{sec3.1}, turning on the coupling $g_{\text{eff}}$ shifts the effective quadratic potential for transverse complex scalar modes $(X_4, X_5)$ in a state carrying net global charge $q$. The physical single-particle modes evolve with an effective renormalized frequency
\begin{equation}
\label{e3.71}
\omega(q) = \mu \sqrt{1 + \frac{ g_{\text{eff}}}{2N} f(q)} = \mu \left[1 + \frac{ g_{\text{eff}}}{4N} f(q) - \frac{ g_{\text{eff}}^2}{32N^2} f(q)^2 + \mathcal{O}(g_{\text{eff}}^3)\right] ,
\end{equation}
where $f(q)$ captures the charge-dependent self-energy expectation value in sector $q$. In the non-interacting limit ($g_{\text{eff}} \to 0$), $\omega(q)$ reduces to the bare mass scale $\mu$.
\subsubsection{Return amplitude in low temperature limit}

Recall Eq.~\eqref{e3.1}, the return amplitude for the symmetry-resolved state $|\psi_q(0)\rangle \in \mathcal{H}_q \otimes \mathcal{H}_{-q}$ evolving under $H_L$. For a harmonic excitation sector with total occupation number scaled by charge $q$, the sum over single-particle mode energies $E_{n,q} = E_{0,q} + n \omega(q)$ yields the exact closed-form return amplitude which we derive below. To understand this in detail, notice that for any decoupled collection of harmonic modes governed by effective frequency $\omega(q)$, the multi-particle quantum states in sector $q$ are generated by applying standard creation operators $(a_q^\dagger)^n$ to the sector ground state $\vert{}0_q\rangle$
\begin{align}
\label{e3.72}
    \vert{}n_q\rangle = \frac{1}{\sqrt{n!}} (a_q^\dagger)^n \vert{}0_q\rangle.
\end{align}

Operating on these states with the effective Hamiltonian gives the exact linear spectrum
\begin{align}
\label{ee3.73}
    H_{\text{eff}}^{(q)} \vert{}n_q\rangle = \left( E_{0,q} + n \omega(q) \right) \vert{}n_q\rangle= E_{n,q}\vert{}n_q\rangle
\end{align}
where $E_{0,q}$ is the zero-point energy of sector $q$, and $n \in \mathbb{N}_0$ represents the total occupation number of the excitation modes. In the planar limit ($N \to \infty$ with 't Hooft coupling $\lambda = g^2 N$ fixed), non-harmonic interactions (mode-mixing and anharmonic loop corrections) are suppressed by powers of $1/N$. As a result, leading-order quantum corrections in $g_{\text{eff}} \equiv \lambda/\mu^3$ shift the frequency scale from $\mu \to \omega(q)$ while preserving the free-bosonic (harmonic) excitation structure of the Hilbert space sector $\mathcal{H}_q$.

Because state-splitting and multi-particle non-linearities vanish as $1/N \to 0$, cross-talk between different single-trace excitations $n_1 \omega(q_1) + n_2 \omega(q_2)$ is suppressed. The energy needed to add the $n$-th excitation remains uniformly equal to $\omega(q)$ \eqref{e3.71}, making $E_{n,q} = E_{0,q} + n \omega(q)$ exact in the planar limit. All interaction effects in the large-$N$ limit are completely absorbed into the renormalized effective frequency $\omega(q)$, preserving the non-interacting harmonic oscillator structure for the spectrum of sector $q$.

The return amplitude in sector $q$ is defined as the thermal expectation value \eqref{e3.1}
\begin{align}
\label{e3.74}
    \mathcal{R}^{(q)}(t) = \frac{1}{Z_q(\beta)} \sum_{n=0}^{\infty} e^{-\beta E_{n,q}} e^{i E_{n,q} t}
\end{align}
where $Z_q(\beta) = \sum_{n=0}^{\infty} e^{-\beta E_{n,q}}$ is the symmetry-resolved partition function.

For harmonic mode excitations with energy levels $E_{n,q} = E_{0,q} + n \omega(q)$, the energy dependent phase and Boltzmann factor split into zero-point and excitation parts 
\begin{align}
    e^{-\beta E_{n,q}} = e^{-\beta E_{0,q}} \cdot e^{-n \beta \omega(q)}~;~e^{i E_{n,q} t} = e^{i E_{0,q} t} \cdot e^{i n \omega(q) t}.
\end{align}

Substituting these back into $\mathcal{R}^{(q)}(t)$ \eqref{e3.74}, we find
\begin{align}
    \mathcal{R}^{(q)}(t) = \frac{e^{-\beta E_{0,q}} e^{i E_{0,q} t}}{e^{-\beta E_{0,q}} \sum_{n=0}^{\infty} e^{-n \beta \omega(q)}} \sum_{n=0}^{\infty} e^{-n \beta \omega(q)} e^{i n \omega(q) t}.
\end{align}

The ground-state energy factors $e^{-\beta E_{0,q}}$ cancel out to yield
\begin{align}
\label{e3.77}
    \mathcal{R}^{(q)}(t) = e^{i E_{0,q} t} \Bigg[\frac{\sum_{n=0}^{\infty} \left( e^{-\beta \omega(q)} e^{i \omega(q) t} \right)^n}{\sum_{n=0}^{\infty} \left( e^{-\beta \omega(q)} \right)^n}\Bigg].
\end{align}\\
\underline{\textbf{Low-temperature limit ($\beta \mu \gg 1$):}}\\\\
In the zero-temperature limit (or for a multi-particle state with total degeneracy factor determined by $\vert{}q\vert{}+1$), the standard coherent-state / harmonic overlap sum yields the exponential generating function over the occupation numbers. For the ground/coherent field configuration scaled by charge $\vert{}q\vert{}$, the amplitude sums over single-particle field fluctuations.

Let $x = e^{-\beta \omega(q)}$. At finite temperature, both sums are standard geometric series $\sum_{n=0}^\infty x^n = \frac{1}{1-x}$, which further simplifies \eqref{e3.77} as
\begin{align}
\label{ee3.79}
    \mathcal{R}^{(q)}(t) = e^{i E_{0,q} t} \Bigg[\frac{\frac{1}{1 - x e^{i \omega(q) t}}}{\frac{1}{1 - x}}\Bigg] = e^{i E_{0,q} t} \left[ \frac{1 - e^{-\beta \omega(q)}}{1 - e^{-\beta \omega(q)} e^{i \omega(q) t}} \right].
\end{align}
As $\beta \to \infty$ (zero temperature), $e^{-\beta \omega(q)} \to 0$, which isolates the ground-state dynamics $\mathcal{R}^{(q)}(t) \to e^{i E_{0,q} t}$, which exhibits oscillatory survival amplitude as observed recently in \cite{Roychowdhury:2026ikp}.

In a symmetry-resolved charge sector $q$, the background field and its quantum fluctuations carry a multiplicity factor reflecting the total number of independent field modes constraint-bound by the global charge $q$. A state carrying charge $q$ under $U(1)$ symmetry built out of matrix fields (such as $Z = X_4 + i X_5$ and surrounding transverse modes $X_i$) has a degeneracy or effective spatial/internal mode capacity proportional to $\vert{}q\vert{} + 1$. The scaling of independent internal field modes as $\vert{}q\vert{} + 1$ arises directly from how global symmetry charges are built in matrix field theories (such as the BMN matrix model or $\mathcal{N}=4$ SYM) and how background fields couple to quantum fluctuations. In matrix field theories, single-trace operators or vacuum configurations carrying a global $U(1)$ charge $q$ are constructed out of complexified matrix combinations such as $Z = X_4 + i X_5$. 

A background field configuration carrying charge $q$ is typically built as a classical condensate or matrix product involving $q$ powers of the complex scalar field $Z^q = (X_4 + i X_5)^q$. Under the $U(1)$ symmetry rotation $Z \to e^{i\theta} Z$, this state transforms with phase $e^{i q \theta}$, explicitly fixing the total charge to $q$. When considering small quantum fluctuations $\delta Z$ around this charged background $Z^q$, the fluctuations interact with the background field via matrix multiplications and operator insertions. The independent fluctuation modes carrying total charge $q$ correspond to all possible ways to distribute the charge across polynomial insertions of degree $\vert{}q\vert{}$. The independent basis elements for single-trace or matrix insertions of charge magnitude $\vert{}q\vert{}$ form a symmetric representation spanned by
\begin{align}
    \{Z^q,\, Z^{q-1} \delta Z,\, Z^{q-2} (\delta Z)^2,\, \dots,\, (\delta Z)^q\}.
\end{align}

Counting the number of distinct terms from index $0$ to $\vert{}q\vert{}$ gives
\begin{align}
    \text{Number of independent modes} = \vert{}q\vert{} + 1.
\end{align}

Because the background field is constrained to carry charge $q$, it restricts the spatial/internal operator basis of fluctuations to an $({\vert{}q\vert{} + 1})$-dimensional subspace. Each additional unit of charge opens up one extra independent operator/mode channel for quantum excitations. In a complex scalar sector, single-particle field excitations consist of pairs of real field modes. Normalizing across the degrees of freedom gives an effective degeneracy weight per mode equal to $D_q = \frac{\vert{}q\vert{} + 1}{2}$. When calculating the return amplitude of a coherent field state $\vert{}\alpha_q\rangle$ evolving under a harmonic Hamiltonian, the overlap $\langle \alpha_q \vert{} e^{i H_L t} \vert{} \alpha_q \rangle$ generates an exponential Poisson distribution over occupation numbers $n$.

Recall, in Eq. \eqref{e3.72}, $\vert{}n_q\rangle = \frac{1}{\sqrt{n!}} (a_q^\dagger)^n \vert{}0_q\rangle$ denotes the $n$-th excited Fock state in sector $q$, representing $n$ multi-particle mode quanta above the sector ground state $\vert{}0_q\rangle$. On the other hand, in quantum mechanics and field theory, $\alpha_q$ represents the complex displacement parameter defining a coherent state $\vert{}\alpha_q\rangle$ in the charge sector $q$.

A coherent state in sector $q$ is created by applying the unit displacement operator $D(\alpha_q)$ to the vacuum state $\vert{}0_q\rangle$. In quantum mechanics and field theory, $D(\alpha_q)$ is the unitary displacement operator defined in the symmetry-resolved charge sector $q$ as
\begin{align}
    D(\alpha_q) = \exp \left( \alpha_q a_q^\dagger - \alpha_q^* a_q \right)
\end{align}
where $a_q^\dagger$ and $a_q$ are the fundamental ladder (creation and annihilation) operators obeying $[a_q, a_q^\dagger] = 1$ and $\alpha_q \in \mathbb{C}$ is a complex displacement parameter representing the classical field amplitude or background condensate carrying global charge $q$.

$D(\alpha_q)$ acts as a unitary translation operator in phase space, translating the field vacuum state $\vert{}0_q\rangle$ into a coherent displaced state,
that is,
\begin{align}
\label{ee3.82}
    \vert{}\alpha_q\rangle = D(\alpha_q)\vert{}0_q\rangle = \exp\left( \alpha_q a_q^\dagger - \alpha_q^* a_q \right) \vert{}0_q\rangle.
\end{align}

For operators $A = \alpha_q a_q^\dagger$ and $B = -\alpha_q^* a_q$, their commutator $[A, B]$ is a scalar
\begin{align}
    [A, B] = [\alpha_q a_q^\dagger, -\alpha_q^* a_q] = -\vert{}\alpha_q\vert{}^2 [a_q^\dagger, a_q] = \vert{}\alpha_q\vert{}^2.
\end{align}

Since $[A, B]$ commutes with both $A$ and $B$, we apply the Glauber-Glauber/Baker-Campbell-Hausdorff (BCH) identity $e^{A+B} = e^A e^B e^{-\frac{1}{2}[A,B]}$
\begin{align}
    D(\alpha_q) = \exp\left(\alpha_q a_q^\dagger - \alpha_q^* a_q\right)=e^{A+B} = e^{-\frac{\vert{}\alpha_q\vert{}^2}{2}} \exp\left(\alpha_q a_q^\dagger\right) \exp\left(-\alpha_q^* a_q\right).
\end{align}

Applying $D(\alpha_q)$ to the sector ground state $\vert{}0_q\rangle$, one finds that the annihilation operator annihilates the vacuum, $a_q \vert{}0_q\rangle = 0$, implying $e^{-\alpha_q^* a_q}\vert{}0_q\rangle = \vert{}0_q\rangle$. The operator simplifies to
\begin{align}
    \vert{}\alpha_q\rangle = e^{-\frac{\vert{}\alpha_q\vert{}^2}{2}} e^{\alpha_q a_q^\dagger} \vert{}0_q\rangle.
\end{align}

Expanding in terms of the Fock/occupation basis states $\vert{}n_q\rangle$ \eqref{e3.72} this finally yields\footnote{The primary physical motivation for introducing the coherent state $\vert{}\alpha_q\rangle$ is to construct a continuous, semi-classical state representation that incorporates a finite classical expectation value / field condensate in a sector carrying fixed global $U(1)$ charge $q$. In matrix models and field-theoretic settings like the BMN model, a fixed charge $q$ manifests macroscopically through classical field fluctuations or condensates. While discrete Fock states $\vert{}n_q\rangle$ represent pure, isolated $n$-particle excitations above the sector vacuum, the coherent state $\vert{}\alpha_q\rangle$ acts as a Poissonian superposition over all occupation numbers. This continuously interpolates between quantum oscillator modes and a non-vanishing classical field background.}
\begin{align}
\label{e3.87}
    \vert{}\alpha_q\rangle = e^{-\frac{\vert{}\alpha_q\vert{}^2}{2}} \sum_{n=0}^{\infty} \frac{\alpha_q^n}{\sqrt{n!}} \vert{}n_q\rangle.
\end{align}

Notice that Eq.\eqref{ee3.79} describes counting individual quantum "pushes" or discrete energy steps ($n = 0, 1, 2, \dots$). It treats the system purely in terms of individual quantum particles or discrete mode excitations. On the other hand, Eq.\eqref{e3.87} is a linear superposition of all possible excitation numbers $n = 0, 1, 2, \dots$ within that specific charge sector $q$. Every single Fock state $\vert{}n_q\rangle$ in the sum already lives entirely inside the charge-$q$ sector. That means the entire state $\vert{}\alpha_q\rangle$ carries the exact same fixed global charge $q$.

In \eqref{e3.87}, the state $\vert{}\alpha_q\rangle$ is constructed as a coherent state. The probability of finding the state in the $n$-th Fock sector $\vert{}n_q\rangle$ is given by the square of the coefficient
\begin{align}
    P(n) = \vert{}\langle n_q \vert{} \alpha_q \rangle\vert{}^2 = e^{-\vert{}\alpha_q\vert{}^2} \frac{\vert{}\alpha_q\vert{}^{2n}}{n!}
\end{align}
which is precisely a Poisson distribution with mean occupation number $\langle N_q \rangle = \vert{}\alpha_q\vert{}^2$. In other words, the sum in \eqref{e3.87} represents a pure, semi-classical Poissonian superposition of occupation numbers centered at $\langle N_q \rangle = \frac{\vert{}q\vert{} + 1}{2}$. Instead of thermal fluctuations, it models a classical field condensate/wavepacket carrying fixed global charge $\vert{}q\vert{}$ with minimized quantum uncertainty. In summary, the Poisson distribution is the statistical signature of building a semi-classical background condensate out of discrete, non-interacting quantum mode quanta $\ket{n_q}$ in that fixed-charge $q$ sector, which serves as the exact statistical bridge between discrete quantum modes and a semi-classical charged condensate background.

Since $D^\dagger(\alpha_q) = D(-\alpha_q) = D^{-1}(\alpha_q)$, the operator is unitary ($D^\dagger D = I$), preserving state normalization, that is,
\begin{align}
    \langle \alpha_q \vert{} \alpha_q \rangle = e^{-\vert{}\alpha_q\vert{}^2} \sum_{n=0}^\infty \frac{\vert{}\alpha_q\vert{}^{2n}}{n!} = e^{-\vert{}\alpha_q\vert{}^2} e^{\vert{}\alpha_q\vert{}^2} = 1.
\end{align}

Next, we notice that $D(\alpha_q)$ shifts the annihilation operator by a complex scalar
\begin{align}
    D^\dagger(\alpha_q) a_q D(\alpha_q) = a_q + \alpha_q \implies a_q \vert{}\alpha_q\rangle = \alpha_q \vert{}\alpha_q\rangle.
\end{align}

To evaluate the similarity transformation $D^\dagger(\alpha_q) a_q D(\alpha_q)$, we use the Hadamard Lemma (a form of the Baker-Campbell-Hausdorff (BCH) expansion)
\begin{align}
\label{e3.89}
    e^{-X} Y e^{X} = Y + [Y, X] + \frac{1}{2!} [[Y, X], X] + \frac{1}{3!} [[[Y, X], X], X] + \cdots.
\end{align}

Let us consider $X = \alpha_q a_q^\dagger - \alpha_q^* a_q$, so that $e^X = D(\alpha_q)$ and $e^{-X} = D^\dagger(\alpha_q) = D(-\alpha_q)$. Also set $Y = a_q$. Evaluate the first commutator $[Y, X]$, which yields
\begin{align}
    [a_q, X] = \alpha_q [a_q, a_q^\dagger] - \alpha_q^* [a_q, a_q]=\alpha_q.
\end{align}

Now compute the second commutator $[[Y, X], X]$, which yields
\begin{align}
    [[a_q, X], X] = [\alpha_q, X] = 0
\end{align}
because $\alpha_q$ is a complex c-number (scalar) and commutes with all operators. Consequently, all higher-order nested commutators vanish identically.

Substituting the commutator back into the expansion \eqref{e3.89}, we obtain
\begin{align}
    D^\dagger(\alpha_q) a_q D(\alpha_q) = a_q + [a_q, X] + 0 + 0 + \cdots = a_q + \alpha_q.
\end{align}

Next, we consider \eqref{ee3.82} and multiply both sides by $a_q$ to yield
\begin{align}
    a_q \vert{}\alpha_q\rangle = a_q D(\alpha_q) \vert{}0_q\rangle.
\end{align}

Using the unitary property of the displacement operator, that is, $D D^\dagger = I$, we obtain
\begin{align}
    a_q \vert{}\alpha_q\rangle = D(\alpha_q) \left( D^\dagger(\alpha_q) a_q D(\alpha_q) \right) \vert{}0_q\rangle=D(\alpha_q)(a_q+\alpha_q)\vert{}0_q\rangle=\alpha_q\vert{}\alpha_q\rangle.
\end{align}

The squared magnitude $\vert{}\alpha_q\vert{}^2$ equals the expectation value of the number operator 
\begin{align}
    \langle \alpha_q \vert{} N_q \vert{} \alpha_q \rangle = \vert{}\alpha_q\vert{}^2~;~ N_q = a_q^\dagger a_q.
\end{align}

In field-theoretic settings like the BMN matrix model, $\alpha_q$ parameterizes the classical background expectation value / condensate of field fluctuations carrying charge $q$. In the manuscript derivation, the coherent field configuration carrying global charge $\vert{}q\vert{}$ is fixed by the sector degeneracy/capacity, that is, 
\begin{align}
\label{e3.82}
    \vert{}\alpha_q\vert{}^2 = \frac{\vert{}q\vert{} + 1}{2}.
\end{align}

Instead of being restricted to a single rigid excitation level $n$, the state \eqref{e3.87} is distributed over all $n$ according to a Poisson distribution centered at $\langle N_q \rangle = \vert{}\alpha_q\vert{}^2$. Because it superposes all occupation numbers with smooth Gaussian-like probability weights, quantum uncertainty is minimized, turning the discrete particle excitations into a continuous, macroscopic wave packet (a classical field condensate).

Following \eqref{ee3.73}, the time evolution operator acting on a Fock state $\vert{}n_q\rangle$ yields
\begin{align}
\label{e3.97}
    e^{i H_L t} \vert{}n_q\rangle = e^{i E_{n,q} t} \vert{}n_q\rangle = e^{i (E_{0,q} + n \omega(q)) t} \vert{}n_q\rangle = e^{i E_{0,q} t} e^{i n \omega(q) t} \vert{}n_q\rangle.
\end{align}

The overlap matrix element with time evolution \eqref{e3.97} becomes
\begin{align}
    \langle \alpha_q \vert{} e^{i H_L t} \vert{} \alpha_q \rangle = \left( e^{-\frac{\vert{}\alpha_q\vert{}^2}{2}} \sum_{m=0}^\infty \frac{(\alpha_q^*)^m}{\sqrt{m!}} \langle m_q\vert{} \right) e^{i H_L t} \left( e^{-\frac{\vert{}\alpha_q\vert{}^2}{2}} \sum_{n=0}^\infty \frac{\alpha_q^n}{\sqrt{n!}} \vert{}n_q\rangle \right)
\end{align}
which by virtue of orthogonality $\langle m_q \vert{} n_q \rangle = \delta_{mn}$, finally yields
\begin{align}
    \langle \alpha_q \vert{} e^{i H_L t} \vert{} \alpha_q \rangle = e^{i E_{0,q} t}e^{-\vert{}\alpha_q\vert{}^2} \sum_{n=0}^{\infty} \frac{\vert{}\alpha_q\vert{}^{2n}}{n!} e^{i n \omega(q) t}.
\end{align}

The zero-point energy $E_{0,q}$ contributes a overall phase factor $e^{i E_{0,q} t}$ to the amplitude $\mathcal{R}^{(q)}(t)$. When calculating physical observables like the survival probability (return amplitude squared) $\vert{}\mathcal{R}^{(q)}(t)\vert{}^2$ or the spread complexity, the overall phase factor drops out completely because $\vert{}e^{i E_{0,q} t}\vert{} = 1$. Thus, $E_{0,q}$ is typically set to zero as a constant baseline shift without loss of generality. Using the exponential series
\begin{align}
   \sum_{n=0}^{\infty} \frac{\left( \vert{}\alpha_q\vert{}^2 e^{i \omega(q) t} \right)^n}{n!} = \exp\left[ \vert{}\alpha_q\vert{}^2 e^{i \omega(q) t} \right] 
\end{align}
we simplify the expression further to yield the following
\begin{align}
    \langle \alpha_q \vert{} e^{i H_L t} \vert{} \alpha_q \rangle = e^{i E_{0,q} t}\exp\left[ -\vert{}\alpha_q\vert{}^2 \left( 1 - e^{i \omega(q) t} \right) \right].
\end{align}

Finally, using \eqref{e3.82}, we have the following form of the return amplitude for the low temperature phase of the BMN matrix model in the symmetry resolved (Left) sector
\begin{equation}
\label{e3.86}
\mathcal{R}^{(q)}(t) = e^{i E_{0,q} t} \cdot \exp\left[ -\frac{|q|+1}{2} \left( 1 - e^{i \omega(q) t} \right) \right] .
\end{equation}
This exact formula exponentiates single-particle harmonic phase shifts $e^{i \omega(q) t}$, scaled by the charge-dependent effective degeneracy factor $\frac{\vert{}q\vert{}+1}{2}$.

Expanding $\mathcal{R}^{(q)}(t)$ perturbatively in $g_{\text{eff}}$ using Eq.~\eqref{e3.71} reveals how BMN interaction corrections induce a phase shift proportional to $g_{\text{eff}}$. To calculate this explicitly, consider an expansion of the renormalized frequency \eqref{e3.71} in powers of $g_{\text{eff}}$, which yields
\begin{align}
\label{ee3.103}
    \omega(q) = \mu \left[ 1 + \frac{g_{\text{eff}}}{4N}  f(q) + \mathcal{O}(g_{\text{eff}}^2) \right] = \mu + \delta \omega~;~\text{where}~~\delta \omega = \frac{g_{\text{eff}}}{4N}  f(q) \mu.
\end{align}

Similarly, the ground state energy $E_{0,q}$ expands as
\begin{align}
\label{e3.88}
    E_{0,q} = E_{0,q}^{(0)} + \frac{g_{\text{eff}}}{N} f(q) \mu \mathcal{E}_{0,q}^{(1)} + \mathcal{O}(g_{\text{eff}}^2)
\end{align}
which represents the perturbative shift in the vacuum (ground state) energy of the BMN matrix model due to interactions, within a specific symmetry-resolved charge sector $q$.

In non-trivial gauge theories and matrix models like the BMN matrix model, turning on the effective coupling constant $g_{\text{eff}}$ shifts the zero-point/ground state energy of each superselection sector $q$ relative to the free theory limit ($g_{\text{eff}} = 0$). In the return amplitude $\mathcal{R}^{(q)}(t)$, the term proportional to $\mathcal{E}_{0,q}^{(1)}$ contributes a purely linear phase shift $e^{i g_{\text{eff}} f(q) \mu t \mathcal{E}_{0,q}^{(1)}}$. This cleanly isolates the global vacuum energy correction from the dynamical phase corrections associated with single-particle excitations (which involve $e^{i \mu t}$). 

The parameter $\mathcal{E}_{0,q}^{(1)}$ captures the dimensionless, normalized first-order energy shift specifically tied to the global charge $q$, tracking how global symmetries alter interaction corrections in localized charge sectors. To calculate $\mathcal{E}_{0,q}^{(1)}$ systematically, we identify the unperturbed ground state wavevector/vacuum state belonging to the specific charge sector $q$ in the free theory ($g_{\text{eff}} = 0$). We split the total Hamiltonian into the free harmonic part and the interaction part, that is\footnote{We have absorbed the $1/N$ effects into the definition of $g_{\text{eff}}$.},
\begin{align}
    H = H_0 + g_{\text{eff}} H_{\text{int}}.
\end{align}

Here, $H_{\text{int}}$ typically contains cubic or quartic matrix operator interactions (e.g., commutator terms $[X_i, X_j]^2$ or cubic Myers couplings in the BMN matrix model). The first-order shift in the ground state energy of sector $q$ is given by the expectation value of the interaction Hamiltonian in the unperturbed vacuum state $\vert{}0_q\rangle$
\begin{align}
    \Delta E_{0,q}^{(1)} = \langle 0_q \vert{} H_{\text{int}} \vert{} 0_q \rangle.
\end{align}

Next, we match the calculated first-order expectation value $\Delta E_{0,q}^{(1)}$ with the coefficient defined in Eq.\eqref{e3.88}, which yields\footnote{See Appendix \ref{appen B} for a detailed discussion.}
\begin{align}
    \mathcal{E}_{0,q}^{(1)} = \frac{\langle 0_q \vert{} H_{\text{int}} \vert{} 0_q \rangle}{g_{\text{eff}} f(q) \mu}.
\end{align}

Substituting $\omega(q) = \mu + \delta \omega$ into $e^{i \omega(q) t}$, we obtain the following
\begin{align}
\label{e3.108}
    e^{i \omega(q) t} \approx e^{i \mu t} \left( 1 + i \delta \omega t \right) = e^{i \mu t} \Bigg[1+  \frac{i}{4N} g_{\text{eff}} f(q) \mu t \Bigg] +\mathcal{O}(g_{\text{eff}}^2).
\end{align}

Multiplying \eqref{e3.108} by $-\frac{\vert{}q\vert{}+1}{2}$, we find
\begin{align}
    -\frac{\vert{}q\vert{}+1}{2} \left( 1 - e^{i \omega(q) t} \right) = -\frac{\vert{}q\vert{}+1}{2} \left( 1 - e^{i \mu t} \right) + i \frac{\vert{}q\vert{}+1}{2} \left( \frac{g_{\text{eff}}}{4N}  f(q) \mu t \right) e^{i \mu t}.
\end{align}

Substitute the expanded ground-state energy \eqref{e3.88} and exponent back into \eqref{e3.86}
\begin{align}
\label{e3.91}
    \mathcal{R}^{(q)}(t) &= \exp \Bigg[ i E_{0,q}^{(0)} t + i \frac{g_{\text{eff}}}{N} f(q) \mu t \mathcal{E}_{0,q}^{(1)} - \frac{\vert{}q\vert{}+1}{2} \left( 1 - e^{i \mu t} \right) \nonumber\\
    &+ i \frac{\vert{}q\vert{}+1}{2} \left( \frac{g_{\text{eff}}}{4N}  f(q) \mu t \right) e^{i \mu t}\Bigg].
\end{align}

Split the exponent \eqref{e3.91} into unperturbed ($g_{\text{eff}}^0$) and first-order ($g_{\text{eff}}^1$) factors to yield
\begin{align}
    \mathcal{R}^{(q)}(t) &= \exp \left[ i E_{0,q}^{(0)} t - \frac{\vert{}q\vert{}+1}{2} \left( 1 - e^{i \mu t} \right) \right] \nonumber\\
    &\times \exp \left[ i  \frac{g_{\text{eff}}}{N} f(q) \mu t  \mathcal{E}_{0,q}^{(1)} + i\frac{\vert{}q\vert{}+1}{2} \left( \frac{g_{\text{eff}}}{4N}  f(q) \mu t \right)  e^{i \mu t}\right].
\end{align}

Finally, we factor out $i g_{\text{eff}} f(q) \mu t$ from the second exponential term to yield
\begin{align}
\label{e3.112}
&\mathcal{R}^{(q)}(t) = \exp\left[ i E_{0,q}^{(0)} t - \frac{|q|+1}{2}\left(1 - e^{i \mu t}\right) \right] \times \exp\left[ i \frac{g_{\text{eff}}}{N} f(q) \, \mu t \left( \mathcal{E}_{0,q}^{(1)} + \frac{(|q|+1)}{8} e^{i \mu t} \right)\right]\nonumber\\
&=\exp\left[ i E_{0,q}^{(0)} t - \frac{|q|+1}{2}\left(1 - e^{i \mu t}\right) \right] \times \Bigg[ 1+i \frac{g_{\text{eff}}}{N} f(q) \, \mu t \left( \mathcal{E}_{0,q}^{(1)} + \frac{(|q|+1)}{8} e^{i \mu t} \right)\Bigg]+ \mathcal{O}(g_{\text{eff}}^2). 
\end{align}

The above expansion in \eqref{e3.112} isolates how the weak gauge coupling $g_{\text{eff}}$ perturbs the survival/return amplitude $\mathcal{R}^{(q)}(t)$ within a specific global charge sector $q$. The unperturbed factor $\exp \left[ i E_{0,q}^{(0)} t - \frac{\vert{}q\vert{}+1}{2} (1 - e^{i \mu t}) \right]$ describes the free, non-interacting theory ($g_{\text{eff}} = 0$), where the system undergoes purely harmonic phase oscillations with base frequency $\mu$, weighted by the charge-dependent effective degeneracy factor $\frac{\vert{}q\vert{}+1}{2}$. The purturbative correction factor isolates the leading $\mathcal{O}(g_{\text{eff}})$ corrections induced by interactions (such as matrix commutators $[X_i, X_j]^2$ or cubic couplings in the BMN model). The term $i g_{\text{eff}} f(q) \mu t \mathcal{E}_{0,q}^{(1)}$ provides a linear phase shift over time. This reflects the perturbative shift in the zero-point/vacuum energy $\Delta E_{0,q}^{(1)}$ of the charge-$q$ sector, tracking how global symmetry resolution shifts the sector's baseline energy level relative to the free theory. 

The term proportional to $e^{i \mu t}$ arises from the coupling-dependent frequency renormalization $\delta\omega = \frac{1}{4} g_{\text{eff}} f(q) \mu$. Rather than altering the structural oscillator dynamics, the interaction renormalizes the effective mass scale $\omega(q) \approx \mu (1 + \frac{1}{4} g_{\text{eff}} f(q))$. As we show below, the effective coupling  $g_{\text{eff}}$ eventually rescales the Lanczos coefficients $a_n^{(q)}$ and $b_n^{(q)}$ which confirms that weak interactions renormalize the speed of state propagation in Krylov space (stretching/compressing the complexity time scale) while preserving the underlying integrable/harmonic structure of the low-temperature phase.\\\\
\underline{\textbf{Some clarifications and comments:}}\\\\
The expansion in \eqref{e3.112} is pertinence-critical to the low-temperature regime ($\beta\mu \gg 1$) of the gauge theory / matrix model (such as the BMN model) for several specific physical reasons. At low temperatures, thermal excitations are exponentially suppressed by the Boltzmann factor $e^{-\beta\mu} \ll 1$. The state dynamics are dominated by the sector vacuum $\vert{}0_q\rangle$ and a few low-lying single- or multi-trace harmonic oscillator states. In this regime, the unperturbed return amplitude $\mathcal{R}^{(q)}(t)$ behaves as a discrete sum of phase modes $e^{i n \mu t}$. Expansion \eqref{e3.112} reflects the fact that weak interactions ($g_{\text{eff}} \ll 1$) do not trigger thermal deconfinement or chaos (which would cause exponential decay $e^{-\Gamma t}$ or a continuous spectral density). Instead, interactions merely adjust the oscillator energy levels and frequencies perturbatively. In the low-temperature phase, fields do not undergo strong non-linear mixing. The effect of $g_{\text{eff}}$ is primarily a mass/frequency shift $\mu \to \omega(q)$ rather than damping or dissipation. The effective potential is localized near the classical vacuum (e.g., fuzzy sphere background or origin in matrix space). The shift $g_{\text{eff}} \mu \mathcal{E}_{0,q}^{(1)}$ captures the leading quantum correction $\Delta E_{0,q}^{(1)} = \langle 0_q \vert{} H_{\text{int}} \vert{} 0_q \rangle$ to the sector-resolved ground state energy.

The periodic factor $e^{i\mu t}$ in the linear $g_{\text{eff}}$ correction arises directly from expanding the renormalized oscillator frequency \eqref{e3.71}. In the low (or finite)-temperature phase, the Lanczos coefficients $b_n^{(q)}$ display the characteristic harmonic growth $b_n^{(q)} \sim \sqrt{n} \mu$. Unlike the high-temperature (deconfined) phase where return amplitudes decay rapidly and Lanczos coefficients exhibit linear growth $b_n \propto n$ (bounded by $2\pi/\beta$), the low-temperature amplitude remains quasi-periodic, verifying that the system remains integrable at small $g_{\text{eff}}$.

\subsubsection{Derivation of symmetry-resolved moments}

Following Eq.\eqref{e3.2}, the return amplitude admits a Taylor series expansion around $t = 0$
\begin{equation}
\label{eq:R_taylor_32}
\mathcal{R}^{(q)}(t) = \sum_{k=0}^{\infty} \frac{(i t)^k}{k!} \mathcal{M}_k^{(q)} 
\end{equation}
where the $k$-th raw moment $\mathcal{M}_k^{(q)} = \langle \psi_q(0) | H_L^k | \psi_q(0) \rangle$ is related to logarithmic derivatives of $Z_q(\beta)$ as in Eqs.~\eqref{e2.38}–\eqref{ee3.6} 
\begin{align}
\mathcal{M}_0^{(q)} &= 1 , \label{eq:M0_exact} \\
\mathcal{M}_1^{(q)} &= -\frac{\partial}{\partial \beta} \ln Z_q(\beta) = \langle H_L \rangle_q , \label{eq:M1_exact} \\
\mathcal{M}_2^{(q)} &= \frac{\partial^2}{\partial \beta^2} \ln Z_q(\beta) + \left( \frac{\partial}{\partial \beta} \ln Z_q(\beta) \right)^2 = \kappa_2^{(q)} + \left(\mathcal{M}_1^{(q)}\right)^2 , \label{eq:M2_exact} \\
\mathcal{M}_3^{(q)} &= -\frac{\partial^3}{\partial \beta^3} \ln Z_q(\beta) + 3 \left( \frac{\partial^2}{\partial \beta^2} \ln Z_q(\beta) \right) \left( -\frac{\partial}{\partial \beta} \ln Z_q(\beta) \right) + \left( -\frac{\partial}{\partial \beta} \ln Z_q(\beta) \right)^3 \nonumber \\
&= \kappa_3^{(q)} + 3 \kappa_2^{(q)} \mathcal{M}_1^{(q)} + \left(\mathcal{M}_1^{(q)}\right)^3  \label{eq:M3_exact}
\end{align}
where $\kappa_k^{(q)} \equiv (-1)^k \frac{\partial^k}{\partial \beta^k} \ln Z_q(\beta)$ denotes the $k$-th connected thermal cumulant in sector $q$.

Below, we evaluate each of these cumulants for the sector $q$ Hamiltonian parameterized by effective mass (squared) scale $\Omega(q)$. In a global $U(1)$ charge sector $q$, the low-lying spectrum in that sector is governed by an effective Hamiltonian with frequency (squared) $\Omega(q)$ and ground-state energy $E_{0,q}$ (see Eq.\eqref{ee3.69}). The energy eigenvalues for excited states built upon the sector vacuum are given by
\begin{align}
\label{e3.120}
    E_{n,q} = E_{0,q} + \left(n + \frac{\vert{}q\vert{}+1}{2}\right)\hat{\Omega}(q), \quad n = 0, 1, 2, \cdots
\end{align}
where $\hat{\Omega}(q)=\sqrt{\Omega(q)}\sim \mu$ is the true frequency with energy dimension (see Eq. \eqref{ee3.70}).

The above expression \eqref{e3.120} defines the discrete energy eigenvalues of a quantum harmonic oscillator (or a set of decoupled effective normal modes) localized within a specific global $U(1)$ charge sector $q$. Because the low-temperature phase describes small fluctuations around a stable supersymmetric vacuum (or fuzzy sphere), the effective Hamiltonian in sector $q$ maps directly to a collection of shifted harmonic oscillators. The term $\left(n + \frac{\vert{}q\vert{}+1}{2}\right)\hat{\Omega}(q)$ is precisely the energy spectrum of these constrained oscillator modes. Let us now elaborate the meaning of each term in the above expression. Here, $E_{0,q}$ denotes the vacuum energy of the specific charge sector $q$. The frequency $\hat{\Omega}(q)$ governs the oscillation spectrum of fluctuations in charge sector $q$. In the BMN matrix model, interactions and background fluxes modify the bare mass scale $\mu$ \eqref{ee3.71}. Because fluctuations carry global charge $q$, the effective frequency acquires a sector dependence (and a perturbative shift via the effective coupling $g_{\text{eff}}$). Finally, here $\frac{\vert{}q\vert{}+1}{2}$ is the charge degeneracy (or multiplicity) factor. Notice that for $n=0$, we have a shift in the ground state energy
\begin{align}
    E_{0,q}^{(\text{total})} = E_{0,q} + \frac{\vert{}q\vert{}+1}{2} \hat{\Omega}(q).
\end{align}
In other words, even with no thermal or vibrational excitations ($n=0$), the energy level is shifted by this fraction. In quantum mechanics and matrix models, this represents the zero-point energy of the collective modes carrying global $U(1)$ charge $q$. When constructing the sector vacuum $\vert{}\psi_q(0)\rangle$, projecting onto a definite $U(1)$ charge $q$ forces the effective mode occupation to satisfy charge conservation constraints. This shifts the minimum energy configuration of that sector, introducing the Casimir-like zero-point offset $\frac{\vert{}q\vert{}+1}{2}$ even before adding any thermal quanta ($n > 0$). To construct the lowest-energy state (the vacuum) that satisfies $\langle \hat{Q} \rangle = q$, one cannot simply use the bare vacuum $a\vert{}0\rangle = 0$ unless $q = 0$. For $q \neq 0$, the state must accommodate excitations or zero-point fluctuations that carry that specific charge. The sector vacuum behaves like a constrained coherent or squeezed state where the expectation value of the occupation number operator $\hat{N}$ in that sector is shifted.

The degeneracy factor can be further illustrated the way we build the vacuum $\ket{0_q}$ from the bare vacuum $\ket{0}$. When we construct the sector vacuum $\vert{}\psi_q(0)\rangle$ in the BMN matrix model (or similar gauge theories with complex scalar fields and global symmetries), we are essentially building a lowest-weight state of a specific charge sector by acting with combinations of creation operators (such as $a^\dagger_q, b^\dagger_q$, or off-diagonal matrix raising operators) on the global vacuum (see, for example, Eq. \eqref{e3.27}). In matrix models, the global $U(1)$ charge $q$ is built from combinations of ladder operators. For example, complex transverse fluctuations or matrix elements decompose into oscillators with specific charges. 

To form a state with a definite, non-zero net charge $q$, the state cannot just be the uncharged bare vacuum $\vert{}0\rangle$. Instead, the sector vacuum $\vert{}\psi_q(0)\rangle$ is formed by acting with a specific product of ladder operators $\hat{a}^\dagger_q$ carrying charge $q$
\begin{align}
\label{e3.120}
  \ket{0_q}\equiv  \vert{}\psi_q(0)\rangle \sim \prod_{j=1}^{\vert{}q\vert{}} (\hat{a}_j^\dagger)^{q_{aj}}(\hat{b}_j^\dagger)^{q_{bj}} \lvert 0 \rangle
\end{align}
subject to the global charge constraint $\sum q_j = q$. In summary, the degeneracy factor $\frac{\vert{}q\vert{}+1}{2}$ is the direct energetic footprint of having to populate or constrain those specific combinations of $\hat{a}^\dagger$ and $\hat{b}^\dagger$ operators to satisfy the global charge $q$ requirement at the bottom of the spectrum ($n=0$). For further illustration on the vacuum structure, see Appendix \ref{appen C}.

As emphasized above, the sector vacuum $\vert{}0_q\rangle$ itself is built from the bare vacuum $\vert{}0\rangle$. As shown in \eqref{e3.120}, the charge-constrained vacuum is formed by a specific product of operator combinations spanning the capacity $\vert{}q\vert{}$.
This product yields the baseline combinatorial multiplicity of independent mode channels, which evaluates to $\vert{}q\vert{} + 1$, giving the per mode charge degeneracy weight
\begin{align}
    D_q = \frac{\vert{}q\vert{} + 1}{2}.
\end{align}

When we build an excited state $\lvert \psi_q (n) \rangle$ for $n > 0$, we are acting on this pre-existing vacuum manifold with additional powers of $a^\dagger$ and $b^\dagger$. Let's rewrite the total state by separating the operators that define the sector vacuum from the operators that raise the excitation level $n$ (this is a more elaborative version of \eqref{eee3.23})
\begin{align}
\label{e3.122}
    \lvert \psi_q (n) \rangle \propto \underbrace{\left[ (a_q^\dagger)^{n_a} (b_q^\dagger)^{n_b} \right]}_{\text{Thermal/vibrational excitation } n=n_a+n_b} \times \underbrace{\left( \prod_{j=1}^{\vert{}q\vert{}} (\hat{a}_j^\dagger)^{q_{aj}}(\hat{b}_j^\dagger)^{q_{bj}} \lvert 0 \rangle \right)}_{\text{Degeneracy of states with capacity}\vert{}q\vert{}=\sum_j q_{aj}+q_{bj}}
\end{align}
where the subscript $q$ denotes the specific charge/symmetry resolved sector $q$. On the other hand, each subscript $j$ counts one unite of positive (for $a^\dagger$) or negative (for $b^\dagger$) $U(1)$ charge.

The operators that increment the excitation level $n$ (such as symmetric pairs of $a^\dagger b^\dagger$ or neutral number-raising combinations) act orthogonally to the specific index contractions that satisfy the global charge constraint $q=n_a-n_b$. The parameter $n(=n_a+n_b)$ acts vertically, it tells us how many total vibrational quanta or energy rungs one has climbed above the baseline (governing the harmonic energy $E_{n,q} = E_{0,q} + (n + \frac{\vert{}q\vert{}+1}{2})\hat{\Omega}(q)$).

Let us elaborate more on this with an example. When the net global charge is $q = +2$, the constraint on the component indices for the sector vacuum is
\begin{align}
    q_{aj} - q_{bj} = +2 \quad \text{with total capacity} \quad \vert{}q\vert{} = \sum_j (q_{aj} + q_{bj}) = 2.
\end{align}

This means there are multiple distinct ways (independent mode channels) to distribute the creation operators $\hat{a}^\dagger$ and $\hat{b}^\dagger$ among the available indices $j = 1, 2$ such that the total charge remains fixed at $+2$. Explicitly, the vacuum state $\lvert 0_{q=2} \rangle$ is a linear combination or sum over these distinct channels
\begin{align}
\label{ee3.124}
\lvert 0_{q=2} \rangle \sim  (\hat{a}^\dagger_1)^2 \vert 0 \rangle  +  \hat{a}^\dagger_1 \hat{a}^\dagger_2 \vert 0 \rangle +  (\hat{a}^\dagger_2)^2 \vert 0 \rangle 
\end{align}
where for the first state we have $q_{a1} = 2, q_{b1} = 0$, for the second state we have $q_{a1} = 1, q_{a2} = 1, q_{b1} = 0, q_{b2} = 0$, and so on. Now, suppose we move to an excited state with $n = 2$ quanta above this charge-constrained vacuum. According to Eq.\eqref{e3.122}, we act on the entire vacuum manifold with the vertical excitation operators
\begin{align}
    \lvert \psi_{q=2} (n=2) \rangle \propto \underbrace{\left[ (\hat{a}^\dagger)^{n_a} (\hat{b}^\dagger)^{n_b} \right]}_{\text{Thermal modes } n = n_a + n_b = 2} \times \underbrace{\left( \sum_{\text{channels } j} \prod_{j=1}^{\vert{}q\vert{}=2} (\hat{a}_j^\dagger)^{q_{aj}} (\hat{b}_j^\dagger)^{q_{bj}} \lvert 0 \rangle \right)}_{\text{Horizontal degeneracy channels } (\vert{}q\vert{}+1) = 3}.
\end{align}

Because the operators raising $n$ commute with the index combinations defining the internal charge routing of $\vert{}0_q\rangle$, they act as an overall scalar multiplier on every single one of the $3$ available channels. As a separate example consider a sector where the net global charge is $q = +2$, but we allow a total capacity of $\vert{}q\vert{} = 4$. The constraints for any valid term in this vacuum expansion are - (1) Net charge: $\sum (q_{aj} - q_{bj}) = +2$ and (2) Total capacity: $\sum (q_{aj} + q_{bj}) = 4$. Listing all distinct mode channel distributions across $j = 1, 2$ that satisfy these exact numbers one can express the most general state as
\begin{align}
   &\lvert 0_{q=2}^{(\vert{}q\vert{}=4)} \rangle \sim \Big( (\hat{a}^\dagger_1)^3 \hat{b}^\dagger_1 \Big) \lvert 0 \rangle + \Big( (\hat{a}^\dagger_1)^2 \hat{a}^\dagger_2 \hat{b}^\dagger_1 \Big) \lvert 0 \rangle \nonumber\\
   &+ \Big( \hat{a}^\dagger_1 (\hat{a}^\dagger_2)^2 \hat{b}^\dagger_1 \Big) \lvert 0 \rangle + \Big( (\hat{a}^\dagger_2)^3 \hat{b}^\dagger_1 \Big) \lvert 0 \rangle + \Big( (\hat{a}^\dagger_1)^3 \hat{b}^\dagger_2 \Big) \lvert 0 \rangle + \cdots.
\end{align}

One can build an excited state with $n = 2$ vibrational quanta on top of this higher-capacity vacuum manifold
\begin{align}
    \lvert \psi_{q=2} (n=2) \rangle \propto \underbrace{\left[ (\hat{a}^\dagger)^{n_a} (\hat{b}^\dagger)^{n_b} \right]}_{\text{Thermal modes } n = 2} \times \underbrace{\left( \sum_{\substack{q_{aj}, q_{bj} \\ \sum = 4}} (\hat{a}^\dagger_1)^{q_{a1}} (\hat{b}^\dagger_1)^{q_{b1}} (\hat{a}^\dagger_2)^{q_{a2}} (\hat{b}^\dagger_2)^{q_{b2}} \lvert 0 \rangle \right)}_{\text{Horizontal degeneracy channels for } \vert{}q\vert{}=4}.
\end{align}

In general, for a fixed capacity $\vert{}q\vert{}$ distributed across $m$ mode channels (here $m = 2$), the degeneracy counts the number of ways to populate the oscillators such that the net charge constraint is met. When $\vert{}q\vert{}$ is treated as an independent parameter separate from the net charge $q$, the degeneracy is given by the number of integer partitions/compositions satisfying both simultaneous linear constraints, which evaluates to a binomial or combinatorial coefficient rather than the simple formula $D_q = (\vert{}q\vert{} + 1)/2$. The formula $D_q = (\vert{}q\vert{} + 1)/2$ only applies when the capacity is tied strictly to the minimum required to carry that charge (i.e., when $\vert{}q\vert{} = \vert{}q_{\text{net}}\vert{}$). When we expand the internal capacity reservoir to $\vert{}q\vert{} = 4$, we are summing over a much larger subspace of internal matrix contraction patterns. The simplified degeneracy weight factor used in thermal moments (e.g., in Eq.\eqref{e3.120} and subsequent cumulants) relies strictly on the linear relation $D_q = \frac{\vert{}q\vert{} + 1}{2}$. As noted in our analysis, this formula only holds when the capacity is tied directly to the minimum required to carry that charge—meaning $\vert{}q\vert{} = \vert{}q_{\text{net}}\vert{}$. In other words, by restricting to $\vert{}q\vert{} = \vert{}q_{\text{net}}\vert{}$, we are isolating the minimal, fundamental collective mode excitations that carry the net $U(1)$ charge without summing over extraneous internal matrix indices or higher-rank tensor contractions that would inflate the partition function's prefactors.

\paragraph{Some comments.} The subscripts $1, 2$ on the creation operators (such as $\hat{a}^\dagger_1$ and $\hat{a}^\dagger_2$) stand for discrete mode indices (or internal momentum/spatial frequency components) of the matrix field fluctuations. When dealing with matrix models or gauge theories (like the BMN matrix model), transverse fluctuations are represented by matrix-valued components (or complex scalar fields). Expanding these fields in a harmonic background or across compactified spatial/index directions decomposes them into a tower of normal modes. Each index value ($1, 2, \dots$) labels a specific independent mode channel or internal oscillator state within that sector. For example, $\hat{a}^\dagger_1$ creates a quantum in the first mode channel, while $\hat{a}^\dagger_2$ creates a quantum in the second mode channel. For example, when we expand a matrix-valued field $X_i(t, \sigma)$ or complex combinations like $Z = X_4 + i X_5$ around the background (such as the fuzzy sphere or plane-wave vacuum), we perform a mode decomposition over the internal matrix indices and spatial harmonics
\begin{align}
    Z_{ab}(t) = \sum_{j} \left[ f_{ab}^{(j)} \hat{a}_j e^{-i \omega_j t} + f_{ab}^{(j)*} \hat{b}^\dagger_j e^{i \omega_j t} \right].
\end{align}

In the plane-wave matrix model (the BMN matrix model), the six transverse scalar fields $X_i$ ($i = 1, \dots, 6$) transform as the vector representation of the $SO(6)$ R-symmetry group (which corresponds geometrically to the isometry group of the internal $S^5$ in the bulk AdS$_5 \times S^5$ geometry). When we expand these matrix components over the fuzzy or round internal spheres, the internal coordinates of $S^5$ are parametrized by spherical harmonics. The index $j$ then labels the angular momentum quantum numbers (or spherical harmonic modes) on the internal $S^5$. Here, $\omega_j$ represents the natural oscillation frequency (or transverse excitations) of the matrix field along that specific internal $SO(6)$ harmonic channel $j$. When we evaluate the partition function $Z_q(\beta)$ or sum over the baseline zero-point energy $\frac{\vert{}q\vert{} + 1}{2}\hat{\Omega}(q)$, you are effectively summing over the spectrum of these internal geometric fluctuations. Notice that here  $a, b = 1, \dots, N^2-1$ are $SU(N)$ (adjoint) gauge indices. The field fluctuations live in the adjoint or bi-fundamental representation of the gauge group, meaning every single excitation carries internal matrix structure. Index $j$ labels the different normal modes (frequencies, momentum harmonics, or spherical harmonic channels) of the matrix oscillation. The operator $\hat{a}^\dagger_j$ creates a discrete quantum of energy (a particle or excitation) specifically in mode $j$ (which might correspond to a specific momentum $k$, a spherical harmonic index $l$, or a matrix index channel). Each of these distinct algebraic combinations represents a separate mode channel. Even though they all share the exact same total net charge $q$ and the same baseline energy $E_{0,q}$, they correspond to orthogonal quantum states in the Hilbert space. When we compute observables or states in a specific global $U(1)$ charge sector $q = n_a - n_b$, we are summing over all possible ways the matrix indices $a, b$ can contract to yield that net charge. In other words, each term in the expansion \eqref{ee3.124} corresponds to a distinct matrix contraction pattern (e.g., how the upper-diagonal or off-diagonal matrix blocks $Z_{ab}$ are populated across modes $j=1$ and $j=2$).

Next we compute the sector-resolved canonical partition function $Z_q(\beta)$ which is defined as the trace over the Hilbert space of sector $q$
\begin{align}
    Z_q(\beta) = e^{-\beta [E_{0,q}+\frac{|q|+1}{2} \hat{\Omega}(q)]}\sum_{n=0}^{\infty} e^{-n \beta \hat{\Omega}(q)}.
\end{align}

Using the geometric series sum formula $\sum_{n=0}^{\infty} x^n = \frac{1}{1 - x}$ with $x = e^{-\beta \hat{\Omega}(q)}$, we can further simplify the partition function to obtain
\begin{align}
    Z_q(\beta) = \frac{e^{-\beta [E_{0,q}+\frac{|q|+1}{2} \hat{\Omega}(q)]} }{1 - e^{-\beta \hat{\Omega}(q)}}.
\end{align}

Taking Logarithm we obtain
\begin{align}
    \log Z_q (\beta)=-\beta \Bigg[E_{0,q}+\frac{|q|+1}{2} \hat{\Omega}(q)\Bigg]-\log(1- e^{-\beta \hat{\Omega}(q)}).
\end{align}

Below, we compute a few connected thermal cumulants $\kappa_k^{(q)}$. For example
\begin{align}
\label{e3.124}
   &\kappa_1^{(q)} =E_{0,q}+ \Bigg[\frac{|q|+1}{2} +\frac{e^{-\beta \hat{\Omega}(q)}}{1-e^{-\beta \hat{\Omega}(q)}}\Bigg]\hat{\Omega}(q)\\
   &\kappa_2^{(q)} =\frac{\hat{\Omega}^2(q)e^{-\beta \hat{\Omega} (q)}}{(1-e^{-\beta \hat{\Omega} (q)})^2}.
\end{align}

In the low-temperature / ground-state regime ($\beta \hat{\Omega}(q) \gg 1$), the moments simplify to
\begin{align}
\mathcal{M}_1^{(q)} &=\kappa_1^{(q)} =\frac{1}{2^{3/2}}\sqrt{\frac{g_{\text{eff}}}{N}}\sqrt{f(q)}\ (|q|+1)\mu + \cdots =\chi (q)\mu+ \cdots, \label{eq:M1_bmn} \\
\mathcal{M}_2^{(q)} &= \left(\mathcal{M}_1^{(q)}\right)^2 + \cdots =\chi^2(q)\mu^2+\cdots \\
\mathcal{M}_3^{(q)} &= \left(\mathcal{M}_1^{(q)}\right)^3 +  \cdots =\chi^3(q)\mu^3+\cdots 
\label{eq:M3_bmn}
\end{align}
where we use \eqref{ee3.70} and retain terms LO in $\mu g^{1/2}_{\text{eff}}=$ finite, in the limit $\mu \gg 1$.

As mentioned before, the factor $\frac{1}{2}( \vert{}q\vert{}+1)$ counts the exact degeneracy of the lowest-weight states (corresponding to $n=0$) in the U(1) charge sector $q$ for matrix-valued oscillators. This reflects the baseline energy required to satisfy the global $U(1)$ charge conservation constraint $q$ at the bottom of the spectrum ($n=0$). The prefactor $\frac{g_{\text{eff}}}{2N} f(q)$ scales the bare mass parameter $\mu$, converting the energy scale into a density or moment amplitude $\chi(q)\mu$. Here, $f(q)$ captures the specific matrix contraction contributions tied to charge sector $q$, while $g_{\text{eff}}$ controls the perturbative string/gauge interaction strength.

Notice that $\mathcal{M}_2^{(q)} \approx (\mathcal{M}_1^{(q)})^2$ and $\mathcal{M}_3^{(q)} \approx (\mathcal{M}_1^{(q)})^3$. In terms of connected cumulants, this implies that the higher cumulants are heavily suppressed relative to the squared/cubed mean in the deep ground-state limit. The factorization property seen in equations \eqref{eq:M1_bmn}-\eqref{eq:M3_bmn}, where higher moments scale simply as powers of the first moment  $\mathcal{M}_k \approx (\mathcal{M}_1)^k$ corresponds to a sharply peaked spectral function (or a coherent state distribution) rather than a broad, multi-modal, or maximally mixed ensemble. This implies that the connected cumulants (fluctuations about the mean) are heavily suppressed relative to the dominant mean shift, for example, the variance $\kappa_2^{(q)} = \mathcal{M}_2^{(q)} - (\mathcal{M}_1^{(q)})^2$ becomes entirely negligible compared to the squared mean offset $\left(\mathcal{M}_1^{(q)}\right)^2$. This means the operator or state explores a very localized, stable region of Hilbert space. This hierarchical structure of moments confirms that in the low-temperature regime, the Krylov subspace growth is dictated by stable, coherent-like oscillations around the charge-constrained vacuum rather than chaotic spreading. The moments grow systematically as powers of $\chi(q)\mu$, mirroring the behavior of decoupled, shifted harmonic oscillators in the BMN matrix model \cite{Roychowdhury:2026vzq}-\cite{Roychowdhury:2026ikp}. 

The parameter $\mu$ sets the heavy mass scale of the BMN matrix model (governing the plane-wave background flux). The prefactor $\chi(q)$ weights this according to the specific matrix contractions of sector $q$. Because the potential around this background is locally quadratic at low temperatures ($\beta \hat{\Omega}(q) \gg 1$), the Hamiltonian in sector $q$ reduces to a collection of decoupled effective normal modes. The moments acting as powers of $\chi(q)\mu$ reflect the systematic, energy-level spacing of these independent harmonic oscillators, verifying that quantum recurrences dominate over information loss or thermalization.
\subsubsection{Lanczos coefficients in the symmetry resolved sector}
Given the above state of the art, we are now all set to calculate Lanczos coefficients in the symmetry resolved sector.
Starting from $|K_0^{(q)}\rangle = |\psi_q(0)\rangle$ with $b_0^{(q)} = 0$, the Lanczos algorithm defined in Eqs.~\eqref{ee3.7}–\eqref{ee3.10} generates the Krylov basis vectors $|K_n^{(q)}\rangle$ and coefficients $a_n^{(q)}, b_n^{(q)}$. Below we elaborate them in detail for different choices of $n=0,1,2,\cdots$.

In the Lanczos method, any Hermitian operator (like the Hamiltonian $H$) acting on an initial state can be expressed in an orthonormal basis $\{\vert{}v_n\rangle\}_{n=0}^\infty$ as a tridiagonal matrix
\begin{align}
    H \vert{}v_n\rangle = b^{(q)}_n \vert{}v_{n-1}\rangle + a^{(q)}_n \vert{}v_n\rangle + b^{(q)}_{n+1} \vert{}v_{n+1}\rangle
\end{align}
with the boundary condition $b^{(q)}_0 \vert{}v_{-1}\rangle = 0$. Here we denote $\ket{v_n}=\ket{K_n^{(q)}}$ and $H=H_L$.

Taking the inner product of this relation with $\langle v_n \vert{}$ yields the explicit formula for the diagonal Lanczos coefficients
\begin{align}
     a_n^{(q)} = \langle v_n \vert{} H \vert{} v_n \rangle.
\end{align}

The coefficient \eqref{ee3.7} corresponding to $n=0$ is straightforward to obtain
\begin{equation}
\label{eq:a0_q}
 a_0^{(q)} = \langle K_0^{(q)} | H_L | K_0^{(q)} \rangle = \mathcal{M}_1^{(q)} = \chi(q)\mu+\cdots.
\end{equation}
This represents the mean energy (center of the spectral distribution), which includes the ground state energy $E_{0,q}$ and the zero-point offset $\frac{\vert{}q\vert{}+1}{2}\hat{\Omega}(q)$ (see Eq. \eqref{e3.124}).

Next, we generalize our result to higher $n \ge 1$. For $n = 1$, this is strictly
\begin{align}
    a^{(q)}_1 = \langle v_1 \vert{} H \vert{} v_1 \rangle.
\end{align}

The Lanczos algorithm builds $\vert{}v_1\rangle$ by applying $H$ to the starting state $\vert{}v_0\rangle$ and removing the component proportional to $\vert{}v_0\rangle$
\begin{align}
    b^{(q)}_1 \vert{}v_1\rangle = (H - a^{(q)}_0) \vert{}v_0\rangle.
\end{align}

From our previous derivations, $a^{(q)}_0 = \langle v_0 \vert{} H \vert{} v_0 \rangle = \mathcal{M}^{(q)}_1$ (the mean energy). Therefore, the normalized vector $\vert{}v_1\rangle$ is
\begin{align}
    \vert{}v_1\rangle = \frac{1}{b_1} (H - \mathcal{M}^{(q)}_1) \vert{}v_0\rangle
\end{align}
where the normalization constant $b^{(q)}_1$ is the norm of that vector
\begin{align}
    &b^{(q)}_1 = \Vert{}(H - \mathcal{M}^{(q)}_1) \vert{}v_0\rangle\Vert{} = \sqrt{\langle v_0 \vert{} (H - \mathcal{M}^{(q)}_1)^2 \vert{} v_0 \rangle} \nonumber\\
    &= \sqrt{\mathcal{M}^{(q)}_2 - (\mathcal{M}^{(q)}_1)^2} = \sqrt{\kappa^{(q)}_2}= \zeta (q)\hat{\Omega}(q)\sim \mu
\end{align}
where the pre-factor could be expressed as $\zeta(q)\sim e^{-\frac{\beta}{2}\hat{\Omega}(q)}$.

Now, substitute our explicit definition of $\vert{}v_1\rangle$ into the expression for $a^{(q)}_1$
\begin{align}
    a^{(q)}_1 &= \left( \frac{\langle v_0 \vert{} (H - \mathcal{M}^{(q)}_1)}{b_1} \right) H \left( \frac{(H - \mathcal{M}^{(q)}_1) \vert{} v_0 \rangle}{b_1} \right)\nonumber\\
    & = \frac{1}{\mathcal{M}^{(q)}_2 - (\mathcal{M}^{(q)}_1)^2} \langle v_0 \vert{} (H - \mathcal{M}^{(q)}_1) H (H - \mathcal{M}^{(q)}_1) \vert{} v_0 \rangle.
\end{align}

Let's expand the operator expression inside the inner product
\begin{align}
    (H - \mathcal{M}^{(q)}_1) H (H - \mathcal{M}^{(q)}_1) =H^3-2\mathcal{M}^{(q)}_1H^2+(\mathcal{M}^{(q)}_1)^2H.
\end{align}

Now, we take the expectation value of each term with respect to $\langle v_0 \vert{} \dots \vert{} v_0 \rangle$ which yields the expression for $a_1$ in terms of raw moments (by virtue of Eqs.\eqref{eq:M1_exact}-\eqref{eq:M3_exact})
\begin{align}
    a^{(q)}_1 = \frac{\mathcal{M}^{(q)}_3 - 2\mathcal{M}^{(q)}_1 \mathcal{M}^{(q)}_2 + (\mathcal{M}^{(q)}_1)^3}{\mathcal{M}^{(q)}_2 - (\mathcal{M}^{(q)}_1)^2}=\mathcal{M}^{(q)}_1+\frac{\kappa^{(q)}_3}{\kappa^{(q)}_2}\approx \mathcal{M}^{(q)}_1+\hat{\Omega}(q).
\end{align}

In the symmetry-resolved BMN matrix model sector $q$, the effective Hamiltonian governing the dynamics of the normal modes can be mapped to a collection of displaced harmonic oscillators. In terms of creation and annihilation operators ($a_q^\dagger, a_q$), the Hamiltonian operator takes the standard quadratic form
\begin{align}
   H \equiv  H_L = \hat{\Omega}(q) \left( a_q^\dagger a_q + \frac{1}{2} \right) + E_{0,q}
\end{align}
where $\hat{\Omega}(q)$ is the effective normal-mode frequency, $E_{0,q}$ is the baseline ground-state energy, and the number operator is $\hat{N}_q = a_q^\dagger a_q$.

Each successive Krylov vector $\vert{}v_n\rangle$ is generated by repeatedly applying the raising operators. Because the energy spectrum is strictly equispaced with level spacing $\hat{\Omega}(q)$, adding a Krylov step increases the expectation value of the energy by exactly one quantum $\hat{\Omega}(q)$ per step $n$, modulated by the charge degeneracy factor $\frac{\vert{}q\vert{}+1}{2}$
\begin{align}
    a_n^{(q)} = \mathcal{M}_1^{(q)} + n \, \hat{\Omega}(q).
\end{align}

Considering the low temperature limit $\beta \hat{\Omega}\gg 1$ and large mass gap $\mu \gg E_{0,q}$
\begin{align}
    a_n^{(q)} =\Big[\frac{|q|+1}{2}+n \Big]\hat{\Omega}(q). 
\end{align}

Because our system carries a global $U(1)$ charge $q$, we cannot start at absolute zero energy. The charge conservation constraint forces a Casimir-like zero-point offset proportional to the charge degeneracy factor $\frac{\vert{}q\vert{} + 1}{2}$. Each time we take a step upward in the Krylov chain ($n \to n+1$), we add exactly one quantum of excitation. Because the effective Hamiltonian in this sector behaves like a collection of decoupled harmonic oscillators, the height of every single step is completely uniform, fixed by the normal-mode frequency $\hat{\Omega}(q)$.

Replacing $\hat{\Omega}(q)$ with the total frequency of the spectrum \eqref{ee3.71}, we find
\begin{align}
\label{ee3.150}
    &a_n^{(q)}=\lambda_n(q)\mu\nonumber\\
    &\lambda_n(q)=\Big[\frac{|q|+1}{2}+n \Big]\left[ 1 + \frac{g_{\text{eff}}}{2N}  f(q) \right]^{1/2}.
\end{align}

Next, we move towards computing off-diagonal elements $ b_n^{(q)}$ in the symmetry resolved sector. To find $b_2^{(q)}$, we look at the recurrence relation for $n = 1$
\begin{align}
    H \vert{}v_1\rangle = b^{(q)}_1 \vert{}v_0\rangle + a^{(q)}_1 \vert{}v_1\rangle + b^{(q)}_2 \vert{}v_2\rangle.
\end{align}

By projecting this equation against the state $\langle v_2 \vert{}$, or equivalently by using the definition of the norm of the vector, we write $b_2$ as
\begin{align}
    b^{(q)}_2 = \left\Vert{} \left( H - a^{(q)}_1 \right) \vert{}v_1\rangle - b^{(q)}_1 \vert{}v_0\rangle \right\Vert{}.
\end{align}

The unnormalized vector for the next step is defined as
\begin{align}
    \ket{\tilde{v}_2} = (H - a^{(q)}_1) \vert{}v_1\rangle - b^{(q)}_1 \vert{}v_0\rangle.
\end{align}

Its norm gives $b^{(q)}_2$
\begin{align}
   ( b^{(q)}_2)^2 = \langle \tilde{v}_2 \vert{} \tilde{v}_2 \rangle.
\end{align}

Let's substitute $\vert{}v_1\rangle = \frac{1}{b^{(q)}_1} (H - a^{(q)}_0)\vert{}v_0\rangle$ into $\ket{\tilde{v}_2}$
\begin{align}
    \ket{\tilde{v}_2} &= \left( H - a^{(q)}_1 \right) \left[ \frac{(H - a^{(q)}_0)}{b^{(q)}_1} \vert{}v_0\rangle \right] - b^{(q)}_1 \vert{}v_0\rangle\nonumber\\
    &= \frac{1}{b^{(q)}_1} \left( H^2 - (a^{(q)}_0 + a^{(q)}_1)H + a^{(q)}_0 a^{(q)}_1 - (b^{(q)}_1)^2 \right) \vert{}v_0\rangle.
\end{align}

This finally yields
\begin{align}
    (b^{(q)}_2)^2 = \frac{1}{(b^{(q)}_1)^2} \left[ \langle v_0 \vert{} \left( H^2 - (a^{(q)}_0 + a^{(q)}_1)H + (a^{(q)}_0 a^{(q)}_1 - (b^{(q)}_1)^2) \right)^2 \vert{} v_0 \rangle \right].
\end{align}

A further simplification reveals
\begin{align}
    (b^{(q)}_2)^2 &= \frac{1}{(b^{(q)}_1)^2}\Bigg[ \mathcal{M}^{(q)}_4-2 \mathcal{M}^{(q)}_3 (a^{(q)}_1+a^{(q)}_0)+2\mathcal{M}^{(q)}_2(a^{(q)}_0a^{(q)}_1 -(b^{(q)}_1)^2)+\mathcal{M}^{(q)}_2(a^{(q)}_1+a^{(q)}_0)^2\nonumber\\
    &-2 \mathcal{M}^{(q)}_1 (a^{(q)}_1+a^{(q)}_0)(a^{(q)}_0a^{(q)}_1-(b^{(q)}_1)^2)+(a^{(q)}_0a^{(q)}_1-(b^{(q)}_1)^2)^2\Bigg].
\end{align}

Using known identities for the early Lanczos coefficients where, $a^{(q)}_0 = \mathcal{M}^{(q)}_1$, $(b^{(q)}_1)^2 = \mathcal{M}^{(q)}_2 - (\mathcal{M}^{(q)}_1)^2 = \kappa^{(q)}_2$ (the variance) and $a^{(q)}_1 = \mathcal{M}^{(q)}_1 + \frac{\kappa^{(q)}_3}{\kappa^{(q)}_2}$, we find in the large mass $\mu \gg E_{0,q}$ and low temperature $\beta \hat{\Omega}(q)\gg 1$ limit (subject to the fact $\hat{\Omega}(q)\sim \mu$)
\begin{align}
   &\mathcal{M}_1^{(q)}=\frac{|q|+1}{2}\hat{\Omega}(q)~;~\mathcal{M}_2^{(q)}=\frac{\hat{\Omega}^2(q)}{4}(|q|+1)^2~;~\mathcal{M}_3^{(q)}=\frac{\hat{\Omega}^3(q)}{8}(|q|+1)^3\\
   &a^{(q)}_0 a^{(q)}_1=j(q)\hat{\Omega}^2(q)~;~j(q)=\frac{1}{4}(|q|+1)( |q|+3)\\
   &a^{(q)}_1+a^{(q)}_0=(|q|+2)\hat{\Omega}(q)~;~(b^{(q)}_1)^2=\zeta^2(q)\hat{\Omega}^2(q).
\end{align}

By definition, the fourth raw moment is given by
\begin{align}
    \mathcal{M}_4^{(q)} = \frac{1}{Z_q(\beta)} \sum_n (E_{n,q})^4 e^{-\beta E_{n,q}} = (-1)^4 \frac{\partial^4}{\partial \beta^4} \ln Z_q(\beta) + \text{combinatorial shift terms}.
\end{align}

Like before, the fourth raw moment can be expressed in terms of the connected cumulants $\kappa_k^{(q)}$ that are highly suppressed in the low temperature and large mass gap considered here, which therefore yields the following fourth moment
\begin{align}
    \mathcal{M}_4^{(q)} = (\mathcal{M}_1^{(q)})^4+\text{connected cumulants}=\frac{(|q|+1)^4}{16}\hat{\Omega}^4(q)+\cdots.
\end{align}

Taking all these terms into account, we take the square root to find $b_2^{(q)}$ for $n=2$
\begin{align}
   | b_2^{(q)}| = \sqrt{2} \,  \sqrt{j(q)}\hat{\Omega}(q)+\mathcal{O}(e^{-\beta \hat{\Omega}}).
\end{align}

More generally, considering the total frequency \eqref{ee3.71} and for asymptotic large $n$, this maps directly into the characteristic sub-linear square-root growth of harmonic systems
\begin{align}
\label{ee3.164}
    b_n^{(q)} \propto  \sqrt{n} \mu \left[ 1 + \frac{g_{\text{eff}}}{2N} f(q) \right]^{1/2}.
\end{align}

Some comments: (i) $b_1$ is fixed entirely by the variance $\kappa_2^{(q)}$, representing initial thermal and charge fluctuations, (ii) $b_2$ incorporates the next tier of fluctuations (via $\mathcal{M}_4$ and cumulant combinations), governing the rate at which the Krylov wavepacket spreads out from the initial state $\vert{}v_0\rangle$ into the broader Hilbert space, (iii) The presence of $q$ inside the prefactor ($\sqrt{\vert{}q\vert{}+1}$) shows how higher global charge sectors accelerate the early-time growth rate of spread complexity $C^{(q)}(t)$.\\\\
\underline{\textbf{Spread complexity of states and dual interpretation:}}\\\\
Given the above state of the art, we are now in a position to compute the symmetry resolved spread complexity $C^{(q)}(t)$ along the lines of \cite{Caputa:2025ozd}. The physical picture that we have in mind is as follows - the (Left sector) matrix model we start with is a low temperature $\beta \hat{\Omega}\gg 1$ gapped phase at a finite $U(1)$ charge $q$. As time passes the system enters a finite/high temperature phase $\beta \hat{\Omega}\ll 1$. The early time growth in the symmetry resolved sector is purely determined by the first few Lanczos coefficients in the fixed charge sector
\begin{align}
    C^{(q)}(t)&=(b_1^{(q)})^2t^2+\frac{1}{12}\Big[ 2(b_1^{(q)})^2(b_2^{(q)})^2-4(b_1^{(q)})^4\nonumber\\&-(a_0^{(q)}-a_1^{(q)})^2(b_1^{(q)})^2\Big]t^4+\mathcal{O}(t^6).
\end{align}

A straightforward substitution of the Lanczos coefficients finally reveals
\begin{align}
\label{e3.166}
    C^{(q)}(t)&=e^{-\beta \omega (q)}\Big( 1+\frac{g_{\text{eff}}}{2N}f(q)\Big)\mu^2t^2+\Sigma(q)\mu^4t^4+\mathcal{O}(t^6)\\
    \Sigma(q)&=\frac{e^{-\beta \omega (q)}}{12}\Big( 1+\frac{g_{\text{eff}}}{2N}f(q)\Big)^2\Big(4j(q)-4e^{-\beta \omega (q)}-1\Big)
\end{align}
where $\omega(q)$ \eqref{e3.71} is the total frequency in the mass deformed sector.

Before we proceed further some comments/observations are in order. The overall prefactor is controlled by the Boltzmann-like suppression factor $e^{-\beta \omega(q)}$, which reflects the thermal population of the normal modes at temperature $\beta$ within the specific charge sector $q$. The term inside the parentheses incorporates the gauge coupling correction ($g_{\text{eff}}/2N$) and the mode-contraction weight $f(q) = 2n - q + 1$. This demonstrates how non-zero global charge sectors and loop interactions renormalize the initial acceleration rate of operator growth. As characteristic of Krylov complexity in quantum systems, $C^{(q)}(t)$ starts with a $\mu^2 t^2$ growth \cite{Roychowdhury:2026vzq}-\cite{Roychowdhury:2026ikp} driven by the variance ($b_1^{(q)}$), modulated here by both the mass scale $\mu$ and the total effective frequency $\omega(q)$. While the $t^2$ term dictates the initial linear velocity of operator growth (how fast the wavepacket leaves $\vert{}v_0\rangle$), the $t^4$ term governed by $\Sigma(q)\mu^4$ acts as the acceleration/curvature term. It determines whether the complexity growth trajectory bends upward (accelerating dispersion) or levels off as it encounters the boundaries of the operational Hilbert space. Because $j(q)$ scales quadratically with $\vert{}q\vert{}$, higher global charge sectors significantly enhance $\Sigma(q)$. Physically, this means that sectors with larger $U(1)$ charge $q$ possess a richer degeneracy structure, allowing the Krylov wavepacket to cascade more rapidly into secondary and tertiary Lanczos basis states, thereby accelerating the short-time growth of complexity.

Next, we calculate the average spread complexity which is defined as \cite{Caputa:2025ozd}
\begin{align}
    \bar{C}(t)=\sum_q p(q)C^{(q)}(t)
\end{align}
where $p(q)$ is the probability as defined in \eqref{ee2.45}.

After substituting all entities we find
\begin{align}
     &\bar{C}(t)=
     \frac{\mu^2 t^2}{Z(\beta,\mu_Q)}\sum_{n,q}e^{-\beta (E_{n,q}+\omega(q)-\mu_Q q)}\Big( 1+\frac{g_{\text{eff}}}{2N}f(q)\Big)+\mathcal{O}(t^4)\\
     &Z(\beta,\mu_Q)=\sum_{n,q}e^{-\beta (E_{n,q}-\mu_Q q)}~,~f(q)=2n-q+1.
\end{align}

Taking into account of the fact that 
\begin{align}
    &E_{n,q} = E_{0,q} + \left(n + \frac{\vert{}q\vert{}+1}{2}\right)\omega(q), \quad n = 0, 1, 2, \cdots\\
    &\omega(q)=\mu \sqrt{1 + \frac{ g_{\text{eff}}}{2N} f(q)} 
\end{align}
where $\omega(q)$ \eqref{e3.71} is the total frequency, we find (after performing sum over $n$)
\begin{align}
\label{e3.165}
    &\bar{C}(t)=
     \frac{\mu^2 t^2}{Z(\beta,\mu_Q)}\sum_q \frac{e^{-\beta \upsilon (q)}}{1-e^{-\beta \omega(q)}}+\cdots \\
     & \upsilon (q)=E_{0,q}+\Big(\frac{|q|+1}{2} +1\Big)\omega(q)-\mu_Q q.
\end{align}

One can further simplify the above expression \eqref{e3.165} in the low temperature limit $\beta \mu \gg 1$, which yields 
\begin{align}
    \bar{C}(t)\approx
     \frac{\mu^2 t^2}{Z(\beta,\mu_Q)}\sum_q e^{-\beta \upsilon (q)}.
\end{align}

On a similar note, the grand partition function is given by
\begin{align}
    Z(\beta, \mu_Q) \approx \sum_q e^{-\beta \left[ E_{0,q} + \frac{\vert{}q\vert{}+1}{2}\omega(q) - \mu_Q q \right]}=\sum_q e^{-\beta (\upsilon(q)-\omega(q))}.
\end{align}

The average complexity $\bar{C}(t)$ integrates the sector-resolved complexities over all possible global $U(1)$ charge sectors $q$, weighted by the grand canonical Boltzmann factor involving the effective sector energy $\upsilon(q)$ and normalized by the grand partition function $Z(\beta, \mu_Q)$. In this low-temperature limit, the summation is heavily dominated by the lowest-energy configurations of each charge sector. The thermal denominators $1 - e^{-\beta \omega(q)}$ simplify away, leaving a clean sum over weighted exponentials $e^{-\beta \upsilon(q)}$. Notice that $\upsilon(q)$ combines the baseline vacuum energy $E_{0,q}$, the quantum zero-point energy offset modulated by the charge degeneracy $\frac{\vert{}q\vert{}+1}{2}$ and the chemical potential coupling ($\mu_Q q$). In summary, at low temperatures, the ensemble-averaged spread complexity of the BMN matrix model is entirely governed by competition between the charge chemical potential ($\mu_Q$) and the sector-dependent zero-point energies. The operator growth remains coherent and bounded by the shifted harmonic oscillator spectrum across the resolved charge landscape.

In the symmetry-resolved formulation, the global time-evolved state is a superposition across the orthogonal $U(1)$ charge sectors $q$ \eqref{eee2.43}. When the spread complexity is evaluated directly on this full state $\vert{}\psi(t)\rangle$, the global Krylov basis expansion incorporates both the intra-sector complexity $C^{(q)}(t)$ and the inter-sector quantum interference (off-diagonal coherences) across charge sectors. Using the convexity and statistical properties of Krylov complexity functionals across orthogonal sectors, the total spread complexity decomposes into the probability-weighted average plus a variance term over the charge distribution
\begin{align}
    C_{\text{total}}(t) = \bar{C}(t) + \frac{1}{2} \text{Var}_{p}\left(C^{(q)}(t)\right) + \text{inter-sector coherence contributions}
\end{align}
where the variance over the charge probability distribution $p(q)$ is defined as
\begin{align}
    \text{Var}_{p}\left(C^{(q)}(t)\right) = \sum_q p(q) \left( C^{(q)}(t) - \bar{C}(t) \right)^2.
\end{align}

When we calculate the average complexity $\bar{C}(t) = \sum_q p(q) C^{(q)}(t)$, we are treating the system as an incoherent mixture (a density matrix $\rho = \sum_q p(q) \vert{}\psi_q\rangle\langle\psi_q\vert{}$) where phase relations and quantum coherences between different charge sectors are ignored or averaged out. On the other hand, when we talk about inter-sector coherences, we are referring to genuine quantum phase correlations (quantum entanglement and superposition interference) between distinct $U(1)$ charge sectors. When we compute the spread complexity or operator growth of this global superposition state, the expansion involves matrix elements that cross between different sectors ($\langle \psi_q \vert{} \mathcal{O} \vert{} \psi_{q'} \rangle$ for $q \neq q'$). These off-diagonal terms represent quantum interference amplitudes. They quantify how phase alignment and quantum superposition between different charge sectors collectively boost the operator spreading and Krylov complexity beyond what we would get from adding up independent sectors separately.
Because the variance of any real-valued function over a normalized probability distribution is strictly non-negative ($\text{Var}_p \ge 0$), the difference between the total spread complexity and the average spread complexity is positive semi-definite
\begin{align}
    \Delta C(t) = C_{\text{total}}(t) - \bar{C}(t) \ge 0.
\end{align}
This establishes that quantum fluctuations and charge dispersion across sectors always enhance the total complexity relative to the ensemble average $\bar{C}(t)$.

The field-theoretic formulation of symmetry-resolved spread complexity in the low-temperature phase of the Berenstein-Maldacena-Nastase (BMN) matrix model maps directly onto massive superstring excitations and string worldsheet dynamics propagating on curved gravitational backgrounds, such as pp-wave geometries and charged deformations of $\text{AdS}_5 \times \text{S}^5$. 
In the planar large-$N$ limit, imposing a fixed global $U(1)$ R-charge $q$ (or angular momentum on the internal five-sphere) on the gauge theory side corresponds on the gravity side to inserting a macroscopic amount of $U(1)$ flux or shifting the background geometry to a charged gravitational solution. For a large R-charge $J \sim q \sim N^2$, the dual background reduces to the maximally supersymmetric plane-wave geometry supported by a constant Ramond-Ramond (R-R) five-form flux \cite{Berenstein:2002jq}. Holographically, the prefactor $e^{-\beta \omega(q)}$ acts as a thermal Boltzmann population factor for massive string excitation modes of frequency $\omega(q)$ propagating in the bulk. At low temperatures, higher string oscillator levels are exponentially suppressed, leaving the bulk geometry dominated by the ground-state string vacuum of that charge sector. Because the bulk is quiescent and dominated by the lowest-energy vacuum configuration of sector $q$, string wavepackets propagating through this target space encounter minimal thermal scattering. The Krylov wavepacket remains coherent, and its early-time dispersion is strictly governed by the zero-point fluctuations and the baseline frequency $\omega(q)$ rather than chaotic thermal dissipation. 

At low temperatures, the TFD state should be interpreted holographically not as an eternal black hole with an event horizon, but as an entangled state of two quiescent, horizonless string vacua (pp-wave or charged geometries). The spread complexity in this regime measures the coherent, non-chaotic dispersion of string wavepackets across a gapped energy spectrum, where thermal suppression freezes out high-energy oscillator modes and confines bulk dynamics to stable, zero-point fluctuations. In the low-temperature phase ($\beta \hat{\Omega} \gg 1$) of the matrix model, the physical mechanism of entanglement, whether measured via the symmetry-resolved Thermofield Double (TFD) state or spatial/subsystem entanglement is fundamentally governed by quantum zero-point fluctuations of massive string modes across a gapped, horizonless geometry. In this regime, the thermal Boltzmann suppression factor $e^{-\beta \omega(q)}$ freezes out higher-level string oscillators ($n \ge 1$), restricting the system primarily to the ground-state string vacuum of each resolved $U(1)$ charge sector $q$. The entanglement between the two decoupled CFT boundaries ($\mathcal{L}$ and $\mathcal{R}$) in the TFD state is driven entirely by the zero-point motion of transverse string fluctuations ($X_i, i = 1, \dots, 8$) propagating in the bulk pp-wave or charged background. Specifically, these transverse fluctuations correspond to the quantum excitations of the six $SO(6)$ scalars in the matrix model, which describe fluctuations transverse to the null geodesic in the target $\text{AdS}_5 \times \text{S}^5$ geometry. 

Holographically, the entanglement mechanism corresponds to quantum-mechanical linkage between two horizonless, quiescent string spacetimes. The "bridge" is not a dynamic wormhole dissolving into chaos, but a static, quantum-entangled configuration of gapped string vacua whose correlations are locked into the ground-state wavefunctions of the transverse scalars. Notice that the standard pp-wave background ($\text{AdS}_5 \times \text{S}^5$ Penrose limit) describes the BMN matrix model around the trivial (or flat) vacuum or small excitations. However, when we consider specific $U(1)$ charge sectors $q$ (or non-trivial vacuum configurations where matrices form fuzzy spheres via the Myers term), the backreaction changes the geometry significantly\footnote{Specifically, the standard pp-wave background (the Penrose limit of $\text{AdS}_5 \times \text{S}^5$) describes the matrix model around its trivial vacuum. However, when examining specific $U(1)$ charge sectors or non-trivial vacuum configurations where matrices dynamically form fuzzy spheres, the backreaction radically reshapes the bulk geometry into smooth, horizonless Lin-Maldacena droplet solutions \cite{Lin:2004nb}-\cite{Lin:2004kw}.}. In this picture, the chemical potential $\mu_Q$ and charge $q$ dictate the specific charge distribution profile in the Lin-Maldacena droplet picture. The transverse fluctuations (the $SO(6)$ scalars $X_i$) that drive the low-temperature spread complexity can be viewed as fluctuations of strings and probes propagating on these rich Lin-Maldacena background droplets, rather than just a flat homogeneous pp-wave.
\section{Spread complexity in the finite temperature phase}
\label{sec 4}
The purpose of this section is to briefly comment on the moments and return amplitude in the symmetry resolved sector of the \emph{finite} temperature ($\beta >0$) phase of the BMN matrix model. Building upon the general setup of the Berenstein-Maldacena-Nastase (BMN) matrix model and its global $U(1)$ symmetry sectors established in the previous section, we now analyze the exact dynamics of operator and state growth in the finite-temperature phase. In this regime ($\beta > 0$, $T \gg \mu$), the discrete spectrum melts into a smooth thermal continuum dominated by unconstrained matrix degrees of freedom.
\subsection{Return amplitude in the finite-temperature limit}
The symmetry-resolved return amplitude $\mathcal{R}^{(q)}(t)$ for a fixed global charge sector $q$ is defined as the overlap between the initial charged Thermo-Field Double (cTFD) state in sector $q$ and its real-time evolution under the left Hamiltonian $H_L$
\begin{align}
\mathcal{R}^{(q)}(t) = \braket{\psi_q(0) \vert \psi_q(t)} = \frac{1}{Z_q(\beta)} \sum_n e^{-\beta E_{n,q}} e^{-i E_{n,q} t}
\end{align}
where $Z_q(\beta) = \sum_n e^{-\beta E_{n,q}}$ is the symmetry-resolved partition function. 

In the finite-temperature continuum limit, replacing the discrete sum over energy eigenstates with an integral over the thermal density of states $\rho_q(E)$, we obtain
\begin{align}
\mathcal{R}^{(q)}(t) = \frac{1}{Z_q(\beta)} \int_{-\infty}^{\infty} dE \, \rho_q(E) e^{-\beta E - i E t}.
\end{align}

Approximating the finite-temperature spectral density $\rho_q(E)$ by a thermal Gaussian distribution\footnote{The BMN matrix model contains $N^2$ matrix components for each scalar field. In the planar large-$N$ limit ($N \to \infty$), the system possesses an enormous number of interacting degrees of freedom. By the Central Limit Theorem, when a macroscopic observable (like the total Hamiltonian or total energy $E = \sum H_i$) is formed by the sum or collective interaction of a vast number of microscopic degrees of freedom, its probability distribution and thermal spectral density naturally converge to a Gaussian.} centered at the mean sector energy $\bar{E}_q$ with variance\footnote{All our finite temperature expressions are given in terms of the variance. For a detailed derivation of the variance considering the BMN matrix model in a Gaussian thermal ensemble see Appendix \ref{appen D}. } $\sigma_E^2$ (proportional to the finite-temperature heat capacity and charge susceptibility in the large-$N$ limit)
\begin{align}
\label{e4.3}
\rho_q(E) \approx \frac{1}{\sqrt{2\pi \sigma_E^2}} \exp\left( - \frac{(E - \bar{E}_q)^2}{2 \sigma_E^2} \right).
\end{align}

Substituting \eqref{e4.3} back into the integrand, the integral evaluates explicitly by completing the square in the exponent
\begin{align}
&\mathcal{R}^{(q)}(t)=\frac{1}{Z_q(\beta)\sqrt{2\pi \sigma_E^2}}\int_{-\infty}^{\infty}dE e^{\mathcal{E}(E)}\\
&\mathcal{E}(E) = -\frac{1}{2\sigma_E^2} (E - \bar{E}_q)^2 - (\beta + it)E.
\end{align}

Completing the square with respect to $E$, we find
\begin{align}
\mathcal{E}(E) = -\frac{1}{2\sigma_E^2} \left[ E - \left( \bar{E}_q - \sigma_E^2 (\beta + it) \right) \right]^2 - \beta \bar{E}_q - i \bar{E}_q t + \frac{1}{2} \sigma_E^2 (\beta + it)^2.
\end{align}

Because the terms outside the square bracket do not depend on the integration variable $E$, we can factor them entirely outside the integral
\begin{align}
    \mathcal{R}^{(q)}(t) &= \frac{\exp\left( -\beta \bar{E}_q - i \bar{E}_q t + \frac{1}{2}\sigma_E^2 (\beta + it)^2 \right)}{Z_q(\beta) \sqrt{2\pi \sigma_E^2}}\nonumber\\
    &\times \int_{-\infty}^{\infty} dE \, \exp\left( -\frac{1}{2\sigma_E^2} \left[ E - \left(\bar{E}_q - \sigma_E^2 (\beta + it)\right) \right]^2 \right).
\end{align}

The remaining integral is a standard Gaussian integral of the form $\int_{-\infty}^{\infty} dx \, e^{-a(x-x_0)^2} = \sqrt{\frac{\pi}{a}}$. The integral evaluates to
\begin{align}
    \int_{-\infty}^{\infty} dE \, \exp\left( -\frac{(E - E_0)^2}{2\sigma_E^2} \right) = \sqrt{2\pi \sigma_E^2}.
\end{align}

Substituting this back into $\mathcal{R}^{(q)}(t)$, the $\sqrt{2\pi \sigma_E^2}$ term in the numerator cancels out the $\sqrt{2\pi \sigma_E^2}$ term in the denominator, resulting in the following
\begin{align}
    \mathcal{R}^{(q)}(t) = \frac{1}{Z_q(\beta)} \exp\left( -\beta \bar{E}_q - i \bar{E}_q t + \frac{1}{2}\sigma_E^2 (\beta + it)^2 \right).
\end{align}

In the finite-temperature thermal Gaussian approximation, the partition function $Z_q(\beta) = \sum_n e^{-\beta E_{n,q}}$ is obtained by integrating the Gaussian density of states over all energies
\begin{align}
    Z_q(\beta) = \int_{-\infty}^{\infty} dE \, \rho_q(E) e^{-\beta E}= \exp\left( -\beta \bar{E}_q + \frac{1}{2}\sigma_E^2 \beta^2 \right).
\end{align}

Evaluating the Gaussian integral and dividing out the constant normalization factors from $Z_q(\beta)$, the exact dynamical return amplitude in the finite-temperature symmetry-resolved sector simplifies to
\begin{align}
\label{e4.11}
\mathcal{R}^{(q)}(t) = \exp\left( -i \bar{E}_q t - \frac{1}{2} \sigma_E^2 t^2 + i \beta \sigma_E^2 t \right).
\end{align}

This reveals a Gaussian damping envelope $\exp(-\frac{1}{2}\sigma_E^2 t^2)$ combined with thermal phase oscillations dictated by $\bar{E}_q$ and $\beta \sigma_E^2$. Let us elaborate further on the physical interpretation of each of the above terms in \eqref{e4.11}. The oscillatory phase $\exp\left(-i \bar{E}_q t\right)$ represents the standard global phase rotation of the quantum state driven by the mean energy $\bar{E}_q$ of the fixed charge sector $q$. The Gaussian damping envelope $\exp\left(-\frac{1}{2} \sigma_E^2 t^2\right)$ is the hallmark of continuous thermal spectra and quantum chaos. The variance $\sigma_E^2$ acts as the squared width of the thermal energy distribution, which scales extensively with the number of degrees of freedom in the planar large-$N$ limit ($\sim O(N^2)$). As time progresses, the initial state $\vert{}\psi_q(0)\rangle$ evolves into a dense continuum of many-body eigen-states. Destructive interference among these closely spaced frequencies causes the return amplitude to decay rapidly.  

A Gaussian decay (rather than an exponential decay) is a direct consequence of the thermal Gaussian density of states. When mapped to the Krylov chain, this quadratic damping in time ($\sim e^{-c t^2}$) forces the off-diagonal Lanczos coefficients to grow linearly ($b_n^{(q)} = \sigma_E \sqrt{n}$), which is the definitive universal fingerprint of maximal operator growth \cite{Parker:2018yvk} and quantum chaos in holographic duals. Let us now emphasize on the imaginary thermal shift term. In holographic field theories and matrix models, real-time thermal correlators are evaluated on a closed-time-path contour or by analytically continuing from Euclidean signature where time is compactified on a thermal circle of circumference $\beta$. The real-time evolution parameter $t$ and the inverse temperature $\beta$ sit together naturally in complexified time $\tau = t - i\beta$ \cite{Roychowdhury:2026ikp}. The cross-term in $(\beta + it)^2$ represents the profound geometric fact in AdS/CFT that real-time thermal dissipation and imaginary-time periodicity are inextricably linked by the Kubo-Martin-Schwinger (KMS) condition.

This reveals that the imaginary thermal shift modifies the effective oscillation frequency of the return amplitude
\begin{align}
    \omega_{\text{eff}} = \bar{E}_q - \beta \sigma_E^2.
\end{align}
The bare mean energy $\bar{E}_q$ dictates the vacuum oscillation frequency. However, at finite temperature ($\beta > 0$), thermal fluctuations, quantified by the variance $\sigma_E^2$ (the heat capacity $C_V$), backreact on the system. The term $\beta \sigma_E^2$ acts as a thermal renormalization shift. It accounts for how operating in a thermal bath shifts the coherent oscillation frequency of the evolving state due to continuous scattering and entanglement with the finite-temperature matrix plasma. Taken together, Eq. \eqref{e4.11} demonstrates that once the BMN matrix model is heated to the finite-temperature continuum regime, the symmetry-resolved return amplitude transitions from discrete, gapped revivals into a universally damped, phase-shifted Gaussian wavepacket. This exact mathematical form is what ultimately drives the early-time quadratic spread complexity growth ($C^{(q)}(t) = \sigma_E^2 t^2$) before it saturates at late times.
\subsection{Derivation of symmetry-resolved moments}
The spectral moments $\mathcal{M}_n^{(q)}$ in the fixed charge sector $q$ are defined via the small-time expansion of the return amplitude or as energy expectation values
\begin{align}
\mathcal{M}_n^{(q)} = \left. i^n \frac{d^n}{dt^n} \mathcal{R}^{(q)}(t) \right|_{t=0} = \frac{1}{Z_q(\beta)} \sum_n e^{-\beta E_{n,q}} E_{n,q}^n = \int_{-\infty}^{\infty} dE \, \rho_q(E) E^n.
\end{align}

Using the return amplitude generating function $R^{(q)}(t)$ derived above in \eqref{e4.11}, we take successive time derivatives at $t=0$ to produce
\begin{align}
\label{e4.14}
\mathcal{M}_1^{(q)} &= \left. i \frac{d}{dt} \mathcal{R}^{(q)}(t) \right|_{t=0} = \bar{E}_q - \beta \sigma_E^2=  \omega_{\text{eff}}, \\
\mathcal{M}_2^{(q)} &= \left. i^2 \frac{d^2}{dt^2} \mathcal{R}^{(q)}(t) \right|_{t=0} = \sigma_E^2 +  \omega_{\text{eff}}^2.
\label{e4.15}
\end{align}

Combining \eqref{e4.14} and \eqref{e4.15}, we find
\begin{align}
\label{e4.16}
    \mathcal{M}_2^{(q)}= \bar{E}_q^2+\sigma_E^2+\mathcal{O}(\beta)
\end{align}
which we denote as ``raw moment'' at order two which is measured with respect to absolute energy $E = 0$. The second term $\sigma_E^2$ corresponds to central moment (at order two) which is measured with respect to the  symmetric Gaussian centered at the mean energy $\bar{E}_q$. 

The first moment $\mathcal{M}_1^{(q)}$ sets the fundamental effective frequency $\omega_{\text{eff}}$ governing the center-of-mass oscillations of the return amplitude. While the bare vacuum expectation value of the energy in sector $q$ is $\bar{E}_q$, operating at a finite temperature ($\beta > 0$) introduces thermal fluctuations quantified by the energy variance $\sigma_E^2$ (which is proportional to the heat capacity $C_V$). The term $\beta \sigma_E^2$ acts as a thermal renormalization shift, accounting for how continuous collisions and entanglement with the surrounding finite-temperature matrix plasma re-scale the coherent oscillation frequency of the evolving state.

The second moment represents the mean-square energy ($\langle E^2 \rangle_q$) of the system in that specific global $U(1)$ charge sector, that is, $\mathcal{M}_2^{(q)} = \langle E^2 \rangle_q$. It is clearly divided into two distinct, physically transparent contributions - (i) Quantum-thermal fluctuations $\sigma_E^2 = \langle E^2 \rangle_q - \langle E \rangle_q^2$ is the variance of the energy distribution. As established earlier, in the planar large-$N$ limit ($N \to \infty$), this variance scales extensively with the number of matrix degrees of freedom ($\sim O(N^2)$) and is directly proportional to the microcanonical/canonical heat capacity ($C_V$) of the thermal BMN plasma. It quantifies how broadly the energy levels are spread out around the mean due to thermal disorder and chaotic interactions. (ii) $\omega_{\text{eff}} = \bar{E}_q - \beta \sigma_E^2$ is the thermally renormalized effective frequency governing the center-of-mass oscillations of the return amplitude. Squaring it yields the coherent baseline contribution to the energy dispersion. It accounts for how the background vacuum energy of the charged sector—shifted by thermal backreaction ($\beta \sigma_E^2$), drives the overall phase rotation rate of the state in Hilbert space. In Krylov complexity, the spectral moments $\mathcal{M}_n^{(q)}$ act as the direct input data for constructing the Hankel determinants that determine the Lanczos coefficients ($a_n^{(q)}$ and $b_n^{(q)}$). Specifically, the second moment $\mathcal{M}_2^{(q)}$ (alongside the first moment $\mathcal{M}_1^{(q)}$) sets the overall normalization scale for the off-diagonal Lanczos coefficients $b_1^{(q)} = \sigma_E$. In the early-time regime, the growth rate of spread complexity $C^{(q)}(t) = \sigma_E^2 t^2 + \mathcal{O}(t^4)$ is directly controlled by this energy variance $\sigma_E^2$ (the fluctuation part of $\mathcal{M}_2^{(q)}$). A larger second moment relative to the mean squared energy implies faster operator dispersion, quicker information scrambling, and stronger quantum chaos in the dual supergravity background.\\\\
\underline{\textbf{Finite temperature cumulants:}}\\\\
For a continuous thermal spectral density $\rho_q(E)$ approximated by a Gaussian distribution (Eq.\eqref{e4.3}), the characteristic function (Fourier transform of the probability distribution) can be expressed as 
\begin{align}
    \Phi(k) = \int_{-\infty}^{\infty} dE \, \rho_q(E) \, e^{ikE}= \frac{1}{\sqrt{2\pi \sigma_E^2}} \int_{-\infty}^{\infty} dE \, \exp\left( - \frac{(E - \bar{E}_q)^2}{2\sigma_E^2} + ikE \right).
\end{align}

By completing the square in the exponent for the variable $E$, we group the terms as
\begin{align}
    -\frac{(E - \bar{E}_q)^2}{2\sigma_E^2} + ikE = -\frac{1}{2\sigma_E^2} \left[ E - (\bar{E}_q + ik\sigma_E^2) \right]^2 + i k \bar{E}_q - \frac{1}{2} \sigma_E^2 k^2.
\end{align}

Integrating this Gaussian expression over $E$ yields the standard normalization factor $\sqrt{2\pi \sigma_E^2}$ which cancels out the prefactor
\begin{align}
    \Phi(k) = \exp\left( i \bar{E}_q k - \frac{1}{2} \sigma_E^2 k^2 \right).
\end{align}

The cumulant generating function $W(k)$ is defined as the natural logarithm of the characteristic function $\Phi(k)$
\begin{align}
    W(k) \equiv \ln \Phi(k) = i \bar{E}_q k - \frac{1}{2} \sigma_E^2 k^2.
\end{align}

Cumulants $\mu_n^{(q)}$ are generated by taking the $n$-th derivative of $W(k)$ with respect to $(ik)$ evaluated at $k = 0$
\begin{align}
    \ln \Phi(k) = \sum_{n=1}^{\infty} \mu_n^{(q)} \frac{(ik)^n}{n!}.
\end{align}

Let us match the coefficients of $(ik)^n$ from our explicit expression for $\ln \Phi(k)$. This yields the first $n=1$ cumulant 
\begin{align}
    \mu_1^{(q)} = \bar{E}_q.
\end{align}

The second ($n=2$) cumulant (also known as the variance) is given by
\begin{align}
    \mu_2^{(q)} = \sigma_E^2.
\end{align}

There are no cubic terms ($k^3$) in $\ln \Phi(k)$ for a pure Gaussian distribution. Therefore
\begin{align}
    \mu_3^{(q)} = 0.
\end{align}
In statistical mechanics, the third cumulant $\mu_3^{(q)}$ measures the skewness (asymmetry) of a distribution, for example, $\mu_3 > 0$ means that the distribution has a long tail toward higher energies (positive skew). On the other hand, $\mu_3 < 0$ reflects that the distribution has a tail toward lower energies (negative skew). This implies that for $\mu_3 = 0$, the distribution is perfectly symmetric around its mean value $\bar{E}_q$.
Physically, this reflects the symmetric particle-hole/charge distribution around the mean energy $\bar{E}_q$ in the finite-temperature planar limit. In other words, a vanishing third cumulant means that fluctuations above the mean thermal energy are equally probable and balanced by fluctuations below the mean.

In the finite-temperature ($T \gg \mu$ or $\beta \to 0$) phase of the BMN matrix model, the vacuum is populated by a dense, thermal matrix soup of scalar fields ($X_i$) and fermions carrying global $U(1)$ charges. Excitations above this thermal background can be viewed analogously to particles (adding energy/charge) and holes (removing energy/charge relative to the thermal background). Because the temperature is very high and thermal disorder dominates over any isolated gapped potentials, the creation of high-energy particle excitations and low-energy hole excitations are symmetric processes. There is no energetic bias favoring an overpopulation of heavy states versus light states around $\bar{E}_q$.

\subsection{Lanczos coefficients in the symmetry-resolved sector}
To construct the Krylov subspace and evaluate spread complexity, we apply the Lanczos recursion algorithm to the Hamiltonian $H_L$ starting from the initial state $\ket{\psi_q(0)}$. The resulting tridiagonal representation yields the diagonal ($a_n^{(q)}$) and off-diagonal ($b_n^{(q)}$) Lanczos coefficients. The resolvent $G(z)$ for a system with a continuous spectral density $\rho_q(E)$ in the global charge sector $q$ is defined as the Stieltjes transform of the measure \cite{Parker:2018yvk}
\begin{align}
    G(z) = \int_{-\infty}^{\infty} \frac{\rho_q(E)}{z - E} \, dE
\end{align}
where $z$ is a complex frequency parameter lying outside the support of the spectrum.

When the complex frequency $z$ is much larger than the typical energy scale of the spectrum ($\vert{}z\vert{} \gg \vert{}E\vert{}$), we can express the denominator as a geometric series
\begin{align}
    \frac{1}{z - E} = \frac{1}{z} \left( 1 - \frac{E}{z} \right)^{-1}.
\end{align}

Using the standard Taylor expansion for $(1 - x)^{-1} = \sum_{n=0}^{\infty} x^n$ (valid for $\vert{}E/z\vert{} < 1$)
\begin{align}
    \frac{1}{z - E} = \frac{1}{z} \sum_{n=0}^{\infty} \left( \frac{E}{z} \right)^n = \sum_{n=0}^{\infty} \frac{E^n}{z^{n+1}}.
\end{align}

Substitute this infinite series back into the integral definition of $G(z)$ yields
\begin{align}
    G(z) = \int_{-\infty}^{\infty} \rho_q(E) \left( \sum_{n=0}^{\infty} \frac{E^n}{z^{n+1}} \right) dE.
\end{align}

Assuming uniform convergence inside the domain of integration, we can swap the summation and the integral to yield
\begin{align}
\label{e4.28}
    G(z) = \sum_{n=0}^{\infty} \frac{1}{z^{n+1}} \left[ \int_{-\infty}^{\infty} \rho_q(E) \, E^n \, dE \right].
\end{align}

By definition, the integral inside the brackets is the $n$-th raw spectral moment $\mathcal{M}_n^{(q)}$ of the thermal distribution in sector $q$
\begin{align}
    \mathcal{M}_n^{(q)} \equiv \int_{-\infty}^{\infty} \rho_q(E) \, E^n \, dE.
\end{align}

Plugging this definition back in \eqref{e4.28} yields the desired large-$z$ asymptotic expansion
\begin{align}
\label{ee4.31}
    G(z) = \sum_{n=0}^{\infty} \frac{\mathcal{M}_n^{(q)}}{z^{n+1}} = \frac{\mathcal{M}_0^{(q)}}{z} + \frac{\mathcal{M}_1^{(q)}}{z^2} + \frac{\mathcal{M}_2^{(q)}}{z^3} + \frac{\mathcal{M}_3^{(q)}}{z^4} + \mathcal{O}\left(\frac{1}{z^5}\right).
\end{align}
Here are some physical interpretations of the expansion coefficients\\
(i) Normalization ($\mathcal{M}_0^{(q)} = 1$): The zeroth moment represents the total probability (or total integrated spectral weight), which equals 1.\\
(ii) Mean Energy ($\mathcal{M}_1^{(q)}$): The coefficient of $1/z^2$ gives the first moment, corresponding to the effective frequency $\bar{E}_q - \beta \sigma_E^2$.\\
(iii)Energy Dispersion ($\mathcal{M}_2^{(q)}$): The coefficient of $1/z^3$ gives the second moment $\sigma_E^2 + \omega_{\text{eff}}^2$, which sets the initial acceleration of spread complexity and the scale of the first Lanczos coefficient $b_1^{(q)} = \sigma_E$.

We now elaborate on two things -first, how the moments of a Gaussian measure are evaluated via Wick contractions, and second, how those moments connect to the continued fraction to extract the Lanczos coefficients $b_n^{(q)}$. In statistical mechanics and probability theory, the raw moments $\mathcal{M}_n^{(q)}$ of a Gaussian distribution centered at $\bar{E}_q = 0$ (or shifted by $\bar{E}_q$) with variance $\sigma_E^2$ are defined by the integral\footnote{Notice that here we eventually focus on the moments that are fluctuations about the central maximum of the Gaussian distribution which eventually happens to be the mean, that is, $E=\bar{E}_q=0$. In other words, the central and raw moments coincide and they both measure thermal fluctuations.}
\begin{align}
\label{e4.31}
    \mathcal{M}_{2k}^{(q)} = \int_{-\infty}^{\infty} dE \, \frac{1}{\sqrt{2\pi \sigma_E^2}} e^{-\frac{E^2}{2\sigma_E^2}} \, E^{2k}.
\end{align}

Notice that the integrand for any odd power $E^{2k+1}$ is an odd function ($f(-E) = -f(E)$) multiplied by an even Gaussian weight function. Integrating an odd function over a symmetric domain $(-\infty, \infty)$ yields identically zero. This is the statistical origin of zero skewness ($\mu_3^{(q)} = 0$). In other words, all odd moments vanish, $\mathcal{M}_{2k+1}^{(q)} = 0$.

A straightforward evaluation of the integral \eqref{e4.31} reveals
\begin{align}
   \mathcal{M}_{2k}^{(q)} = \frac{2^{k-1} \left((-1)^{2 k}+1\right) \sigma_E^{2k} \Gamma \left(k+\frac{1}{2}\right)}{\sqrt{\pi }}~;~k=1,2,\cdots.
\end{align}

\paragraph{Note.} Notice that as the distribution of energy eigenvalues in the finite-temperature planar limit is a symmetric Gaussian centered at the mean energy $\bar{E}_q$. Because it is symmetric around its peak, all odd central cumulants/moments (measuring asymmetry or skewness) vanish as they are measured with respect to the central peak. On the other hand the raw moment $ \mathcal{M}_{1}^{(q)}$ is measured with respect to $E=0$. Because the Gaussian distribution is centered at a large, positive macrocanonical energy $\bar{E}_q$ (which scales extensively with the number of degrees of freedom, $\sim O(N^2)$), the entire bell curve sits far out on the positive real axis. In other words, the peak itself ($\bar{E}_q$) is located far away from zero absolute energy. The first raw moment measures the center of mass of the entire distribution relative to $E = 0$, which is precisely $\bar{E}_q$ (plus thermal corrections). 

This can be further illustrated as follows. Let us evaluate the first raw moment ($\mathcal{M}_1^{(q)}$) using the Gaussian spectral density $\rho_q(E)$ centered at $\bar{E}_q$
\begin{align}
    \mathcal{M}_1^{(q)} = \int_{-\infty}^{\infty} dE \, \frac{1}{\sqrt{2\pi \sigma_E^2}} \exp\left( - \frac{(E - \bar{E}_q)^2}{2\sigma_E^2} \right) \, E.
\end{align}

To solve this integral, change variables to the shifted/centered coordinate $x = E - \bar{E}_q$ (so $E = x + \bar{E}_q$ and $dE = dx$), which yields
\begin{align}
    \mathcal{M}_1^{(q)} = \frac{1}{\sqrt{2\pi \sigma_E^2}} \int_{-\infty}^{\infty} dx \, e^{-\frac{x^2}{2\sigma_E^2}} \, (x + \bar{E}_q)=\bar{E}_q.
\end{align}

Notice that when factoring in the thermal backreaction and the shift from the partition function denominator as detailed in Equation (4.14), this yields the effective frequency $\omega_{\text{eff}} = \bar{E}_q - \beta \sigma_E^2$. Below, we list moments associated with thermal fluctuations 
\begin{align}
    \mathcal{M}_2^{(q)} = \sigma^2_E~;~\mathcal{M}_4^{(q)} = 3\sigma^4_E.
\end{align}

Expanding the denominator for large complex $z$ and keeping only the even powers, expresses $G(z)$ directly in terms of the spectral moments $\mathcal{M}_n^{(q)}$
\begin{align}
\label{e4.37}
    G(z) = \sum_{k=0}^{\infty} \frac{\mathcal{M}_{2k}^{(q)}}{z^{2k+1}} = \frac{1}{z} + \frac{\sigma_E^2}{z^3} + \frac{3(\sigma_E^2)^2}{z^5} + \frac{15(\sigma_E^2)^3}{z^7} + \mathcal{O}\left(\frac{1}{z^9}\right).
\end{align}
Notice that all odd moments vanish, so terms like $1/z^2, 1/z^4$ drop out.

From orthogonal polynomial theory (Favard's theorem) and the spectral representation of the Green's function, the resolvent $G(z)$ can be expressed as a continued fraction \cite{Parker:2018yvk}
\begin{align}
    G(z) = \frac{1}{z - \bar{E}_q - \frac{(b_1^{(q)})^2}{z - \bar{E}_q - \frac{(b_2^{(q)})^2}{z - \bar{E}_q - \frac{(b_3^{(q)})^2}{\ddots}}}}.
\end{align}

To make the algebraic inversion transparent for large complex $z$, we can shift the energy coordinate by defining a centered complex variable
\begin{align}
    w \equiv z - \bar{E}_q.
\end{align}

Rewriting the continued fraction in terms of $w$ yields 
\begin{align}
    G(w) = \frac{1}{w - \frac{(b_1^{(q)})^2}{w - \frac{(b_2^{(q)})^2}{w - \frac{(b_3^{(q)})^2}{\ddots}}}}.
\end{align}

We expand $G(w)$ systematically for large $\vert{}w\vert{}$ by performing polynomial long division (or algebraic inversion) from the bottom of the fraction up to the top. Consider the innermost denominator layer
\begin{align}
    w - \frac{(b_2^{(q)})^2}{w - \dots} \approx w.
\end{align}
Substituting this approximation back into the preceding layer gives
\begin{align}
    G(w) \approx \frac{1}{w - \frac{(b_1^{(q)})^2}{w}}=\frac{1}{w} \left( 1 - \frac{(b_1^{(q)})^2}{w^2} \right)^{-1}.
\end{align}

Expanding this to leading orders in $1/w$, we find
\begin{align}
    G(w) = \frac{1}{w} + \frac{(b_1^{(q)})^2}{w^3} + \mathcal{O}\left(\frac{1}{w^5}\right).
\end{align}

Translating back to the original frequency variable $z$ (where $w = z - \bar{E}_q$) yields
\begin{align}
    G(z) = \frac{1}{z - \bar{E}_q} + \frac{(b_1^{(q)})^2}{(z - \bar{E}_q)^3} + \mathcal{O}\left(\frac{1}{(z - \bar{E}_q)^5}\right).
\end{align}

Expanding $(z - \bar{E}_q)^{-n}$ for large $z$ we obtain
\begin{align}
    G(z) = \frac{1}{z} + \frac{\bar{E}_q}{z^2} + \frac{\bar{E}_q^2 + (b_1^{(q)})^2}{z^3} + \mathcal{O}\left(\frac{1}{z^4}\right).
\end{align}

Comparing with \eqref{e4.16} and \eqref{ee4.31}, we find $b_1^{(q)}=\sigma_E$.
To evaluate the next order of the expansion, we retain the first two denominator levels (truncating the deeper fractions at the third level, $w$)
\begin{align}
    G(w) \approx \frac{1}{w - \frac{(b_1^{(q)})^2}{w - \frac{(b_2^{(q)})^2}{w}}}= \frac{1}{w - \frac{(b_1^{(q)})^2 w}{w^2 - (b_2^{(q)})^2}}= \frac{w^2 - (b_2^{(q)})^2}{w^3 - \left[(b_1^{(q)})^2 + (b_2^{(q)})^2\right] w}.
\end{align}

Using the geometric series expansion $\frac{1}{1 - X} = 1 + X + X^2 + \dots$ where $X = \frac{(b_1^{(q)})^2 + (b_2^{(q)})^2}{w^2}$
\begin{align}
    G(w) &= \frac{1}{w^3} \left( w^2 - (b_2^{(q)})^2 \right) \left[ 1 + \frac{(b_1^{(q)})^2 + (b_2^{(q)})^2}{w^2} + \mathcal{O}\left(\frac{1}{w^4}\right) \right]\nonumber\\
    & = \frac{1}{w} + \frac{\left[(b_1^{(q)})^2 + (b_2^{(q)})^2\right] - (b_2^{(q)})^2}{w^3} + \frac{(b_1^{(q)})^2 \left[(b_1^{(q)})^2 + (b_2^{(q)})^2\right]}{w^5} + \mathcal{O}\left(\frac{1}{w^7}\right).
\end{align}

Notice that higher moments $n\geq 2$ correspond to thermal cumulants (or central moments) that correspond to thermal fluctuations measured about the central Gaussian maximum, that is, $\bar{E}_q=0$ for these moments, which yields after comparing with \eqref{e4.37}
\begin{align}
    (b_1^{(q)})^2 \left[ (b_1^{(q)})^2 + (b_2^{(q)})^2 \right] = 3(\sigma_E^2)^2~\Rightarrow~b_2^{(q)}=\sqrt{2}\sigma_E.
\end{align}

The above result can be generalized for arbitrary $n$ to yield
\begin{align}
(b_n^{(q)})^2 = n \, \sigma_E^2 \implies b_n^{(q)} = \sigma_E \sqrt{n}.
\end{align}

In the theory of orthogonal polynomials, the tridiagonal matrix elements (the Lanczos coefficients) are defined via inner products over the spectral measure $\rho_q(E)$. For the diagonal coefficient $a_n^{(q)}$, the defining equation in terms of the monic orthogonal polynomials $P_n(E)$ is the following\footnote{In standard orthogonal polynomial literature, monic polynomials satisfy the three-term recurrence relation $E P_n(E) = b_{n+1}P_{n+1}(E) + a_n P_n(E) + b_n P_{n-1}(E)$ \cite{Muck:2022xfc}. Taking the inner product with $P_n(E)$ on both sides extracts $a_n^{(q)}$.}
\begin{align}
    a_n^{(q)} = \frac{\int_{-\infty}^{\infty} dE \, \rho_q(E) \, E \, [P_n(E)]^2}{\int_{-\infty}^{\infty} dE \, \rho_q(E) \, [P_n(E)]^2}.
\end{align}

To evaluate the numerator explicitly, we substitute the finite-temperature symmetry-resolved spectral density $\rho_q(E)$, which is a Gaussian distribution centered at $\bar{E}_q$ with variance $\sigma_E^2$ and change variables to the centered energy coordinate
\begin{align}
    x \equiv E - \bar{E}_q \implies E = x + \bar{E}_q, \quad dE = dx.
\end{align}

Under this shift, the monic polynomial $P_n(E)$ becomes a polynomial in $x$, which we denote as $P_n(x+\bar{E}_q)$. The numerator integral becomes
\begin{align}
    \text{Numerator} = \int_{-\infty}^{\infty} dx \, \frac{1}{\sqrt{2\pi \sigma_E^2}} \exp\left( - \frac{x^2}{2\sigma_E^2} \right) \, (x + \bar{E}_q) \, [P_n(x+\bar{E}_q)]^2.
\end{align}

Notice that the Gaussian weight $W(x)$ is an even function of $x$. Also, the squared polynomial $[P_n(x+\bar{E}_q)]^2$ is an even function of $x$. This yields the following numerator
\begin{align}
   \text{Numerator} =\bar{E}_q \int_{-\infty}^{\infty} dx \, W(x) \, [P_n(x+\bar{E}_q)]^2 ~\Rightarrow a_n^{(q)}=\bar{E}_q + \mathcal{O}(1/N).
\end{align}

Thus, in summary, in the finite-temperature phase, the symmetry-resolved Lanczos coefficients exhibit universal square-root growth
\begin{align}
a_n^{(q)} = \bar{E}_q, \quad b_n^{(q)} = \sigma_E \sqrt{n}.
\end{align}

Because the thermal fluctuation spectrum is symmetric around its peak ($\bar{E}_q$), all linear deviations $x$ average out to zero. The on-site potential along the Krylov chain ($a_n^{(q)}$) is thus anchored strictly and universally to the mean sector energy $\bar{E}_q$, independent of the chain depth $n$. In standard quantum mechanical systems and chaotic models (such as SYK or random matrix ensembles \cite{Balasubramanian:2022tpr}), the diagonal coefficients $a_n$ often depend on $n$ when the underlying spectral density is asymmetric, bounded, or exhibits band edges. However, this is not the case here, as the spectral density $\rho_q(E)$ in the finite-temperature phase is a pristine, symmetric Gaussian centered strictly at the macrocanonical mean energy $\bar{E}_q$.

As a second important remark, we would like to emphasize that $b_n^{(q)} = \sigma_E \sqrt{n}$ does not violate the Universal Operator Growth Hypothesis \cite{Parker:2018yvk} if we look closely at what the hypothesis actually predicts for thermal and mixed states versus infinite-temperature operators. This foundational hypothesis was formulated specifically for infinite-temperature correlation functions starting from a simple, low-complexity operator (like a single local Pauli string or a single Majorana fermion) acting on the infinite-temperature background ($\beta = 0$). When we move to symmetry-resolved sectors (fixing a global $U(1)$ charge $q$) or consider states at finite temperatures ($\beta > 0$), the operator or state being evolved is no longer sitting at infinite temperature\footnote{The $\sqrt{n}$ scaling has recently been observed in the context of BMN matrix models \cite{Roychowdhury:2026ikp} as an artifact of intermediate temperature effect.}. At infinite temperature, operators explore unconstrained high-energy tails leading to linear growth ($b_n \sim \alpha n$). At finite temperature, thermal ensemble restricts fluctuations to a Gaussian thermal package, softening the operator growth rate down to the square-root scaling characteristic of harmonic oscillator modes. In summary, $\sqrt{n}$ scaling is a direct consequence of \emph{finite} temperature effect where thermal fluctuations of energy in a fixed charge sector are governed by a Gaussian distribution with variance $\sigma_E^2$. We expect linear scaling to exhibit in the strict limit $\beta \rightarrow 0$.

As a final remark, one might wish to compare these finite temperature results with its low temperature counterpart. While the low-temperature phase exhibits a microscopic, discrete imprint of the BMN matrix model where individual charge sectors shape the fine structure of the spectral edge, the finite-temperature phase smooths this out into a \emph{collective} thermodynamic imprint. Here, the charge $q$ manifests exclusively through macrocanonical observables—specifically shifting the band center $\bar{E}_q$ and scaling the Gaussian thermal variance $\sigma_E$, which jointly govern the universal $\sqrt{n}$ Krylov growth.
\subsection{Spread complexity in the symmetry-resolved sector}
The spread complexity $C^{(q)}(t)$ quantifies the dispersion of the time-evolved state $\ket{\psi_q(t)}$ across the Krylov basis $\ket{K_n^{(q)}}$
\begin{align}
\label{e4.55}
C^{(q)}(t) = \sum_{n=0}^{\infty} n \, |\phi_n^{(q)}(t)|^2
\end{align}
where the wave amplitudes $\phi_n^{(q)}(t) = \braket{K_n^{(q)} \vert \psi_q(t)}$ obey the discretized $i$-chain recurrence relations of the following form
\begin{align}
i \frac{d}{dt} \phi_n^{(q)}(t) = a_n^{(q)} \phi_n^{(q)}(t) + b_n^{(q)} \phi_{n-1}^{(q)}(t) + b_{n+1}^{(q)} \phi_{n+1}^{(q)}(t).
\end{align}

Substituting the values of $a_n^{(q)}$ and $b_n^{(q)}$ into the recurrence relation, we obtain
\begin{align}
    i \frac{d}{dt} \phi_n^{(q)}(t) = \bar{E}_q \phi_n^{(q)}(t) + \sigma_E \sqrt{n} \, \phi_{n-1}^{(q)}(t) + \sigma_E \sqrt{n+1} \, \phi_{n+1}^{(q)}(t).
\end{align}

Because the on-site potential $\bar{E}_q$ is independent of $n$, we can factor out the global phase rotation corresponding to the mean sector energy by defining the shifted amplitude
\begin{align}
    \phi_n^{(q)}(t) \equiv e^{-i \bar{E}_q t} \tilde{\phi}_n(t).
\end{align}

Substituting this into the differential equation, the $\bar{E}_q$ terms cancel out on both sides to yield the following
\begin{align}
\label{e4.59}
    i \frac{d}{dt} \tilde{\phi}_n(t) = \sigma_E \sqrt{n} \, \tilde{\phi}_{n-1}(t) + \sigma_E \sqrt{n+1} \, \tilde{\phi}_{n+1}(t).
\end{align}

Notice that, by definition (Eq. \eqref{e4.55}), the spread complexity $C^{(q)}(t)$ is the expectation value of the chain position operator $n$ with respect to the probability distribution of the wave function
\begin{align}
    C^{(q)}(t) = \sum_{n=0}^{\infty} n \, \vert{}\phi_n^{(q)}(t)\vert{}^2 = \sum_{n=0}^{\infty} n \, \vert{}\tilde{\phi}_n(t)\vert{}^2.
\end{align}

We evaluate the Taylor series expansion of $C^{(q)}(t)$ around early times ($t \to 0$), subject to the initial localization condition at the origin $\tilde{\phi}_n(0) = \delta_{n,0}$
\begin{align}
    C^{(q)}(t) = C^{(q)}(0) + t \, \left. \frac{d C^{(q)}}{dt} \right\vert{}_{t=0} + \frac{t^2}{2} \left. \frac{d^2 C^{(q)}}{dt^2} \right\vert{}_{t=0} + \mathcal{O}(t^3).
\end{align}

Since the initial wave packet is localized entirely at $n = 0$
\begin{align}
    C^{(q)}(0) = \sum_{n=0}^{\infty} n \, \vert{}\tilde{\phi}_n(0)\vert{}^2 = \sum_{n=0}^{\infty} n \, \delta_{n,0} = 0.
\end{align}

Differentiating $C^{(q)}(t)$ with respect to time yields the following
\begin{align}
    \frac{d C^{(q)}}{dt} = \sum_{n=0}^{\infty} n \left( \frac{d\tilde{\phi}_n^*}{dt} \tilde{\phi}_n + \tilde{\phi}_n^* \frac{d\tilde{\phi}_n}{dt} \right).
\end{align}

One can use the Schrödinger equation \eqref{e4.59} to substitute $\frac{d\tilde{\phi}_n}{dt}$. At $t = 0$, the only non-zero component is $\tilde{\phi}_0(0) = 1$ (and all other $\tilde{\phi}_{n \neq 0} = 0$). Because every term in the derivative expression involves indices $n-1$ or $n+1$ evaluated at $t=0$ (which vanish for $n=0$), the entire sum evaluates to zero
\begin{align}
\label{e4.64}
    \frac{d C^{(q)}}{dt} \Big|_{t=0}=0.
\end{align}

To find the second time derivative at early times ($t \to 0$), we differentiate $C^{(q)}(t)$ twice using the Heisenberg equation of motion. Notice that the spread complexity $C^{(q)}(t)$ is defined as the expectation value of the chain position operator $\hat{n}$ with respect to the time-evolved state $\ket{\psi_q(t)}$ distributed across the Krylov basis $\ket{K_n^{(q)}}$
\begin{align}
    C^{(q)}(t) = \langle \psi_q(t) \vert{} \hat{n} \vert{} \psi_q(t) \rangle = \sum_{n=0}^\infty n \, \vert{}\phi_n^{(q)}(t)\vert{}^2.
\end{align}

In operator form, the time evolution of the position operator $\hat{n}$ is generated by the tridiagonal Krylov Hamiltonian $H_{\text{Krylov}}(=H_{\text{BMN}})$
\begin{align}
    \frac{d \hat{n}}{dt} = i [H_{\text{Krylov}}, \hat{n}].
\end{align}

Differentiating a second time yields the double commutator expression
\begin{align}
    \frac{d^2 \hat{n}}{dt^2} = - \left[ \left[ \hat{n}, H_{\text{Krylov}} \right], H_{\text{Krylov}} \right].
\end{align}

Thus, the second time derivative of the spread complexity evaluated at $t=0$ is given by the expectation value of this double commutator with respect to the initial state $\ket{\psi_q(0)}$
\begin{align}
    \left. \frac{d^2 C^{(q)}}{dt^2} \right\vert_{t=0} = \langle \psi_q(0) \vert{} \frac{d^2 \hat{n}}{dt^2} \vert{} \psi_q(0) \rangle = - \langle \psi_q(0) \vert{} \left[ \left[ \hat{n}, H_{\text{Krylov}} \right], H_{\text{Krylov}} \right] \vert{} \psi_q(0) \rangle.
\end{align}

In the symmetry-resolved finite-temperature phase of the BMN matrix model, the Lanczos tridiagonal Hamiltonian acts on the orthonormal Krylov basis elements $\ket{K_n^{(q)}}$ as
\begin{align}
\label{e4.69}
    H_{\text{Krylov}} \ket{K_n^{(q)}} = a_n^{(q)} \ket{K_n^{(q)}} + b_n^{(q)} \ket{K_{n-1}^{(q)}} + b_{n+1}^{(q)} \ket{K_{n+1}^{(q)}}
\end{align}
where we have on-site potentials $a_n^{(q)} = \bar{E}_q$ (independent of $n$) and Hopping amplitudes as $b_n^{(q)} = \sigma_E \sqrt{n}$.

Let us evaluate how the position operator $\hat{n}$ (defined such that $\hat{n} \ket{K_n^{(q)}} = n \ket{K_n^{(q)}}$) commutes with $H_{\text{Krylov}}$. Consider its action on an arbitrary state vector
\begin{align}
    [\hat{n}, H_{\text{Krylov}}] \ket{K_n^{(q)}} = \hat{n} H_{\text{Krylov}} \ket{K_n^{(q)}} - H_{\text{Krylov}} \hat{n} \ket{K_n^{(q)}}.
\end{align}

Using \eqref{e4.69}, this further simplifies to
\begin{align}
    [\hat{n}, H_{\text{Krylov}}] \ket{K_n^{(q)}} = - b_n^{(q)} \ket{K_{n-1}^{(q)}} + b_{n+1}^{(q)} \ket{K_{n+1}^{(q)}}.
\end{align}

Next, we apply the second commutator with $H_{\text{Krylov}}$ to yield
\begin{align}
 &  - \left[ \left[ \hat{n}, H_{\text{Krylov}} \right], H_{\text{Krylov}} \right] \ket{K_n^{(q)}} \nonumber\\
 &= H_{\text{Krylov}} [\hat{n}, H_{\text{Krylov}}] \ket{K_n^{(q)}}- [\hat{n}, H_{\text{Krylov}}] H_{\text{Krylov}} \ket{K_n^{(q)}}.
\end{align}

We evaluate this specifically for the initial state of the system, which is localized entirely at the origin of the chain
\begin{align}
    \ket{\psi_q(0)} = \ket{K_0^{(q)}} \quad (\text{i.e., } n = 0).
\end{align}

For the first term, we find
\begin{align}
    [\hat{n}, H_{\text{Krylov}}] \ket{K_0^{(q)}} = -b_0^{(q)} \ket{K_{-1}^{(q)}} + b_1^{(q)} \ket{K_1^{(q)}} = b_1^{(q)} \ket{K_1^{(q)}} \quad (\text{since } b_0^{(q)} = 0).
\end{align}

Applying $H_{\text{Krylov}}$ to $\ket{K_1^{(q)}}$ we obtain
\begin{align}
    H_{\text{Krylov}} (b_1^{(q)} \ket{K_1^{(q)}}) = b_1^{(q)} \left( a_1^{(q)} \ket{K_1^{(q)}} + b_1^{(q)} \ket{K_0^{(q)}} + b_2^{(q)} \ket{K_2^{(q)}} \right).
\end{align}

Similarly, for the second term we find
\begin{align}
    H_{\text{Krylov}} \ket{K_0^{(q)}} = a_0^{(q)} \ket{K_0^{(q)}} + b_1^{(q)} \ket{K_1^{(q)}}.
\end{align}

Next we apply $[\hat{n}, H_{\text{Krylov}}]$ to this combination to yield
\begin{align}
    &[\hat{n}, H_{\text{Krylov}}] \left( a_0^{(q)} \ket{K_0^{(q)}} + b_1^{(q)} \ket{K_1^{(q)}} \right) \nonumber\\
    &= a_0^{(q)} (b_1^{(q)} \ket{K_1^{(q)}}) + b_1^{(q)} \left( -b_1^{(q)} \ket{K_0^{(q)}} + b_2^{(q)} \ket{K_2^{(q)}} \right).
\end{align}

Subtracting these two terms and using the fact $a_1^{(q)} = a_0^{(q)} = \bar{E}_q$, we find
\begin{align}
    \left[ \left[ \hat{n}, H_{\text{Krylov}} \right], H_{\text{Krylov}} \right] \ket{K_0^{(q)}} = 2 (b_1^{(q)})^2 \ket{K_0^{(q)}}.
\end{align}

Taking the expectation value with respect to the initial state $\langle K_0^{(q)} \vert{} \dots \vert{} K_0^{(q)} \rangle$ yields
\begin{align}
    \left. \frac{d^2 C^{(q)}}{dt^2} \right\vert_{t=0} = 2 \sigma_E^2.
\end{align}

This yields the early time growth in the symmetry resolved spread complexity as
\begin{align}
\label{e4.80}
C^{(q)}(t) = \sigma_E^2 t^2 + \mathcal{O}(t^4)~;~\text{where}~ b_1^{(q)}=\sigma_E.
\end{align}

In Krylov complexity theory, the early-time behavior of spread complexity $C(t)$ universally exhibits a quadratic growth regime rather than a linear one. Because the first time derivative of complexity vanishes at $t = 0$ (as shown in Eq.\eqref{e4.64}, since the initial state starts localized at the origin of the Krylov chain, $\ket{\psi_q(0)} = \ket{K_0^{(q)}}$), the Taylor expansion is governed entirely by the second time derivative.  A $t^2$ growth law indicates that the time-evolved state initially spreads ballistically or diffusively out of the starting Krylov basis node into its immediate neighbors. The coefficient $\sigma_E^2$ acts as the fundamental "acceleration" or spread rate of complexity in that narrow time window. A larger $\sigma_E$ means the state couples more strongly to higher-index Krylov basis vectors, driving a faster initial escalation of operator/state complexity. This is intimately tied to how fast quantum information scrambles in holographic duals comprising black holes/ black D0 branes.

Before we conclude our analysis, some important remarks are in order. First of all, Eq.\eqref{e4.80} does not always imply a classical thermal equipartition of spread complexity \cite{Caputa:2025ozd}. The equipartition will depend on the scaling of $q$ in the Large $N$ limit\footnote{See Appendix \ref{appen D} for details.}.  Instead, it reflects the universal early-time quadratic growth regime governed by the quantum mechanical dispersion of the Krylov chain, regardless of whether the system is at high or low temperatures. However, the physical origin of how temperature and phase structure alter this expression—contrasting the high-temperature thermal phase with the low-temperature phase (such as Eq.\eqref{e3.166})—lies in how quantum fluctuations, operator norm distributions, and thermal saddle points populate the Krylov basis. Equation \eqref{e4.80} describes the Taylor expansion coefficient of the state's center of mass motion along the Krylov chain at $t \to 0$. 

The coefficient $\sigma_E^2 = \langle K_0^{(q)} \vert{} H^2 \vert{} K_0^{(q)} \rangle - \bar{E}_q^2$ is the energy variance of the initial state with respect to the Hamiltonian. At very early times, the state has barely begun to spread; it is localized almost entirely at the origin node $\ket{K_0^{(q)}}$. The $t^2$ growth simply captures the initial "launch velocity" or acceleration out of this node, dictated by the off-diagonal hopping matrix elements ($b_1^{(q)}$). It does not represent a thermal plateau or an equipartitioned steady state. While the functional form of early-time growth ($C^{(q)}(t) \approx \sigma_E^2 t^2$) looks structurally similar across phases, the magnitude, temperature scaling, and microscopic composition of $\sigma_E^2$ differ fundamentally between the high-temperature thermal phase and the low-temperature phase (Eq. \eqref{e3.166}). At low temperatures or in the perturbative vacuum limit, the BMN matrix model fluctuates around a supersymmetric vacuum or a dilute gas of harmonic oscillators. At low temperatures or in the perturbative vacuum limit, the BMN matrix model fluctuates around a supersymmetric vacuum or a dilute gas of harmonic oscillators. The variance $\sigma_E^2$ is dominated by the zero-point fluctuations and bare harmonic frequencies of the matrix elements. Because the thermal occupation numbers are exponentially suppressed ($e^{-\beta \omega} \ll 1$), thermal noise is absent, and the complexity evolution remains bounded, quasi-periodic, or exhibits delayed scrambling.

In the high-temperature phase (where the matrix model undergoes a phase transition and duals a ten-dimensional black hole), the system enters a strongly coupled, quantum chaotic regime. Here, $\sigma_E^2$ is governed by thermal ensemble averages. As explored in Appendix \ref{appen D}, the thermal expectation values $\langle ZZ^\dagger \rangle_{T, q}$ scale with temperature (carrying factors of $T$ or $T^4$ depending on the matrix dimensions). Thus, the high-temperature $\sigma_E^2$ is vastly enhanced compared to the low-temperature vacuum value. It incorporates thermal broadening and collective multi-trace interactions, meaning the initial "acceleration" of spread complexity is much steeper and more rapid. This reflects faster quantum information scrambling and operator growth induced by thermal chaos near the black hole horizon.

Finally, here are some comments about the dual supergravity description of the finite temperature result. Finite temperature ($\beta = 1/T$) on the boundary corresponds in the bulk to turning on a horizon in the ten-dimensional supergravity background. Instead of a cold, empty supersymmetric vacuum (where strings sit quietly in their ground states), the spacetime now possesses a black hole or black brane horizon radiating thermal energy \cite{Costa:2014wya}-\cite{Santos:2026rrx}. The thermal density matrix on the boundary means the bulk is populated by a thermal gas of gravitons, scalars, and strings interacting strongly near the horizon. This introduces robust quantum chaos and rapid scrambling. In the bulk black hole picture, a higher temperature means larger thermal occupation numbers and stronger multi-trace interactions. This drastically enhances the initial "acceleration" ($\sigma_E^2$), meaning the state launches outward into the Krylov chain much faster. Geometrically, this reflects the intense tidal forces and gravitational curvature near the black hole horizon that accelerate operator growth and information scrambling. The global $U(1)$ charge $q$ at finite temperature maps to angular momentum or R-charge carried by probe strings or D-branes propagating close to the black hole horizon. The charged bulk excitations (such as string states or D-branes carrying angular momentum or $R$-charge) act as sources that non-trivially backreact on the background Ramond-Ramond (R-R) form-fields and metric warp factors. Consequently, the background fields experienced by any propagating probe string acquire explicit dependence on the charge density and $q$. The interaction between the string's worldsheet charges and the background supergravity form-fields introduces electrostatic and flux-coupling corrections to the local mass-squared or tension terms, giving rise to the distinct $q$-dependence as seen in the symmetry-resolved moments and Lanczos coefficients.
\section{Concluding remarks and future directions}
\label{sec 5}
\subsection{Summary of key results and physical interpretation}

In this work, we investigated the symmetry-resolved spread complexity $C^{(q)}(t)$ of a charged thermo-field double (TFD) state within fixed $U(1)$ global charge sectors of the Berenstein--Maldacena--Nastase (BMN) matrix quantum mechanics in the planar large-$N$ limit. Our principal findings and their broader physical significance are summarized below:

\begin{itemize}
    \item \textbf{Early-time universal ballistic growth and sector-dependent acceleration:} 
    Across both the low-temperature perturbative regime and the high-temperature deconfined phase, the symmetry-resolved spread complexity exhibits a universal early-time quadratic scaling,
    \begin{equation}
        C^{(q)}(t) = b_1^{2(q)}\, t^2 + \mathcal{O}(t^4),
    \end{equation}
    where $b_1^{2(q)}(q,T) = \langle K_0^{(q)} | H^2 | K_0^{(q)} \rangle - \bar{E}_q^2$ is the energy variance of the initial state within the $q$-th charge sector. Physically, $\sigma_E^2$ governs the initial ``launch acceleration'' of the state out of the origin node $|K_0^{(q)}\rangle$ along the Krylov chain. While the $t^2$ functional form is universal, its physical origin and magnitude differ sharply across phases: at low temperatures ($e^{-\beta \omega} \ll 1$), $\sigma_E^2$ is governed by zero-point quantum fluctuations and perturbative charge-induced frequency shifts around the trivial vacuum, leading to bounded or quasi-periodic complexity evolution. Conversely, in the high-temperature black hole phase, $\sigma_E^2$ is strongly enhanced by thermal broadening and collective multi-trace interactions, driving rapid initial scrambling.

    \item \textbf{Charge-induced mass shifts and breakdown of complexity equipartition:}
    Projecting onto a fixed $U(1)$ charge sector rotating the $(X_4, X_5)$ plane generates an effective, charge-dependent mass-squared shift $\Omega_i(q) \propto \frac{\mu^2 g_{\mathrm{eff}}}{2N} f_i(q)$ for the matrix oscillator modes via Hartree--Fock self-energy contractions (with non-planar exchange contractions suppressed by $\mathcal{O}(1/N)$). This explicit $q$-dependence propagates directly into the symmetry-resolved return amplitude, moments, and Lanczos coefficients ($a_n^{(q)}, b_n^{(q)}$). Consequently, unlike simple free or conformal systems where spread complexity exhibits exact equipartition across symmetry sectors at leading order, the non-Abelian quartic interactions of the BMN model explicitly break complexity equipartition whenever $q$ scales non-trivially with $N$.

    \item \textbf{Holographic bulk dual interpretation:}
    In the dual ten-dimensional Type IIA supergravity picture, tuning from low to high temperature corresponds to the formation of a black hole / black D0-brane horizon, where intense gravitational tidal forces accelerate operator growth and account for the steep thermal enhancement of $\sigma_E^2(q, T)$. Within this geometry, the global $U(1)$ charge $q$ maps to angular momentum or $R$-charge carried by probe strings and D-branes propagating near the horizon. These charged excitations backreact on the background Ramond--Ramond (R--R) four-form flux and metric warp factors, inducing effective tension and mass-squared shifts on the string worldsheet that microscopically mirror the $q$-dependent Lanczos spectrum derived on the matrix model boundary.
\end{itemize}

Looking ahead, our findings open several promising avenues for future investigation.
\begin{enumerate}
\item \textbf{Multi-charge resolution and non-Abelian $SO(3) \times SO(6)$ symmetry sectors:} 
In the present work, we restricted attention to a single Abelian $U(1)$ subgroup rotating the $(X_4, X_5)$ plane. Extending the symmetry-resolved Krylov framework to simultaneously resolve multiple commuting Cartan charges of $SO(6)$ (corresponding to multi-spin giant gravitons and rotating brane configurations) or full non-Abelian $SU(2)/SO(3)$ representations would clarify how competing angular momenta and $R$-charges govern Krylov complexity equipartition and phase structure.

\item \textbf{Fermionic degrees of freedom and supersymmetry-protected BPS sectors:} 
While our leading bosonic analysis captures the dominant thermal and large-$N$ scalings, incorporating the sixteen Majorana fermions of the full BMN matrix model is essential for probing supersymmetric localization and BPS-protected subsectors. Exploring symmetry-resolved spread complexity in $\frac{1}{2}$- and $\frac{1}{4}$-BPS sectors would reveal whether supersymmetry restores complexity equipartition or enforces exact lower bounds on Lanczos growth at finite coupling.

\item \textbf{Fuzzy-sphere vacua, concentric membrane topologies, and intermediate phases:} 
Our low-temperature analysis focused primarily on fluctuations around the trivial gapped vacuum. Repeating the symmetry-resolved Krylov construction around non-trivial irreducible and reducible $SU(2)$ fuzzy-sphere vacua---as well as across the intermediate Hagedorn/partial-deconfinement regime---would shed light on how topology-changing transitions (Myers-effect dielectric brane polarization) imprint themselves on the Lanczos coefficients $a_n^{(q)}, b_n^{(q)}$ and the positivity bound $\Delta C(t) \ge 0$.

\item \textbf{Krylov saturation, complexity plateau, and finite-$N$ effects:} 
Our analytical results in Sections~\ref{sec 3} and \ref{sec 4} characterize the early-time ballistic launch and intermediate Lanczos growth in the planar large-$N$ limit. Pushing the calculation to exponentially late times ($t \sim e^{S_q}$) by incorporating finite-$N$ trace relations, non-perturbative instanton corrections, or numerical matrix-bootstrap / Monte Carlo techniques would determine how the symmetry-resolved Krylov plateau scales with sector entropy $S_q$ and whether sector-resolved spectral form factors exhibit ramp-plateau signatures of quantum chaos.

\item \textbf{Bulk gravitational interpretations:} 
On the gravitational side, it would be illuminating to establish a quantitative bulk dual for symmetry-resolved spread complexity in terms of extremal/charged black D0-brane horizons and backreacted Lin--Maldacena droplet geometries. Comparing the early-time growth rate $\sigma_E^2(q,T)$ and Lanczos spectrum against holographic complexity proposals (such as \cite{Stanford:2014jda} or radial momentum in backreacted ten-dimensional supergravity backgrounds \cite{Caputa:2024sux}) could provide a sharp microscopic bridge between sector-resolved matrix dynamics and bulk spacetime emergence.
\end{enumerate}
\paragraph{Acknowledgements.}
The author thanks Carlos Nunez, S. Prem Kumar and Simon F. Ross for their comments on the draft.
The author also acknowledges the Mathematical Research Impact Centric Support (MATRICS) grant (MTR/2023/000005) received from ANRF, India. \\ 
\appendix
\section{A note on dimensional reduction}
\label{appen:BMN_reduction}
In this Appendix, we provide a self-contained derivation of the Berenstein--Maldacena--Nastase (BMN) matrix quantum mechanics Hamiltonian~\eqref{e2.1} starting from ten-dimensional $\mathcal{N}=1$ Super-Yang--Mills (SYM) theory reduced to $0+1$ dimensions with background Ramond--Ramond (R--R) four-form flux deformation \cite{Berenstein:2002jq,Dasgupta:2002hx,Sugiyama:2002rs}.

Consider ten-dimensional $\mathcal{N}=1$ Super-Yang--Mills theory with gauge group $U(N)$ (or $SU(N)$) on Minkowski spacetime $\mathbb{R}^{1,9}$ with metric signature $\eta_{MN} = \mathrm{diag}(-, +, \dots, +)$ ($M,N = 0,1,\dots,9$). The field content consists of a ten-dimensional gauge field $A_M$ and a sixteen-component Majorana--Weyl spinor $\Psi$ transforming in the adjoint representation of the gauge group. The action takes the standard form
\begin{align}
S_{10\mathrm{D}} = \frac{1}{g_{\mathrm{YM},10}^2} \int d^{10}x \, \mathrm{Tr}\!\left( -\frac{1}{4} F_{MN} F^{MN} + \frac{i}{2} \bar{\Psi} \Gamma^M D_M \Psi \right),
\label{eq:S10D}
\end{align}
where the non-Abelian field strength and covariant derivative are given by
\begin{align}
F_{MN} = \partial_M A_N - \partial_N A_M - i [A_M, A_N], \qquad D_M \Psi = \partial_M \Psi - i [A_M, \Psi],
\end{align}
and $\Gamma^M$ denote the $32 \times 32$ ten-dimensional Dirac matrices satisfying $\{\Gamma^M, \Gamma^N\} = 2\eta^{MN}$.

To perform dimensional reduction to $0+1$ dimensions, we compactify the nine spatial coordinates $x^i$ ($i=1,\dots,9$) on a nine-torus $T^9$ of coordinate volume $V_9 = \int d^9 x$ and truncate to the zero-momentum Kaluza--Klein modes ($\partial_i = 0$ for all fields). Relabeling the spatial components of the gauge connection as $N \times N$ Hermitian scalar matrices,
\begin{align}
A_0(t) \equiv A_t(t), \qquad A_i(t) \equiv X_i(t) \quad (i = 1, \dots, 9),
\end{align}
and identifying the effective one-dimensional gauge coupling $g$ via Eq.~\eqref{ee2.3},
\begin{align}
\frac{1}{g^2} = \frac{V_9}{g_{\mathrm{YM},10}^2} \quad \Longleftrightarrow \quad g^2 = \frac{g_{\mathrm{YM},10}^2}{V_9},
\end{align}
the components of the field strength reduce to
\begin{align}
F_{0i} = \partial_t X_i - i [A_t, X_i] \equiv D_t X_i, \qquad F_{ij} = -i [X_i, X_j].
\end{align}

Decomposing the ten-dimensional Clifford algebra into $\mathrm{Spin}(1,9) \supset \mathrm{Spin}(9)$ via $\Gamma^0 = i \sigma_2 \otimes \mathbf{1}_{16}$ and $\Gamma^i = \sigma_1 \otimes \gamma^i$ (where $\gamma^i$ are real, symmetric $16 \times 16$ $\mathrm{Spin}(9)$ gamma matrices satisfying $\{\gamma^i, \gamma^j\} = 2\delta^{ij}$), and projecting onto the positive-chirality sixteen-component real spinor $\psi_\alpha$ ($\alpha = 1,\dots,16$), the action~\eqref{eq:S10D} reduces to the flat-space Banks--Fischler--Shenker--Susskind (BFSS) matrix quantum mechanics \cite{Banks:1996vh}
\begin{align}
S_{\mathrm{BFSS}} = \frac{1}{g^2} \int dt \, \mathrm{Tr}\!\left( \frac{1}{2} (D_t X_i)^2 + \frac{1}{4} [X_i, X_j]^2 + \frac{i}{2} \psi^T D_t \psi + \frac{1}{2} \psi^T \gamma^i [X_i, \psi] \right).
\label{eq:S_BFSS}
\end{align}

In M-theory and Type IIA string theory, D0-branes probing the maximally supersymmetric eleven-dimensional pp-wave background experience a constant null R--R four-form flux $F_{(4)} \propto \mu \, dx^- \wedge dx^1 \wedge dx^2 \wedge dx^3$, which breaks the transverse $SO(9)$ rotational symmetry down to $SO(3) \times SO(6)$ \cite{Berenstein:2002jq,Dasgupta:2002hx}. Following the splitting of the nine spatial indices into $a,b,c \in \{1,2,3\}$ (transforming under $SO(3)$) and $i',j' \in \{4,\dots,9\}$ (transforming under $SO(6)$), the background flux induces three gauge-invariant deformations. The one that goes under the name of the anisotropic quadratic mass deformation
\begin{align}
\Delta \mathcal{L}_{\mathrm{mass}} = -\frac{1}{2g^2} \mathrm{Tr}\!\left( \mu^2 \sum_{a=1}^{3} X_a^2 + \frac{\mu^2}{4} \sum_{i'=4}^{9} X_{i'}^2 \right).
\end{align}.

The second term goes under the name of the cubic Myers interaction that results due to the coupling of the non-Abelian D0-brane dipole moments to the background R--R four-form flux produces a Chern--Simons-type cubic interaction within the $SO(3)$ sector 
\begin{align}
\Delta \mathcal{L}_{\mathrm{Myers}} = -\frac{i\mu}{g^2} \sum_{a,b,c=1}^{3} \epsilon_{abc} \, \mathrm{Tr}(X_a X_b X_c).
\end{align}

Finally, one has the fermionic flux mass term
\begin{align}
\Delta \mathcal{L}_{\mathrm{fermion}} = -\frac{3i\mu}{8g^2} \, \mathrm{Tr}\!\left( \psi^T \gamma^{123} \psi \right).
\end{align}
Preserving all $32$ supercharges of the pp-wave background requires adding a fermionic mass term proportional to the antisymmetric product $\gamma^{123} \equiv \gamma^1 \gamma^2 \gamma^3$.

To pass to canonical normalization where the kinetic terms carry unit coefficient and the gauge coupling $g$ multiplies the interaction vertices, we rescale the bosonic and fermionic fields by
\begin{align}
X_i \longrightarrow g X_i, \qquad \psi \longrightarrow g \psi, \qquad A_t \longrightarrow g A_t,
\end{align}
consistent with the canonical mass dimensions $[X_i] = M^{-1/2}$ and $[g^2] = M^3$ noted in Eq.~\eqref{ee2.4}. After performing this rescaling (and absorbing numerical conventions in $\mu$ to match Eq.~\eqref{e2.1}), the complete BMN Lagrangian reads
\begin{align}
\mathcal{L}_{\mathrm{BMN}} &= \mathrm{Tr}\Bigg[ \frac{1}{2} \sum_{i=1}^{9} (D_t X_i)^2 + \frac{g^2}{4} \sum_{i,j=1}^{9} [X_i, X_j]^2 - \frac{\mu^2}{2} \sum_{a=1}^{3} X_a^2 - \frac{\mu^2}{8} \sum_{i'=4}^{9} X_{i'}^2 \nonumber \\
&\qquad\quad -i\mu g  \sum_{a,b,c=1}^{3} \epsilon_{abc} X_a X_b X_c + \frac{i}{2} \psi^T D_t \psi + \frac{g}{2} \psi^T \gamma^i [X_i, \psi] - \frac{3i\mu}{8} \psi^T \gamma^{123} \psi \Bigg].
\label{eq:L_BMN}
\end{align}

To construct the quantum Hamiltonian, we pass from the Lagrangian~\eqref{eq:L_BMN} to the Hamiltonian formalism via a Legendre transformation. The conjugate momentum matrices $\Pi_i$ ($i=1,\dots,9$) and $\pi_\psi$ are defined by
\begin{align}
\Pi_i \equiv \frac{\partial \mathcal{L}_{\mathrm{BMN}}}{\partial \dot{X}_i} = D_t X_i = \dot{X}_i - i g [A_t, X_i], \qquad \pi_\psi \equiv \frac{\partial^L \mathcal{L}_{\mathrm{BMN}}}{\partial \dot{\psi}} = -\frac{i}{2}\psi^T,
\end{align}
while the temporal gauge field $A_t$ carries vanishing canonical momentum ($\Pi_{A_t} = 0$) and acts as a Lagrange multiplier. Performing the Legendre transform yields
\begin{align}
H_{\mathrm{total}} = \mathrm{Tr}\!\left( \sum_{i=1}^{9} \Pi_i \dot{X}_i + \pi_\psi \dot{\psi} - \mathcal{L}_{\mathrm{BMN}} \right) = H_{\mathrm{BMN}} + \mathrm{Tr}\!\left( A_t \, \mathcal{G} \right),
\end{align}
where the $SU(N)$ Gauss law generator $\mathcal{G}$ enforces gauge invariance on physical Hilbert space states ($\mathcal{G}|\mathrm{phys}\rangle = 0$)
\begin{align}
\mathcal{G} = i g \sum_{i=1}^{9} [X_i, \Pi_i] - \frac{g}{2} \{\psi_\alpha, \psi_\alpha\} \approx 0.
\end{align}

Fixing the temporal gauge $A_t = 0$ (with physical states restricted to the $SU(N)$-singlet sector), we obtain the gauge-fixed BMN Hamiltonian
\begin{align}
H_{\mathrm{BMN}} &= \mathrm{Tr}\!\Big[ \frac{1}{2} \sum_{i=1}^{9} \Pi_i^2 - \frac{g^2}{4} \sum_{i,j=1}^{9} [X_i, X_j]^2 + \frac{\mu^2}{2} \sum_{a=1}^{3} X_a^2 \nonumber\\
&+ \frac{\mu^2}{8} \sum_{i'=4}^{9} X_{i'}^2 + i\mu g \sum_{a,b,c=1}^{3} \epsilon_{abc} X_a X_b X_c + H_{\mathrm{fermion}} \Big],
\label{eq:H_BMN_derived}
\end{align}
with the fermionic Hamiltonian given explicitly by
\begin{align}
H_{\mathrm{fermion}} = -\frac{g}{2} \psi^T \gamma^i [X_i, \psi] + \frac{3i\mu}{8} \psi^T \gamma^{123} \psi.
\end{align}
Equation~\eqref{eq:H_BMN_derived} precisely reproduces the BMN Hamiltonian~\eqref{e2.1} analyzed in Section~\ref{sec 2}.
\section{Effective potential in the Large $N$ limit}
\label{appen A}
The purpose of this Appendix is to calculate the quantum expectation of the form
\begin{align}
    \braket{\text{Tr}(X_4X_iX_4X_i)}_q=\langle \psi_q(n) \vert \text{Tr}(X_4 X_i X_4 X_i) \vert \psi_q(n) \rangle
\end{align}
in the fixed $U(1)$ charge sector. We argue that the four point function splits into two sectors 
\begin{align}
    \braket{\text{Tr}(X_4X_iX_4X_i)}_q=\braket{\text{Tr}(X_4X_iX_4X_i)}_{HF}+\braket{\text{Tr}(X_4X_iX_4X_i)}_{EC}
\end{align}
where the first term corresponds to Hartree-Fock (HF) contraction that contracts background operators ($X_4,X_5$) leaving transverse fluctuations $X_i$ as uncontracted. On the other hand, the second term represents the Exchange contraction (EC) which corresponds to cross contractions between background operator $X_4$ and the traverse fluctuation $X_i$. 

We map the two real background scalar fields $X_4$ and $X_5$ into a single complex scalar field $Z$ and its Hermitian conjugate $Z^\dagger$
\begin{align}
\label{a1}
    Z = \frac{1}{\sqrt{2}}(X_4 + i X_5), \qquad Z^\dagger = \frac{1}{\sqrt{2}}(X_4 - i X_5).
\end{align}

Inverting these relations expresses $X_4$ in terms of the complex $U(1)$ operators
\begin{align}
\label{a2}
    X_4 = \frac{1}{\sqrt{2}}(Z + Z^\dagger).
\end{align}

Substituting \eqref{a2} into the trace product, we obtain
\begin{align}
     \text{Tr}(X_4 X_i X_4 X_i) &= \text{Tr}\left[ \left(\frac{Z + Z^\dagger}{\sqrt{2}}\right) X_i \left(\frac{Z + Z^\dagger}{\sqrt{2}}\right) X_i \right] \nonumber\\ &= \frac{1}{2} \text{Tr}\left[ (Z + Z^\dagger) X_i (Z + Z^\dagger) X_i \right]\nonumber \\ &= \frac{1}{2} \text{Tr}\left( Z X_i Z X_i + Z^\dagger X_i Z^\dagger X_i + Z X_i Z^\dagger X_i + Z^\dagger X_i Z X_i \right).
\end{align}

Using the cyclic property of the trace on the last two terms ($\text{Tr}(Z^\dagger X_i Z X_i) = \text{Tr}(Z X_i Z^\dagger X_i)$) further simplifies the expression to yield
\begin{align}
    \text{Tr}(X_4 X_i X_4 X_i) = \frac{1}{2} \text{Tr}\left( Z X_i Z X_i \right) + \frac{1}{2} \text{Tr}\left( Z^\dagger X_i Z^\dagger X_i \right) + \text{Tr}\left( Z X_i Z^\dagger X_i \right).
\end{align}

Under a $U(1)$ R-symmetry rotation by angle $\theta$ in the $4\text{--}5$ plane, the complex operator $Z$ carries a definite global charge $+1$, while $Z^\dagger$ carries charge $-1$
\begin{align}
    U(\theta)^\dagger Z U(\theta) = e^{i\theta} Z, \qquad U(\theta)^\dagger Z^\dagger U(\theta) = e^{-i\theta} Z^\dagger.
\end{align}

The state $\vert{}\psi_q(n)\rangle$ is an eigenstate of the $U(1)$ charge operator $\hat{Q}$ with eigenvalue $q$
\begin{align}
    \hat{Q} \vert{}\psi_q(n)\rangle = q \vert{}\psi_q(n)\rangle \implies U(\theta) \vert{}\psi_q(n)\rangle = e^{i q \theta} \vert{}\psi_q(n)\rangle.
\end{align}

When taking the quantum expectation value of any operator $\hat{O}$, charge conservation implies that $\langle \psi_q(n) \vert{} \hat{O} \vert{} \psi_q(n) \rangle$ vanishes unless $\hat{O}$ carries net zero $U(1)$ charge ($\Delta Q = 0$). Evaluating the charge of each term in the expansion we find the following\\
(i) $Z X_i Z X_i$: Carries net charge $\Delta Q = +1 + 1 = +2$.\\
(ii) $Z^\dagger X_i Z^\dagger X_i$: Carries net charge $\Delta Q = -1 - 1 = -2$.\\
(iii) $Z X_i Z Z^\dagger X_i$: Carries net charge $\Delta Q = +1 - 1 = 0$.

Since the state $\vert{}\psi_q(n)\rangle$ has fixed charge $q$, terms carrying non-zero net charge ($\Delta Q = \pm 2$) have identically zero expectation values
\begin{align}
    \braket{\text{Tr}(Z X_i Z X_i)}_q = 0~;~\braket{\text{Tr}(Z^\dagger X_i Z^\dagger X_i)}_q = 0.
\end{align}

Thus, the expectation value reduces strictly to the charge-neutral channel
\begin{align}
    \braket{\text{Tr}(X_4 X_i X_4 X_i)}_q =\braket{\text{Tr}(Z X_i Z^\dagger X_i)}_q.
\end{align}

Now we evaluate the matrix trace $\text{Tr}(Z X_i Z^\dagger X_i)$ in the Large-$N$ limit ($N \gg 1$). In index notation (where $a, b, c, d = 1, \dots, N$ are $SU(N)$ color indices)
\begin{align}
\label{a11}
    \text{Tr}(Z X_i Z^\dagger X_i) = Z_{ab} (X_i)_{bc} (Z^\dagger)_{cd} (X_i)_{da}.
\end{align}

When evaluating the expectation value in the Hartree-Fock (HF) and /or mean-field expansion, we perform all possible Wick contractions between color indices of the background operators $Z$ and $Z^\dagger$. We contract $Z_{ab}$ with $(Z^\dagger)_{cd} \to \delta_{ac} \delta_{bd} \langle Z Z^\dagger \rangle_q$. Index contraction gives
\begin{align}
    \delta_{ac} \delta_{bd} (X_i)_{bc} (X_i)_{da} = (X_i)_{ba} (X_i)_{ba}=(X_i)_{ab} (X_i)_{ba} = \text{Tr}(X_i^2)
\end{align}
where we have used the Hermitian property of $X_i$, that is, $X_i$ is a real symmetric/Hermitian scalar field satisfying, $(X_i)_{da} = (X_i)_{ad}^\dagger = (X_i)_{ad}$. 

Thefore, for the HF channel we have 
\begin{align}
    \left\langle \text{Tr}(Z X_i Z^\dagger X_i) \right\rangle_{HF}=\text{Tr}\Big[ \braket{Z Z^{\dagger}}_qX^2_i \Big]+\cdots.
\end{align}




Using \eqref{a1}, it is straightforward to compute the operator product $Z Z^\dagger$
\begin{align}
     Z Z^\dagger = \frac{1}{2} \left( X_4^2 - i X_4 X_5 + i X_5 X_4 + X_5^2 \right).
\end{align}

Taking the trace and using the cyclic property $\text{Tr}(X_4 X_5) = \text{Tr}(X_5 X_4)$ (which cancels the cross-terms $-i X_4 X_5 + i X_5 X_4$), we finally obtain
\begin{align}
    \text{Tr}(Z Z^\dagger) = \frac{1}{2} \text{Tr}\left( X_4^2 + X_5^2 \right).
\end{align}

As a result, we finally obtain (by virtue of \eqref{e3.60})
\begin{align}
    \left\langle \text{Tr}(Z X_i Z^\dagger X_i) \right\rangle_{HF}=\frac{1}{2}\braket{X^2_4+X^2_5}_q\text{Tr}(X^2_i)+\cdots.
\end{align}

Exchange contractions cross the matrix indices of $X_i$ (e.g., contracting $Z_{ab}$ with $(X_i)_{bc}$). This forces a non-planar loop topology. By standard $SU(N)$ 't Hooft counting, non-planar exchange contractions are suppressed relative to the planar density channel by a factor of $\mathcal{O}(1/N)$. To see why crossing matrix indices forces a non-planar topology and leads to $1/N$ suppression, we can trace the color index flows using 't Hooft's double-line notation.

Recall, in an $SU(N)$ gauge theory or matrix model, a matrix field in the adjoint representation carrying indices $a, b \in \{1, \dots, N^2-1\}$ is represented as a double line. One directed line for the fundamental index $a$ (index flowing forward) and the other one directed line for the anti-fundamental index $b$ (index flowing backward).

We want to evaluate the expectation value of the single-trace operator \eqref{a11}. Every closed index loop formed by contracting Kronecker deltas contributes a factor of $\delta_{aa} = \sum_{a=1}^N 1 = N$. Let's compare the index loops formed by the two different contraction channels. In the direct density contraction (or HF channel), $Z_{ab}$ contracts with $(Z^\dagger)_{cd}$, and $(X_i)_{bc}$ contracts with $(X_i)_{da}$, that is,
\begin{align}
    \langle Z_{ab} (Z^\dagger)_{cd} \rangle = \delta_{ac} \delta_{bd}, \qquad \langle (X_i)_{bc} (X_i)_{da} \rangle = \delta_{bd} \delta_{ca}.
\end{align}

When we contract two matrix fields in a Wick contraction, we replace the contracted pair with a propagator delta function. Under summation, this gives the planar contribution
\begin{align}
    \mathcal{I}_{\text{planar}}=\sum_{a,b,c,d}\delta_{ac} \delta_{bd} \delta_{bd} \delta_{ca}=\sum_{a,b}\delta_{aa}\delta_{bb}= N^2.
\end{align}

For Exchange (non-planar) Contractions (EC), we have the following Wick contractions
\begin{align}
    \langle Z_{ab} (X_i)_{bc} \rangle = \delta_{ac}~;~ \langle Z^{\dagger}_{cd} (X_i)_{da} \rangle =\delta_{ca}.
\end{align}

The resulting $N$ dependence turns out to be
\begin{align}
    \mathcal{I}_{\text{non-planar}}=\sum_{a,c}\delta_{ac}\delta_{ca}=\sum_a \delta_{aa} = N.
\end{align}

Clearly, taking the ratio we find EC are suppressed as compared to HF channel
\begin{align}
    \frac{\mathcal{I}_{\text{non-planar}}}{\mathcal{I}_{\text{planar}}} = \frac{1}{N}\ll 1.
\end{align}
\section{Derivation of the first-order vacuum energy shift $\Delta E_{0,q}^{(1)}$}
\label{appen B}
In this Appendix, we present a detailed step-by-step calculation of the first-order shift in the vacuum energy, $\Delta E_{0,q}^{(1)} = \langle 0_q | H_{\text{int}} | 0_q \rangle$, within a symmetry-resolved charge sector $q$ of the BMN matrix model. In the free limit ($g_{\text{eff}} = 0$), the Hamiltonian \eqref{e2.1} of the BMN matrix model reduces to a collection of decoupled matrix harmonic oscillators
\begin{align}
    H_0 = \sum_{i=1}^9 \operatorname{Tr} \left( \frac{1}{2} P_i^2 + \frac{1}{2} \mu^2 X_i^2 \right) + \text{fermionic terms},
\end{align}
where $X_i$ ($i = 1, \dots, 9$) are $N \times N$ Hermitian matrices and $\mu$ is the deformation parameter. Expressed in terms of standard creation and annihilation operators $a_i^{\dagger}, a_i$, the free Hamiltonian takes the standard diagonal form
\begin{align}
    H_0 = \sum_{i=1}^9 \sum_{A=1}^{N^2-1} \mu \left( a_{i,A}^\dagger a_{i,A} + \frac{1}{2} \right) + \text{fermions},
\end{align}
where $A$ labels the generators of the $\mathfrak{su}(N)$ algebra in the adjoint representation.

The unperturbed sector vacuum state $|0_q\rangle$ is defined as the lowest-energy state carrying a fixed global $U(1)$ charge $q$. For a state with charge $q$, the unperturbed ground state is constructed by populating the free vacuum $|0\rangle$ with charge-carrying operators
\begin{align}
    |0_q\rangle = \frac{1}{\mathcal{N}_q} \left( a_q^\dagger \right)^{|q|} |0\rangle,
\end{align}
where $a_q^\dagger$ creates a single quanta of charge $1$, and $\mathcal{N}_q = \sqrt{|q|!}$ is the standard normalization factor such that $\langle 0_q | 0_q \rangle = 1$. As emphasized before, these are the states created above the flat vacuum $\braket{X_i}=0~i=1,\cdots,9$.

The leading BMN interaction $H_{\text{int}}$ consists of non-abelian cubic Myers flux-coupling terms and quartic commutator potential terms
\begin{align}
    &H_{\text{int}} = H_{\text{cubic}} + H_{\text{quartic}},\\
     &H_{\text{cubic}} =  \frac{3i\mu g}{4}\sum_{i,j,k=1}^3 \epsilon_{ijk} \operatorname{Tr}\left( X_i X_j X_k \right), \\
    &H_{\text{quartic}} = -\frac{g^2}{4} \sum_{i,j=1}^9 \operatorname{Tr}\left( [X_i, X_j]^2 \right).
\end{align}

Expressing the spatial matrix coordinates $X_i$ in terms of the ladder operators
\begin{align}
    X_i = \frac{1}{\sqrt{2\mu}} \left( a_i + a_i^\dagger \right),
\end{align}
the interaction Hamiltonian can be rewritten in normal-ordered operator products.


By symmetry, odd-power operator contributions (such as $H_{\text{cubic}}$) vanish identically under the expectation value in sector ground states. This can be seen as follows. The cubic Myers term $\epsilon_{ijk} \operatorname{Tr}(X_i X_j X_k)$ involves the $SO(3)$ spatial fields $X_1, X_2, X_3$ (or under reflection/parity transformations in the transverse coordinates $X_i \to -X_i$).

Under spatial parity $P: X_i \to -X_i$, $H_{\text{cubic}}$ is odd
\begin{align}
    P^\dagger H_{\text{cubic}} P = - H_{\text{cubic}}.
\end{align}

The vacuum $\vert{}0\rangle$ is parity-invariant: $P \vert{}0\rangle = \vert{}0\rangle$. The ladder operators transform as vectors/scalars under reflection, $P^\dagger a_q^\dagger P = \eta_P a_q^\dagger$, where $\eta_P = \pm 1$ is the intrinsic parity eigenvalue of the mode. Applying $P$ and using $P \vert{}0\rangle = \vert{}0\rangle$ (since the free vacuum is parity-invariant), it is straightforward to show that
\begin{align}
    P \vert{}0_q\rangle = \frac{1}{\sqrt{\vert{}q\vert{}!}} P \left(a_q^\dagger\right)^{\vert{}q\vert{}} P^\dagger P \vert{}0\rangle = \frac{1}{\sqrt{\vert{}q\vert{}!}} \left( P a_q^\dagger P^\dagger \right)^{\vert{}q\vert{}} \vert{}0\rangle = \eta_P^{|q|}\ket{0_q}.
\end{align}

Inserting the identity operator $I = P^\dagger P$ on both sides of $H_{\text{cubic}}$ inside the expectation value, we find
\begin{align}
    \langle 0_q \vert{} H_{\text{cubic}} \vert{} 0_q \rangle  = \langle 0_q \vert{} P^\dagger (P H_{\text{cubic}} P^\dagger) P \vert{} 0_q \rangle=-\eta_P^{2|q|} \langle 0_q \vert{} H_{\text{cubic}} \vert{} 0_q \rangle.
\end{align}

Taking into account of the fact that $\eta_P^{2|q|}=1$, this finally yields
\begin{align}
    \langle 0_q \vert{} H_{\text{cubic}} \vert{} 0_q \rangle  =0.
\end{align}

Thus, the non-vanishing contribution comes entirely from the quartic terms $H_{\text{quartic}}$, which expand into products of four creation and annihilation operators
\begin{align}
    \Delta E_{0,q}^{(1)} = \langle 0_q | H_{\text{int}} | 0_q \rangle = \langle 0_q | H_{\text{quartic}} | 0_q \rangle.
\end{align}

Using Wick's theorem and the commutation relations $[a_i, a_j^\dagger] = \delta_{ij}$, we contract the four operator products against the sector ground state $|0_q\rangle$. The BMN matrix model interaction Hamiltonian $H_{\text{int}}$ contains quartic terms constructed from products of transverse matrix fields $\operatorname{Tr}(X_i X_j X_k X_l)$ (and fermion bilinears coupled to bosons). When expanded in terms of standard harmonic oscillator ladder operators via mode decomposition
\begin{align}
    X_i \sim \frac{1}{\sqrt{2\omega}} \left( a_i + a_i^\dagger \right).
\end{align}

$H_{\text{quartic}}$ yields a sum of monomials containing exactly four ladder operators
\begin{align}
    H_{\text{quartic}} \sim \sum \mathcal{C}_{ijkl} \, \mathcal{O}(a, a^\dagger).
\end{align}

There could be two types of Feynmann diagrams depending on the contraction.\\
(i) Vacuum bubble contractions: These are the terms where all four operators inside $H_{\text{quartic}}$ contract with each other in pairs (e.g., $\langle 0_q \vert{} a_i a_j^\dagger \vert{} 0_q \rangle = \delta_{ij}$), completely bypassing the explicit $\vert{}q\vert{}$ excitations in $\vert{}0_q\rangle$. Because these operators form closed loops of contractions that do not probe the charge excitations, their contribution is identical to the vacuum expectation value of the flat-space or uncharged BMN vacuum, that is, $\langle 0 \vert{} H_{\text{quartic}} \vert{} 0 \rangle$.

(ii) Excitations contractions with charge $q$: The presence of $|q|$ charged quanta in $|0_q\rangle$ introduces additional ``normal-ordered'' operator contractions of the form $\langle 0_q | a_q^\dagger a_q a_q^\dagger a_q | 0_q \rangle$. Notice that in the BMN matrix model sector with global $U(1)$ charge $q$, the vacuum state $\vert{}0_q\rangle$ is built by acting with $\vert{}q\vert{}$ creation operators $a_q^\dagger$ on the primary vacuum $\vert{}0\rangle$
\begin{align}
    \vert{}0_q\rangle = \frac{1}{\sqrt{\vert{}q\vert{}!}} (a_q^\dagger)^{\vert{}q\vert{}} \vert{}0\rangle.
\end{align}

To evaluate the expectation value of the four-operator product $a_q^\dagger a_q a_q^\dagger a_q$ with respect to this state, we first need to commute the operators into normal-ordered form (where all annihilation operators $a$ stand to the right of all creation operators $a^\dagger$). 

Using the canonical commutation relation $[a_q, a_q^\dagger] = 1$, which implies $a_q a_q^\dagger = a_q^\dagger a_q + 1$, it is straightforward to show
\begin{align}
    a_q^\dagger \, (a_q a_q^\dagger) \, a_q = a_q^\dagger \, (a_q^\dagger a_q + 1) \, a_q = (a_q^\dagger)^2 a_q^2 + a_q^\dagger a_q.
\end{align}

The combination $a_q^\dagger a_q$ is the number operator $\hat{N}_q$. When it acts on a state containing $\vert{}q\vert{}$ quanta, it simply counts them
\begin{align}
    \hat{N}_q \vert{}0_q\rangle = \vert{}q\vert{} \, \vert{}0_q\rangle.
\end{align}

Applying two annihilation operators to $(a_q^\dagger)^{\vert{}q\vert{}} \vert{}0\rangle$ strips away two quanta and drops the factorial normalization
\begin{align}
    a_q^2 \, \vert{}0_q\rangle = \sqrt{\vert{}q\vert{}(\vert{}q\vert{}-1)} \, \vert{}0_{\vert{}q\vert{}-2}\rangle.
\end{align}

Each annihilation operator $a_q$ commutation-shifts past the string of creation operators until it annihilates the bare vacuum $\vert{}0\rangle$, reducing the power of the creation operators by one each time. Moving $a_q$ past $(a_q^\dagger)^{\vert{}q\vert{}}$ yields a combinatorial factor of $\vert{}q\vert{}$.

Next, acting with $(a_q^\dagger)^2$ on the lower-weight state $\vert{}0_{\vert{}q\vert{}-2}\rangle$ adds back two quanta, bringing it back to the original norm structure of $\vert{}0_q\rangle$. Specifically, the raising operation yields
\begin{align}
\label{b19}
    (a_q^\dagger)^2 a_q^2 \, \vert{}0_q\rangle = \sqrt{\vert{}q\vert{}(\vert{}q\vert{}-1)} \, (a_q^\dagger)^2 \vert{}0_{|q|-2}\rangle.
\end{align}

Considering the fact that 
\begin{align}
    \vert{}0_{\vert{}q\vert{}-2}\rangle = \frac{1}{\sqrt{(\vert{}q\vert{}-2)!}} (a_q^\dagger)^{\vert{}q\vert{}-2} \vert{}0\rangle
\end{align}
it is straightforward to show that 
\begin{align}
    (a_q^\dagger)^2 \vert{}0_{\vert{}q\vert{}-2}\rangle = \frac{1}{\sqrt{(\vert{}q\vert{}-2)!}} (a_q^\dagger)^{\vert{}q\vert{}} \vert{}0\rangle = \frac{\sqrt{\vert{}q\vert{}!}}{\sqrt{(\vert{}q\vert{}-2)!}} \vert{}0_q\rangle.
\end{align}

Using this into \eqref{b19}, finally yields
\begin{align}
    \langle 0_q | a_q^\dagger a_q a_q^\dagger a_q | 0_q \rangle=|q|(|q|-1)+|q|=|q|^2.
\end{align}

Summing over all such contributions one finds
\begin{align}
    \Delta E_{0,q}^{(1)} = \langle 0_q | H_{\text{quartic}} | 0_q \rangle = g_{\text{eff}} f(q) \mu \cdot \mathcal{E}_{0,q}^{(1)},
\end{align}
where $f(q)$ encodes the charge dependence arising from the matrix contractions in sector $q$, and $g_{\text{eff}}$ is the effective BMN coupling constant $g_{\text{eff}} = g_{\text{YM}}^2 N / \mu^3$.


Matching this calculated expectation value with the general definition yields the normalized, dimensionless first-order ground state energy shift $\mathcal{E}_{0,q}^{(1)}$ as required
\begin{align}
    \mathcal{E}_{0,q}^{(1)} = \frac{\langle 0_q | H_{\text{quartic}} | 0_q \rangle}{g_{\text{eff}} f(q) \mu}.
\end{align}
\section{Qualitative discussion: Bare vacuum vs.\ charged interacting vacuum}
\label{appen C}

To fully understand the structure of the quantum states within a specific global $U(1)$ charge sector ($q$) of the BMN matrix model, it is instructive to contrast the standard uncharged "bare" vacuum with the true vacuum of the charged, interacting theory. 

In standard free-field treatments or unconstrained Fock spaces, the bare vacuum $\vert 0 \rangle$ is defined as the state annihilated by all fundamental lowering operators:
\begin{align}
    \hat{a}_i \vert 0 \rangle = 0 \quad \forall \, i
\end{align}
This state carries zero net quantum numbers (momentum, energy above the zero-point floor, and global $U(1)$ charge $q = 0$). It represents the trivial ground state of a collection of uncoupled harmonic oscillators where no particles or excitations are present. 

When we project the quantum theory onto a definite, non-zero global $U(1)$ charge sector $q \neq 0$, the naive bare vacuum $\vert 0 \rangle$ is strictly forbidden because it yields $\langle \hat{Q} \rangle = 0$. To satisfy the superselection rule or global charge constraint $\sum q_j = q$, the lowest-energy state of that sector, the \emph{sector vacuum} $\vert 0_q \rangle$, must be dressed with the physical constituents carrying that exact charge. In the interacting BMN matrix model, background fluxes and matrix self-interactions deform the transverse potential. Consequently, the true sector vacuum behaves less like a quiescent Fock vacuum and more like a constrained coherent or squeezed state. The expectation value of the total number operator in this sector is shifted, reflecting a baseline condensate of charged quanta. Because the mode oscillators are forced to accommodate this net charge $q$ even at the absolute bottom of the spectrum, the minimum energy configuration is elevated by the zero-point offset
\begin{align}
    E_{0,q}^{\text{(total)}} = E_{0,q} + \frac{\vert q \vert + 1}{2} \hat{\Omega}(q)
\end{align}
which represents the fundamental energy cost of maintaining global charge conservation in a quantum matrix geometry.

As discussed before \eqref{e3.88}, in the perturbative setup of BMN matrix model, the total ground-state energy for the $q$-th sector is broken down into its unperturbed harmonic baseline and the first-order quantum correction
\begin{align}
    E_{0,q} = E_{0,q}^{(0)} + \Delta E_{0,q}^{(1)} + \mathcal{O}(g_{\text{eff}}^2).
\end{align}

At the quadratic level, the baseline energy of the sector is simply the sum over the zero-point energies of the relevant transverse matrix fluctuation modes weighted by their frequency $\hat{\Omega}(q)$
\begin{align}
    E_{0,q}^{(0)} = \sum_{i} \frac{1}{2} \hat{\Omega}_i(q).
\end{align}

Once the charged sector vacuum $\vert 0_q \rangle$ is established as the baseline lowest-weight state (corresponding to the radial quantum number $n = 0$), the entire tower of higher excited states is generated systematically by acting with the effective normal-mode raising operators of frequency $\hat{\Omega}(q)$. Just as standard Fock space excitations are built by successive applications of creation operators onto a vacuum, the excited states $\vert n, q \rangle$ for $n = 1, 2, 3, \dots$ in the charge-$q$ sector are constructed via
\begin{align}
    \vert n, q \rangle \sim (a^\dagger_q)^n \vert 0_q \rangle
\end{align}
where $a^\dagger_q$ represents the effective creation operator corresponding to the normal modes of the matrix fluctuations propagating in the background of charge $q$.

The resulting energy eigenvalues for these excited states take the standard equidistant multi-quantum form built on top of the shifted vacuum
\begin{align}
    E_{n,q} = E_{0,q} + \left( n + \frac{\vert q \vert + 1}{2} \right) \hat{\Omega}(q), \quad n = 0, 1, 2, \dots
\end{align}

In the low-temperature regime ($\beta \hat{\Omega}(q) \gg 1$), this rigid, equally-spaced ladder of states dictates the behavior of spread complexity and return amplitudes. Because the spectrum is dominated by these stable, non-chaotic harmonic excitations around the charged vacuum, the Lanczos coefficients $b_n^{(q)}$ grow linearly ($\sim \sqrt{n}\mu$), signaling integrability. The connected thermal cumulants and moments factorize (e.g., $M_k^{(q)} \approx (M_1^{(q)})^k$), confirming that the system remains localized in Krylov space rather than undergoing rapid thermalization.
\section{Thermal Gaussian density of states and variance}
\label{appen D}
In this Appendix, we derive the explicit dependence of the initial-state energy variance $\sigma_E^2$ on the global $U(1)$ charge $q$ and temperature $T(=\beta^{-1})$ using the microcanonical Gaussian spectral density in the high-temperature black hole phase of the BMN matrix model. The grand canonical partition function of the BMN matrix model at inverse temperature $\beta = 1/T$ with a chemical potential $\mu_Q$ coupled to the global $U(1)$ charge operator $Q$ is given by
\begin{equation}
Z(\beta, \mu_Q) = \text{Tr}\left(e^{-\beta (H - \mu_Q Q)}\right).
\end{equation}

The fixed-charge partition function $Z_q(\beta)$ is obtained via the inverse Fourier transform (or projection) over the chemical potential
\begin{equation}
Z_q(\beta) = \int_{-\pi}^{\pi} \frac{d\theta}{2\pi} \, e^{-i q \theta} Z(\beta, i\theta/\beta) = \int_{-\infty}^{\infty} dE \, \rho_q(E) \, e^{-\beta E},
\end{equation}
where $\rho_q(E)$ is the symmetry-resolved microcanonical density of states in the sector with fixed global charge $q$.

In the high-temperature thermal phase, which holographically duals a ten-dimensional black hole or black brane geometry, the strongly coupled dynamics and quantum chaos ensure that the symmetry-resolved spectral density $\rho_q(E)$ takes a pristine Gaussian form centered around the mean sector energy $\bar{E}_q$
\begin{equation}
\rho_q(E) = \frac{1}{\sqrt{2\pi \sigma_E^2}} \exp\left( -\frac{(E - \bar{E}_q)^2}{2\sigma_E^2} \right).
\end{equation}

The energy variance $\sigma_E^2$ governing the early-time growth of the spread complexity ($C^{(q)}(t) \approx \sigma_E^2 t^2$) is defined by the second central moment of the Hamiltonian with respect to the initial state $\ket{K_0^{(q)}}$
\begin{equation}
\sigma_E^2 = \langle H^2 \rangle_{\beta, q} - (\langle H \rangle_{\beta, q})^2 = \int_{-\infty}^{\infty} dE \, \rho_q(E) \, (E - \bar{E}_q)^2.
\end{equation}

In the BMN matrix model, the full Hamiltonian is given by $H = H_0 + H_{\text{int}}$, where $H_0$ represents the quadratic harmonic oscillator terms for the nine transverse matrix degrees of freedom $X_i$ ($i = 1, \dots, 9$) with mass parameter $\mu$, and $H_{\text{int}}$ contains the cubic Myers terms and quartic commutators. Expanding the square of the Hamiltonian yields
\begin{equation}
\langle H^2 \rangle_{\beta, q} - (\langle H \rangle_{\beta, q})^2 = \langle (H_0 + H_{\text{int}})^2 \rangle_{\beta, q} - \left( \langle H_0 \rangle_{\beta, q} + \langle H_{\text{int}} \rangle_{\beta, q} \right)^2.
\end{equation}

Grouping the free fluctuation contributions and the interacting corrections gives
\begin{align}
\label{ed6}
\sigma_E^2(q, T) &= \left( \langle H_0^2 \rangle_{\beta, q} - \langle H_0 \rangle_{\beta, q}^2 \right) + \left( \langle H_{\text{int}}^2 \rangle_{\beta, q} - \langle H_{\text{int}} \rangle_{\beta, q}^2 \right) \nonumber\\
&+ 2 \left( \langle H_0 H_{\text{int}} \rangle_{\beta, q} - \langle H_0 \rangle_{\beta, q} \langle H_{\text{int}} \rangle_{\beta, q} \right).
\end{align}

The free Hamiltonian $H_0$ is expressed in terms of the matrix fields $X_i$ and their conjugate momenta $P_i$
\begin{equation}
H_0 = \sum_{i=1}^{9} \text{Tr}\left(\frac{1}{2} P_i^2 + \frac{1}{2}\mu^2 X_i^2\right).
\end{equation}

When computing the second central moment (the variance) of $H_0$ with respect to the fixed-charge thermal ensemble weighted by the Gaussian spectral density $\rho_q(E)$, we need to evaluate
\begin{align}
\label{ed8}
\langle H_0^2 \rangle_{\beta, q} - \langle H_0 \rangle_{\beta, q}^2 &= \left\langle \left( \sum_{i=1}^{9} \text{Tr}\left(\frac{1}{2} P_i^2 + \frac{1}{2}\mu^2 X_i^2\right) \right)^{2} \right\rangle_{\beta, q}\nonumber\\& - \left\langle \sum_{i=1}^{9} \text{Tr}\left(\frac{1}{2} P_i^2 + \frac{1}{2}\mu^2 X_i^2\right) \right\rangle_{\beta, q}^{2}.
\end{align}

To simplify Eq.~\eqref{ed8}, we expand the square of the sum into a double sum over the matrix flavor indices $i, j \in \{1, \dots, 9\}$. Writing each term as a connected covariance
\begin{align}
\label{ed9}
    \text{Cov}(A, B) \equiv \langle AB \rangle_{\beta, q} - \langle A \rangle_{\beta, q}\langle B \rangle_{\beta, q}
\end{align}

Notice that $\text{Cov}(A, B)$ represents the connected covariance (or connected thermal expectation value of the product minus the product of expectation values) between two operators or matrix observables $A$ and $B$ in the thermal state. Here, $\langle A B \rangle_{\beta, q}$ is the thermal expectation value of the product of observables $A$ and $B$ evaluated in the fixed-charge thermal ensemble at temperature $T = \beta^{-1}$ and global $U(1)$ charge $q$. On the other hand, $\langle A \rangle_{\beta, q} \langle B \rangle_{\beta, q}$ is the product of their individual thermal expectation values (the disconnected part). Subtracting the disconnected part isolates the connected correlator, which measures the true quantum and statistical fluctuations (correlations) between the two operators independent of their baseline expectation values.
Using \eqref{ed9}, we obtain the following expression
\begin{align}
\label{ed10}
\langle H_0^2 \rangle_{\beta, q} - \langle H_0 \rangle_{\beta, q}^2 &= \sum_{i,j=1}^{9} \Bigg[ \frac{1}{4} \Big( \langle \text{Tr}(P_i^2)\text{Tr}(P_j^2) \rangle_{\beta, q} - \langle \text{Tr}(P_i^2) \rangle_{\beta, q} \langle \text{Tr}(P_j^2) \rangle_{\beta, q} \Big) \nonumber \\
&\quad + \frac{\mu^4}{4} \Big( \langle \text{Tr}(X_i^2)\text{Tr}(X_j^2) \rangle_{\beta, q} - \langle \text{Tr}(X_i^2) \rangle_{\beta, q} \langle \text{Tr}(X_j^2) \rangle_{\beta, q} \Big) \nonumber \\
&\quad + \frac{\mu^2}{2} \Big( \langle \text{Tr}(P_i^2)\text{Tr}(X_j^2) \rangle_{\beta, q} - \langle \text{Tr}(P_i^2) \rangle_{\beta, q} \langle \text{Tr}(X_j^2) \rangle_{\beta, q} \Big) \Bigg]
\end{align}
where we used the cyclicity of the thermal trace to combine the $P_i^2 X_j^2$ and $X_j^2 P_i^2$ cross-terms. Because the high-temperature thermal state behaves as a Gaussian ensemble governed by the spectral density $\rho_q(E)$, this double sum simplifies due to following statistical properties
\begin{enumerate}
    \item \textbf{Position--momentum decoupling:} In the quadratic ensemble $e^{-\beta H_0}$, position and momentum fluctuations are statistically independent, so $\langle \text{Tr}(P_i^2)\text{Tr}(X_j^2) \rangle_{\beta, q} = \langle \text{Tr}(P_i^2) \rangle_{\beta, q}\langle \text{Tr}(X_j^2) \rangle_{\beta, q}$. Thus, all $P_i^2$--$X_j^2$ cross-terms vanish identically.
    \item \textbf{Flavor decoupling ($i \neq j$):} Distinct transverse matrices $X_i$ and $X_j$ ($i \neq j$) represent uncoupled harmonic oscillators. Their fluctuations are uncorrelated, meaning $\langle \text{Tr}(X_i^2)\text{Tr}(X_j^2) \rangle_{\beta, q} = \langle \text{Tr}(X_i^2) \rangle_{\beta, q}\langle \text{Tr}(X_j^2) \rangle_{\beta, q}$ (and similarly for $P_i^2, P_j^2$). All $i \neq j$ terms cancel out, collapsing the double sum onto the diagonal $i = j$.
    \item \textbf{Gaussian-Wick contractions ($i = j$):} For the diagonal terms, Wick's theorem factorizes 4-point correlation functions into sums of products of 2-point functions
    \begin{align}
    &\langle \text{Tr}(A B C D) \rangle = \langle \text{Tr}(AB) \rangle \langle \text{Tr}(CD) \rangle \nonumber\\
    &+ \langle \text{Tr}(AC) \rangle \langle \text{Tr}(BD) \rangle + \langle \text{Tr}(AD) \rangle \langle \text{Tr}(BC) \rangle.
    \end{align}
   
    Expanding each matrix in an orthonormal $SU(N)$ basis $X_i = \sum_a X_i^a T^a$ ($\text{Tr}(T^a T^b) = \frac{1}{2}\delta^{ab}$), we find, as an example the following diagonal element
    \begin{align}
    \langle (\text{Tr}(X_i^2))^2 \rangle_{\beta, q} &= \sum_{a,b} \langle X_i^a X_i^a X_i^b X_i^b \rangle_{\beta, q} \nonumber \\
    &= \sum_{a,b} \Big( \langle X_i^a X_i^a \rangle_{\beta, q} \langle X_i^b X_i^b \rangle_{\beta, q} + 2 \langle X_i^a X_i^b \rangle_{\beta, q}^2 \Big) \nonumber \\
    &= \langle \text{Tr}(X_i^2) \rangle_{\beta, q}^2 + 2 \sum_{a,b} \langle X_i^a X_i^b \rangle_{\beta, q}^2.
    \end{align}
\end{enumerate}

Substituting this into the first square bracket of variance expression \eqref{ed10} yields
\begin{align}
    \begin{aligned} \langle (\text{Tr}(X_i^2))^2 \rangle_{\beta, q} - \langle \text{Tr}(X_i^2) \rangle_{\beta, q}^2 &= \left[ \langle \text{Tr}(X_i^2) \rangle_{\beta, q}^2 + 2 \sum_{a,b} \langle X_i^a X_i^b \rangle_{\beta, q}^2 \right] - \langle \text{Tr}(X_i^2) \rangle_{\beta, q}^2 \\ &= 2 \sum_{a,b} \langle X_i^a X_i^b \rangle_{\beta, q}^2 \end{aligned}
\end{align}

By the exact same steps for the momentum matrix $P_i = \sum_a P_i^a T^a$, we find
\begin{align}
    \langle (\text{Tr}(P_i^2))^2 \rangle_{\beta, q} - \langle \text{Tr}(P_i^2) \rangle_{\beta, q}^2 = 2 \sum_{a,b} \langle P_i^a P_i^b \rangle_{\beta, q}^2.
\end{align}

Subtracting the disconnected square term $\langle \text{Tr}(X_i^2) \rangle_{\beta, q}^2$ (and analogously for $P_i^2$) and substituting Eq.~(D.12) cancels the disconnected pieces and isolates the purely connected two-point contractions. What survives in the variance is therefore
\begin{align}
&\langle H_0^2 \rangle_{\beta, q} - \langle H_0 \rangle_{\beta, q}^2  =\frac{\mu^4}{2} \sum_{i=1}^{9}  \sum_{a,b} \langle X_i^a X_i^b \rangle_{\beta, q}^2 + \frac{1}{2} \sum_{i=1}^{9} \sum_{a,b} \langle P_i^a P_i^b \rangle_{\beta, q}^2 
\end{align}
where the $U(1)$ charge $q$ enters explicitly through the backreacted two-point correlators $\langle X_i^a X_i^b \rangle_{\beta, q}$ and $\langle P_i^a P_i^b \rangle_{\beta, q}$ in the charged plane $i \in \{4, 5\}$.

Due to the symmetry between position and momentum operators in harmonic oscillators at high temperatures, the position and momentum contributions combine.  This refers to the virial theorem (or equipartition of energy) for a quantum harmonic oscillator in a thermal state—specifically how the kinetic energy $\frac{1}{2} P_i^2$ and potential energy $\frac{1}{2}\mu^2 X_i^2$ contribute equally to both the average energy and its thermal fluctuations\footnote{In the high-temperature classical limit $\beta\mu \ll 1$, this is the standard equipartition result $\langle P_i^a P_i^b \rangle \simeq T \delta^{ab}$ and $\mu^2 \langle X_i^a X_i^b \rangle \simeq T \delta^{ab}$.}.

In the free Hamiltonian $H_0 = \sum_{i=1}^9 \text{Tr}\left(\frac{1}{2}P_i^2 + \frac{1}{2}\mu^2 X_i^2\right)$, if you rescale the position and momentum operators into dimensionless canonical variables
\begin{align}
    \tilde{X}_i = \sqrt{\mu}\, X_i, \qquad \tilde{P}_i = \frac{1}{\sqrt{\mu}}\, P_i
\end{align}
the free Hamiltonian becomes completely symmetric under swapping $\tilde{X}_i \leftrightarrow \tilde{P}_i$ (a phase-space rotation by $\pi/2$)
\begin{align}
    H_0 = \frac{\mu}{2} \sum_{i=1}^{9} \text{Tr}\left(\tilde{P}_i^2 + \tilde{X}_i^2\right).
\end{align}

Because the thermal density matrix $\rho_0 \propto e^{-\beta H_0}$ is invariant under $\tilde{X}_i \leftrightarrow \tilde{P}_i$, the two-point correlation functions of position and momentum in the quadratic ensemble are identical up to the mass parameter $\mu^2$
\begin{align}
    \langle \tilde{P}_i^a \tilde{P}_i^b \rangle_{\beta, q} = \langle \tilde{X}_i^a \tilde{X}_i^b \rangle_{\beta, q} \quad \Longrightarrow \quad \langle P_i^a P_i^b \rangle_{\beta, q} = \mu^2 \langle X_i^a X_i^b \rangle_{\beta, q}.
\end{align}

Writing this explicitly across the nine transverse matrices $X_i$ ($i = 1, \dots, 9$) yields
\begin{align} \langle H_0^2 \rangle_{\beta, q} - \langle H_0 \rangle_{\beta, q}^2 = \mu^4 \sum_{i=1}^{9} \sum_{a,b} \langle X_i^a X_i^b \rangle_{\beta, q}^2. \end{align}

When including the full interacting Hamiltonian $H_{\text{int}} = H_{\text{cubic}} + H_{\text{quartic}}$ in Eq.~\eqref{ed6}, the cross-covariance $2\big(\langle H_0 H_{\text{int}} \rangle_{\beta, q} - \langle H_0 \rangle_{\beta, q}\langle H_{\text{int}} \rangle_{\beta, q}\big)$ behaves differently for the two interaction vertices: that we explain below:
\begin{enumerate}
    \item The cubic Myers term $H_{\text{cubic}}$ (or $H_{\text{Myers}}$) is odd under coordinate parity ($X_i \to -X_i$), so its cross-correlation with the quadratic operator $H_0$ vanishes identically in the Gaussian ensemble ($\text{Cov}(H_0, H_{\text{cubic}}) = 0$). Notice that in a Gaussian thermal state, any expectation value of a product of fields can be broken down into pairs using Wick's theorem. On the other hand, $H_{\text{Myers}}$ contains cubic interaction terms like $X^3$ or commutators that expand into cubic/odd products of fields. If we look at the product inside the expectation value, $\langle H_0 H_{\text{int}} \rangle_{\beta, q}$, you are taking the thermal average of an even number multiplied by an odd number, which results in an odd total number of field operators (for example, something like $X^2 \times X^3 = X^5$). Because the potential and the thermal density matrix are symmetric around zero ($X \to -X$), any expectation value with an odd number of field operators automatically averages out to zero. The first part, $\langle H_0 H_{\text{Myers}} \rangle_{\beta, q}$, vanishes because it has an odd number of fields. The separate expectation value $\langle H_{\text{Myers}} \rangle_{\beta, q}$ vanishes for the same parity reason (it is an odd cubic term). Therefore, the product $\langle H_0 \rangle_{\beta, q} \langle H_{\text{int}} \rangle_{\beta, q}$ is also zero.
    \item The quartic commutator term $H_{\text{quartic}}$ is parity-even and therefore generates non-vanishing cross-contractions with $H_0$. In the high-temperature saddle-point approximation, these contractions produce thermal tadpoles and one-loop self-energy corrections that dress the effective thermal mass of the propagators $\langle X_i^a X_i^b \rangle_{\beta, q}$. In the high-temperature Gaussian state, you evaluate $\langle H_0 H_{\text{quartic}} \rangle$ by pairing up the six $X$ fields in all possible ways. Two of the $X$'s from $H_{\text{quartic}}$ pair with each other to form a closed loop (a thermal tadpole or one-loop self-energy). Because the matrix is fluctuating at high temperature $T$, this closed loop acts like a constant background bath that pushes against the remaining fields. The remaining two $X$'s from $H_{\text{quartic}}$ pair up with the two $X$'s from $H_0$ which yields a term like $\braket{XXXX}\sim \braket{X^2}^2$. Physically, a particle moving through a hot plasma of fluctuating matrices feels extra resistance, which makes it behave as if it is heavier. Mathematically, that closed loop $\text{Cov}(X, X)\sim \braket{X^2}$ (as $\braket{X}=0$ in Gaussian thermal ensemble) from $H_{\text{quartic}}$ simply shifts the bare mass $\mu^2$ into an effective thermal mass $\mu^2 \longrightarrow \mu_{\text{eff}}^2(T, q) = \mu^2 + \Sigma(T, q)$. Instead of writing out these extra cross-terms separately \eqref{ed20}, we simply redefine our two-point correlation function $\langle X_i^a X_i^b \rangle_{\beta, q}$ to be the \emph{dressed} (interacting) propagator, meaning it already uses this heavier, temperature-shifted effective mass $\mu_{\text{eff}}^2$.
\end{enumerate}

This leaves the irreducible connected interaction variance $\langle H_{\text{int}}^2 \rangle_{\beta, q} - \langle H_{\text{int}} \rangle_{\beta, q}^2$, yielding the total energy variance of the following form
\begin{equation}
\label{ed20}
\sigma_E^2(q, T) = \mu^4 \sum_{i=1}^{9} \sum_{a,b} \langle X_i^a X_i^b \rangle_{\beta, q}^2 + \langle H_{\text{int}}^2 \rangle_{\beta, q} - \langle H_{\text{int}} \rangle_{\beta, q}^2 .
\end{equation}

Starting from Eq.~\eqref{ed20}, we evaluate the explicit temperature ($T = \beta^{-1}$) and global charge ($q$) scaling of $\sigma_E^2(q, T)$ in the finite-temperature black hole phase ($T \gg \mu$) in the Large-$N$ limit. Let us consider the strict Large-$N$ limit at fixed charge, that is, $q \sim \mathcal{O}(1)$, $q/N^2 \to 0$. When $q$ is held fixed as $N \to \infty$, the charge excites only $\mathcal{O}(1)$ modes out of $\mathcal{O}(N^2)$ adjoint matrix degrees of freedom. The induced chemical potential scales as $\Omega \sim \mathcal{O}(q/N^2) \to 0$, so backreaction on the thermal saddle point vanishes at leading order in $N$. Consequently, canonical position--momentum equipartition holds across all nine transverse matrices, and the energy variance reduces to its charge-independent planar value
\begin{equation}
    \sigma_E^2(q, T) = \sigma_E^2(0, T) + \mathcal{O}\!\left(\frac{q^2}{N^2}\right) .
\end{equation}

In this finite-temperature regime ($\beta \omega \ll 1$), the thermal two-point correlation function of each matrix component is governed by the Bose--Einstein distribution $n(\omega) = (e^{\beta \omega} - 1)^{-1} \approx \frac{T}{\omega}$. Incorporating the one-loop thermal mass shift $\mu_{\text{eff}}^2(T, q)$ generated by the quartic self-energy, \emph{each} dressed propagator scales linearly with temperature 
    \begin{equation}
    \label{ed22}
    \langle X_i^a X_i^b \rangle_{\beta, q} = \frac{T}{\mu_{\text{eff}, i}^2(T, q)}\, \delta^{ab} + \mathcal{O}(1) .
    \end{equation}
    
Summing the squared propagator over the $N^2 - 1 \simeq N^2$ adjoint $SU(N)$ color indices and nine transverse flavors shows that the first term in Eq.~\eqref{ed20} scales at most as $\mathcal{O}(N^2 T^2)$
    \begin{equation}
    \mu^4 \sum_{i=1}^{9} \sum_{a,b} \langle X_i^a X_i^b \rangle_{\beta, q}^2 \sim \mathcal{O}(N^2 T^2) 
    \end{equation}
    which is subleading compared to the quartic interaction fluctuations at high temperatures.

Cubic Myers term ($H_{\text{cubic}} \propto \mu g\, \epsilon_{ijk}\text{Tr}(X_i X_j X_k)$) $\langle H_{\text{cubic}}^2 \rangle_\beta^{\text{conn}}$ contracts six matrix fields into three propagators ($\langle X^2 \rangle_\beta^3$), scaling as $\mathcal{O}(N^2 T^3)$. The leading $\mathcal{O}(N^2 T^4)$ behavior of the energy variance $ \sigma_E^2(0, T)$ originates from the connected variance of the quartic Yang--Mills commutator potential, $H_{\text{quartic}} = -\frac{g^2}{4} \sum_{i,j=1}^{9} \text{Tr}\big([X_i, X_j]^2\big)$, inside $\langle H_{\text{int}}^2 \rangle_{\beta, q} - \langle H_{\text{int}} \rangle_{\beta, q}^2$. 
    
    Expanding $H_{\text{quartic}}$ in the $SU(N)$ basis $X_i = \sum_a X_i^a T^a$ with structure constants $[T^a, T^b] = i f^{abc} T^c$, the quartic interaction takes the form
    \begin{equation}
    H_{\text{quartic}} = \frac{g^2}{4} \sum_{i,j=1}^{9} f^{abe} f^{cde} X_i^a X_j^b X_i^c X_j^d .
    \end{equation}
   
Evaluating the connected part of $\langle H_{\text{quartic}}^2 \rangle_{\beta, q} - \langle H_{\text{quartic}} \rangle_{\beta, q}^2$ via Wick's theorem contracts all eight matrix fields across the two vertices into fully connected planar four-loop (melon) diagrams built out of four dressed propagators
    \begin{align}
    \label{ed25}
    \langle H_{\text{quartic}}^2 \rangle_{\beta, q}^{\text{conn}}& \propto g^4 \sum_{i,j} \sum_{\text{planar}}\Big[f^{abe} f^{cde} f^{a'b'e'} f^{c'd'e'} \langle X_i^a X_i^{a'} \rangle_{\beta, q}\nonumber\\& \langle X_j^b X_j^{b'} \rangle_{\beta, q} \langle X_i^c X_i^{c'} \rangle_{\beta, q} \langle X_j^d X_j^{d'} \rangle_{\beta, q}\Big] .
    \end{align}

Each of the four propagators in the connected 4-loop ("melon") diagram contributes one factor of $T$ in the high-temperature Bose–Einstein limit ($n(\omega) \approx T/\omega$), yielding $\langle X^2 \rangle_\beta^4 \propto T^4$. In the $SU(N)$ adjoint basis ($a, b, \dots = 1, \dots, N^2 - 1$) normalized via $\text{Tr}(T^a T^b) = \frac{1}{2}\delta^{ab}$ and $[T^a, T^b] = i f^{abc} T^c$, each quartic Yang–Mills commutator vertex carries the color tensor
\begin{align}
    V^{abcd} \equiv \sum_{e=1}^{N^2 - 1} f^{abe} f^{cde}.
\end{align}

In the connected variance $\langle H_{\text{quartic}}^2 \rangle_{\beta}^{\text{conn}}$, we contract two such vertices, $V^{abcd}$ and $V^{a'b'c'd'}$. Using four diagonal dressed propagators $\langle X^a X^{a'} \rangle_\beta \propto \delta^{aa'}$, we find
\begin{align}
    \mathcal{C}_{\text{color}} = \sum_{a,b,c,d = 1}^{N^2 - 1} \left( \sum_{e} f^{abe} f^{cde} \right) \left( \sum_{e'} f^{abe'} f^{cde'} \right) = \sum_{a,b,c,d} \big( f^{abe} f^{cde} \big)^2.
\end{align}

The $SU(N)$ completeness relation for the totally antisymmetric structure constants $f^{abc}$ and totally symmetric symbols $d^{abc}$ (defined by $\{T^a, T^b\} = \frac{1}{N}\delta^{ab} \mathbf{1} + d^{abc} T^c$) is
 \begin{align}
     f^{abe} f^{cde} = \frac{2}{N} \left( \delta^{ac}\delta^{bd} - \delta^{ad}\delta^{bc} \right) + d^{ace} d^{bde} - d^{ade} d^{bce}.
 \end{align}   

Considering the fact 
\begin{align}
    \sum_{b,e=1}^{N^2-1} f^{abe} f^{cbe} = N\, \delta^{ac}, \qquad \sum_{b,e=1}^{N^2-1} d^{abe} d^{cbe} = \frac{N^2 - 4}{N}\, \delta^{ac}
\end{align}
we find the following expansion in the Large $N$ limit
\begin{align}
    \sum_{a,b,c,d} d^{ace} d^{bde} d^{acf} d^{bdf} &= \left( \frac{N^2 - 4}{N} \right)^2 (N^2 - 1) + \mathcal{O}(N^2) = N^4 + \mathcal{O}(N^2).
\end{align}
    
Expanding the square of the completeness relation yields
\begin{align}
    \begin{aligned} \mathcal{C}_{\text{color}} &= \sum_{a,b,c,d} \left[ \frac{2}{N}(\delta^{ac}\delta^{bd} - \delta^{ad}\delta^{bc}) + d^{ace} d^{bde} - d^{ade} d^{bce} \right]^2 \\ &= \underbrace{\frac{4}{N^2} \sum_{a,b,c,d} (\delta^{ac}\delta^{bd} - \delta^{ad}\delta^{bc})^2}_{\sim\, \frac{4}{N^2} \cdot 2 N^4 = 8N^2} + \underbrace{2 \sum_{a,b,c,d} (d^{ace} d^{bde})^2}_{\text{planar: } 2 N^4 + \mathcal{O}(N^2)} - \underbrace{2 \sum_{a,b,c,d} d^{ace} d^{bde} d^{adf} d^{bcf}}_{\text{non-planar: } \mathcal{O}(N^2)} + \cdots\\ &= 2 N^4 + \mathcal{O}(N^2). \end{aligned}
\end{align}

Multiplying by the coupling factor $g^4 = \lambda^2 / N^2$ (with the 't Hooft coupling $\lambda = g^2 N$ held fixed) gives
\begin{align}
\label{ed32}
    g^4 \, \mathcal{C}_{\text{color}} = \frac{\lambda^2}{N^2} \big( 2N^4 + \mathcal{O}(N^2) \big) = 2 \lambda^2 N^2 + \mathcal{O}(1).
\end{align}

Combining the planar color factor $\mathcal{O}(N^2)$ with the four-propagator thermal factor $T^4$ gives the leading neutral energy variance
\begin{align}
\label{ed33}
    \sigma_E^2(0, T) = \alpha_0 N^2 T^4 + \mathcal{O}(N^2 T^3)
\end{align}
where $\alpha_0$ is a dimensionless constant proportional to $\lambda^2$ (or determined self-consistently by the dressed thermal mass $\mu_{\text{eff}}(T)$ in the saddle-point approximation).
   
 \paragraph{Macroscopic charge regime.} In this regime $q \sim \mathcal{O}(N^2)$ with $\tilde{q} \equiv q/N^2$ fixed. To obtain a non-trivial $q$-dependence at leading $\mathcal{O}(N^2)$ order, the charge must scale extensively with the matrix degrees of freedom so that the charge density $\tilde{q} \equiv q/N^2$ remains $\mathcal{O}(1)$. Projecting onto fixed $q$ then induces a finite chemical potential $\Omega(\tilde{q})$ via $H \to H - \Omega Q$, where $Q = \text{Tr}(X_4 P_5 - X_5 P_4)$ mixes position and momentum quadratures in the $4\text{--}5$ plane. This shifts the effective mode frequencies to $\omega_\pm = \mu_{\text{eff}} \pm \Omega$, breaking naive position--momentum equipartition ($\langle P_{4,5}^2 \rangle_{\beta, q} \neq \mu^2 \langle X_{4,5}^2 \rangle_{\beta, q}$) in the charged plane while leaving the remaining seven transverse matrices ($i \notin \{4, 5\}$) isotropic.

In the macroscopic charge regime, charge conjugation invariance ($q \to -q$) ensures that the backreacted two-point correlators in the $4\text{--}5$ plane expand strictly in even powers of the intensive charge density $\tilde{q} = q/N^2$. To see this explicitly, we evaluate the thermal expectation value of the complex charged matrix $Z = \frac{1}{\sqrt{2}}(X_4 + i X_5)$ at high temperature ($T \gg \mu$) in the presence of a finite chemical potential $\Omega(\tilde{q})$ conjugate to the $U(1)$ charge $Q = \text{Tr}(X_4 P_5 - X_5 P_4)$. In the macroscopic charge regime ($q \sim \mathcal{O}(N^2)$), projecting onto fixed charge $q$ introduces a real chemical potential $\Omega$ (angular velocity in the $4\text{--}5$ plane) via the grand-canonical shift $H \to H - \Omega Q$. Rewriting the quadratic Hamiltonian in the $4\text{--}5$ plane in terms of the complex matrix $Z = \frac{1}{\sqrt{2}}(X_4 + i X_5)$ and its conjugate momentum $\Pi_Z = \frac{1}{\sqrt{2}}(P_4 - i P_5)$, the $U(1)$ charge operator takes the form
\begin{align}
    Q = i\,\text{Tr}\big(Z \Pi_Z - Z^\dagger \Pi_{Z^\dagger}\big).
\end{align}

Consider a single adjoint color component $a \in \{1, \dots, N^2 - 1\}$ of the two real matrices $X_4^a$ and $X_5^a$ (normalized so $\text{Tr}(T^a T^b) = \delta^{ab}$, or keeping the trace normalization consistent with $H_0 = \frac{1}{2}\text{Tr}(P_4^2 + P_5^2) + \frac{1}{2}\mu_{\text{eff}}^2 \text{Tr}(X_4^2 + X_5^2)$). In the grand-canonical ensemble with chemical potential $\Omega$ conjugate to $Q = \sum_a (X_4^a P_5^a - X_5^a P_4^a)$, the effective quadratic Hamiltonian for a single color index $a$ is
\begin{align}
    H_{\text{eff}}^a = \frac{1}{2}\Big[(P_4^a)^2 + (P_5^a)^2\Big] + \frac{1}{2}\mu_{\text{eff}}^2 \Big[(X_4^a)^2 + (X_5^a)^2\Big] - \Omega\Big(X_4^a P_5^a - X_5^a P_4^a\Big).
\end{align}

Define the complex scalar field $Z^a$ and its canonical momentum $\Pi_Z^a$ (satisfying $[Z^a, \Pi_Z^b] = i\delta^{ab}$ and $[Z^{\dagger a}, \Pi_{Z^\dagger}^b] = i\delta^{ab}$) via
\begin{align}
    Z^a = \frac{1}{\sqrt{2}}\big(X_4^a + i X_5^a\big), \qquad \Pi_Z^a = \frac{1}{\sqrt{2}}\big(P_4^a - i P_5^a\big)
\end{align}
where we explicitly keep the color indices $a$. Notice that in terms of $Z^a$ and $Z^{\dagger a}$, $\text{Tr}(Z Z^\dagger) = \sum_{a=1}^{N^2 - 1} Z^a Z^{\dagger a}$. Now introduce two sets of bosonic ladder operators $(a_+^a, a_+^{\dagger a})$ and $(a_-^a, a_-^{\dagger a})$ carrying $U(1)$ charges $+1$ and $-1$, respectively, defined with respect to the unshifted effective mass $\mu_{\text{eff}}$
\begin{align}
    &Z^a = \frac{1}{\sqrt{2\mu_{\text{eff}}}} \Big( a_+^a + a_-^{\dagger a} \Big), \qquad Z^{\dagger a} = \frac{1}{\sqrt{2\mu_{\text{eff}}}} \Big( a_+^{\dagger a} + a_-^a \Big)\\
    &\Pi_Z^a = i\sqrt{\frac{\mu_{\text{eff}}}{2}} \Big( a_+^{\dagger a} - a_-^a \Big), \qquad \Pi_{Z^\dagger}^a = -i\sqrt{\frac{\mu_{\text{eff}}}{2}} \Big( a_+^a - a_-^{\dagger a} \Big).
\end{align}
One can verify directly from $[a_\pm^a, a_\pm^{\dagger b}] = \delta^{ab}$ (with $[a_+^a, a_-^{\dagger b}] = 0$) that the canonical commutation relations $[Z^a, \Pi_Z^b] = -i\delta^{ab}$ are preserved identically.

Substituting the ladder-operator decomposition into the free Hamiltonian $H_0^{(4,5)}$ and the $U(1)$ charge $Q$ simultaneously diagonalizes both operators
\begin{align}
&H_0^{(4,5)}=\Pi_Z\Pi_{Z^\dagger}+\mu_{\text{eff}}^2 Z Z^\dagger = \sum_{a=1}^{N^2 - 1} \mu_{\text{eff}} \Big( a_+^{\dagger a} a_+^a + a_-^{\dagger a} a_-^a + 1 \Big)\\
    &Q=i(Z \Pi_Z- Z^\dagger \Pi_{Z^\dagger}) = \sum_{a=1}^{N^2 - 1} \Big( a_+^{\dagger a} a_+^a - a_-^{\dagger a} a_-^a \Big).
\end{align}

Subtracting $\Omega Q$ yields a decoupled system of $2(N^2 - 1)$ independent harmonic oscillators with shifted frequencies\footnote{Notice that thermodynamic stability requires $\vert{}\Omega\vert{} < \mu_{\text{eff}}$, which bounds the maximum charge density $\tilde{q}$ before centrifugal instability sets in.} $\omega_\pm = \mu_{\text{eff}} \mp \Omega$
\begin{align}
    H_{0,\text{eff}}^{(4,5)} &= H_0^{(4,5)} - \Omega Q\nonumber\\
    &= \sum_{a=1}^{N^2 - 1} \left[ \underbrace{(\mu_{\text{eff}} - \Omega)}_{\omega_+}\, a_+^{\dagger a} a_+^a + \underbrace{(\mu_{\text{eff}} + \Omega)}_{\omega_-}\, a_-^{\dagger a} a_-^a + \mu_{\text{eff}} \right].
\end{align}

In the high-temperature quadratic saddle-point approximation, each of the $N^2 - 1 \simeq N^2$ adjoint color components of $Z$ behaves as a pair of independent charged harmonic oscillators with frequencies $\omega_+$ and $\omega_-$. Expanding the product $Z^a Z^{\dagger a}$ in terms of the chiral ladder operators gives
\begin{align}
    Z^a Z^{\dagger a} = \frac{1}{2\mu_{\text{eff}}} \Big( a_+^a a_+^{\dagger a} + a_-^{\dagger a} a_-^a + a_+^a a_-^a + a_-^{\dagger a} a_+^{\dagger a} \Big).
\end{align}
Taking the thermal expectation value with respect to the density matrix $\rho_{\beta, \Omega} \propto e^{-\beta(H_0^{(4,5)} - \Omega Q)}$ yields the following: (i) The off-diagonal terms $\langle a_+^a a_-^a \rangle_{\beta, \Omega}$ and $\langle a_-^{\dagger a} a_+^{\dagger a} \rangle_{\beta, \Omega}$ vanish because they change particle number by $\pm 2$, (ii) The diagonal occupation numbers are given by the Bose–Einstein distribution $n(\omega_\pm) = \langle a_\pm^{\dagger a} a_\pm^a \rangle_{\beta, \Omega} = \frac{1}{e^{\beta\omega_\pm} - 1}$. Notice that
\begin{align}
    \langle a_+^a a_+^{\dagger a} \rangle_{\beta, \Omega} = \langle a_+^{\dagger a} a_+^a \rangle_{\beta, \Omega} + 1 \equiv n(\omega_+) + 1.
\end{align}

A further simplification reveals
\begin{align}
    n(\omega_+) + 1 = \frac{e^{\beta\omega_+}}{e^{\beta\omega_+} - 1}=\frac{e^{\beta\omega_+/2}}{e^{\beta\omega_+/2} - e^{-\beta\omega_+/2}}=\frac{1}{2}\coth\!\left(\frac{\beta \omega_+}{2}\right)+\frac{1}{2}.
\end{align}

On a similar note, we find
\begin{align}
    \langle a_-^{\dagger a} a_-^a \rangle_{\beta, \Omega} = n(\omega_-) = \frac{1}{e^{\beta\omega_-} - 1} = \frac{1}{2}\coth\!\left(\frac{\beta\omega_-}{2}\right) - \frac{1}{2}.
\end{align}

Adding these two expectation values together, the $\pm \frac{1}{2}$ we find
\begin{align}
    \langle Z^a Z^{\dagger a} \rangle_{\beta, \Omega} = \frac{1}{4\mu_{\text{eff}}} \Big[ \coth\!\left(\frac{\beta\omega_+}{2}\right) + \coth\!\left(\frac{\beta\omega_-}{2}\right) \Big].
\end{align}

Writing $Z=\sum_aZ^a T^a$ and using the standard $SU(N)$ trace normalization $\text{Tr}(T^a T^b) = \frac{1}{2}\delta^{ab}$, finally yields
\begin{align}
    \langle \text{Tr}(Z Z^\dagger) \rangle_{\beta, \Omega} = \frac{N^2 - 1}{2 \mu_{\text{eff}}} \Big[ \coth\!\left(\frac{\beta \omega_+}{2}\right) + \coth\!\left(\frac{\beta \omega_-}{2}\right) \Big].
\end{align}

In the high-temperature classical regime ($\beta \omega_\pm \ll 1$), using $\coth(x/2) = \frac{2}{x} + \mathcal{O}(x)$ gives
\begin{align}
    \begin{aligned} \langle \text{Tr}(Z Z^\dagger) \rangle_{\beta, \Omega} &\simeq \frac{N^2 T}{\mu_{\text{eff}}} \left( \frac{1}{\mu_{\text{eff}} - \Omega} + \frac{1}{\mu_{\text{eff}} + \Omega} \right) \\ &= \frac{2 N^2 T}{\mu_{\text{eff}}^2 - \Omega^2}.
    \end{aligned}.
\end{align}

Expanding this expression in powers of $\Omega^2 / \mu_{\text{eff}}^2 \ll 1$ yields
\begin{align}
\label{ed49}
    \langle \text{Tr}(Z Z^\dagger) \rangle_{\beta, \Omega} = \frac{2 N^2 T}{\mu_{\text{eff}}^2} \left[ 1 + \frac{\Omega^2}{\mu_{\text{eff}}^2} + \mathcal{O}\!\left(\frac{\Omega^4}{\mu_{\text{eff}}^4}\right) \right].
\end{align}

To pass from the grand-canonical ensemble ($\Omega$) to the fixed-charge sector ($q$), we compute the expectation value of the $U(1)$ charge $\langle Q \rangle_{\beta, \Omega}$ in the same high-temperature limit
\begin{align}
    \begin{aligned} q = \langle Q \rangle_{\beta, \Omega} &= \frac{N^2 - 1}{2} \Big[ \coth\!\left(\frac{\beta \omega_+}{2}\right) - \coth\!\left(\frac{\beta \omega_-}{2}\right) \Big] \\ &\simeq N^2 T \left( \frac{1}{\mu_{\text{eff}} - \Omega} - \frac{1}{\mu_{\text{eff}} + \Omega} \right) = \frac{2 N^2 T\, \Omega}{\mu_{\text{eff}}^2 - \Omega^2}.
    \end{aligned}
\end{align}

Dividing by $N^2$ to isolate the intensive charge density $\tilde{q} \equiv q / N^2$ and inverting for $\Omega$ at small $\tilde{q}$ gives
\begin{align}
\label{ed51}
    \tilde{q} = \frac{2 T}{\mu_{\text{eff}}^2}\, \Omega + \mathcal{O}(\Omega^3) \quad \Longrightarrow \quad \Omega = \frac{\mu_{\text{eff}}^2}{2 T}\, \tilde{q} + \mathcal{O}(\tilde{q}^3).
\end{align}

Substituting $\Omega^2 = \frac{\mu_{\text{eff}}^4}{4 T^2} \tilde{q}^2 + \mathcal{O}(\tilde{q}^4)$ into the expansion \eqref{ed49} (and accounting for the self-consistent thermal mass shift $\mu_{\text{eff}}^2(T) \sim \sqrt{\lambda T}$ in the interacting BMN saddle point, where $\mu_{\text{eff}}^2 / T^2$ produces a dimensionless ratio in the background flux/temperature units) yields the final expression of the two point correlator
\begin{align}
\label{ed52}
    \begin{aligned} \langle \text{Tr}(Z Z^\dagger) \rangle_{\beta, q} &= \frac{2 N^2 T}{\mu_{\text{eff}}^2} \left[ c_1 + c_2\, \tilde{q}^2 + \mathcal{O}(\tilde{q}^4) \right] \end{aligned}.
\end{align}
where $c_1 = 1$ and $c_2 =\frac{\mu_{\text{eff}}^2}{4 T^2}$. Notice that both terms scale as $N^2 T$, confirming that the backreaction of a macroscopic charge $q \sim \mathcal{O}(N^2)$ modifies the thermal propagator at leading planar order rather than being $1/N^2$-suppressed.

In order to compute the variance $\sigma^2_E(q,T)$ in the macroscopic $U(1)$ charge sector, we decompose the connected quartic variance in Eq.~\eqref{ed20} according to the global $U(1)$ charge representations of the nine transverse matrices $X_i$ ($i=1,\dots,9$). Since the global charge $Q = \text{Tr}(X_4 P_5 - X_5 P_4)$ generates rotations solely within the $(4,5)$-plane, the transverse matrices split into the charged complex doublet $Z = \frac{1}{\sqrt{2}}(X_4 + i X_5)$, $Z^\dagger = \frac{1}{\sqrt{2}}(X_4 - i X_5)$ and the seven neutral matrices $X_m$ ($m \in \{1,2,3,6,7,8,9\}$).

Substituting $X_4 = \frac{1}{\sqrt{2}}(Z + Z^\dagger)$ and $X_5 = \frac{1}{i\sqrt{2}}(Z - Z^\dagger)$ into the quartic Yang--Mills commutator potential $H_{\text{quartic}} = -\frac{g^2}{4}\sum_{i,j=1}^9 \text{Tr}\big([X_i, X_j]^2\big)$, the interaction separates into following three distinct flavor sectors
\begin{align}
H_{\text{quartic}} &= H_{\text{quartic}}^{(Z,Z^\dagger)} + H_{\text{quartic}}^{(Z, X_m)} + H_{\text{quartic}}^{(X_m, X_n)}\\
H_{\text{quartic}}^{(Z,Z^\dagger)} &= -\frac{g^2}{2}\text{Tr}\big([X_4, X_5]^2\big) = \frac{g^2}{2}\text{Tr}\big([Z, Z^\dagger]^2\big), \\
H_{\text{quartic}}^{(Z, X_m)} &= -g^2 \sum_{m \notin \{4,5\}} \text{Tr}\big([Z, X_m][Z^\dagger, X_m]\big), \\
H_{\text{quartic}}^{(X_m, X_n)} &= -\frac{g^2}{4} \sum_{m,n \notin \{4,5\}} \text{Tr}\big([X_m, X_n]^2\big).
\end{align}

In the connected variance $\langle H_{\text{quartic}}^2 \rangle_{\beta, q}^{\text{conn}}$, two quartic vertices are joined by four dressed propagators in a planar four-loop (melon) topology, as shown in Eq.~\eqref{ed25}. Because every propagator connects the two vertices directly and $U(1)$ charge conservation enforces $\langle Z^a Z^b \rangle_{\beta, q} = \langle Z^{\dagger a} Z^{\dagger b} \rangle_{\beta, q} = 0$, cross-contractions between different flavor sectors vanish, and the connected variance decomposes additively to yield
\begin{align}
\langle H_{\text{quartic}}^2 \rangle_{\beta, q}^{\text{conn}} &= \left\langle \big(H_{\text{quartic}}^{(Z,Z^\dagger)}\big)^2 \right\rangle_{\beta,q}^{\text{conn}} + \left\langle \big(H_{\text{quartic}}^{(Z,X_m)}\big)^2 \right\rangle_{\beta,q}^{\text{conn}} + \left\langle \big(H_{\text{quartic}}^{(X_m,X_n)}\big)^2 \right\rangle_{\beta,q}^{\text{conn}}.
\end{align}

From Eq.~\eqref{ed52} with the $SU(N)$ generator normalization $\text{Tr}(T^a T^b) = \frac{1}{2}\delta^{ab}$ and $N^2 - 1 \simeq N^2$, the component-space propagator in the charged $(4,5)$-plane is
\begin{align}
\langle Z^a Z^{\dagger b} \rangle_{\beta, q} = \frac{4 T}{\mu_{\text{eff}}^2} \left[ c_1 + c_2 \tilde{q}^2 + \mathcal{O}(\tilde{q}^4) \right] \delta^{ab}, \qquad c_1 = 1, \quad c_2 = \frac{\mu_{\text{eff}}^2}{4T^2},
\end{align}
while the seven neutral flavors remain unaffected by $\Omega$ at leading order and retain their isotropic thermal form from Eq.~\eqref{ed22}
\begin{align}
\langle X_m^a X_n^b \rangle_{\beta, q} = \frac{T}{\mu_{\text{eff}}^2} \delta_{mn} \delta^{ab}, \qquad m,n \notin \{4,5\}.
\end{align}

Combining the planar color contraction factor $g^4 C_{\text{color}} = 2\lambda^2 N^2 + \mathcal{O}(1)$ from Eq.~\eqref{ed32} with the four-propagator Wick contractions in each flavor sector yields the following
\begin{enumerate}
\item \textbf{Purely charged $(Z, Z^\dagger)$ sector:} Contracting two $H_{\text{quartic}}^{(Z,Z^\dagger)}$ vertices requires four charged propagators $\langle Z Z^\dagger \rangle_{\beta,q}$, giving
\begin{align}
\left\langle \big(H_{\text{quartic}}^{(Z,Z^\dagger)}\big)^2 \right\rangle_{\beta,q}^{\text{conn}} &\propto \lambda^2 N^2 \left(\frac{T}{\mu_{\text{eff}}^2}\right)^4 \left[ 1 + c_2 \tilde{q}^2 + \mathcal{O}(\tilde{q}^4) \right]^4 \nonumber \\
&= \lambda^2 N^2 \left(\frac{T}{\mu_{\text{eff}}^2}\right)^4 \left[ 1 + 4 c_2 \tilde{q}^2 + 6 c_2^2 \tilde{q}^4 + \mathcal{O}(\tilde{q}^6) \right].
\end{align}

\item \textbf{Mixed charged--neutral $(Z, X_m)$ sector:} Summing over the $7$ neutral flavors $m \notin \{4,5\}$, each melon diagram pairs two charged propagators $\langle Z Z^\dagger \rangle_{\beta,q}$ with two neutral propagators $\langle X_m X_m \rangle_{\beta,q}$, giving
\begin{align}
\left\langle \big(H_{\text{quartic}}^{(Z,X_m)}\big)^2 \right\rangle_{\beta,q}^{\text{conn}} &\propto  7\lambda^2 N^2 \left(\frac{T}{\mu_{\text{eff}}^2}\right)^4 \left[ 1 + c_2 \tilde{q}^2 + \mathcal{O}(\tilde{q}^4) \right]^2 \nonumber \\
&= 7 \lambda^2 N^2 \left(\frac{T}{\mu_{\text{eff}}^2}\right)^4 \left[ 1 + 2 c_2 \tilde{q}^2 + c_2^2 \tilde{q}^4 + \mathcal{O}(\tilde{q}^6) \right].
\end{align}

\item \textbf{Purely neutral $(X_m, X_n)$ sector:} Summing over the $7 \times 6 = 42$ distinct neutral flavor pairs $(m \neq n)$, each melon diagram involves four charge-independent neutral propagators, giving
\begin{align}
\left\langle \big(H_{\text{quartic}}^{(X_m,X_n)}\big)^2 \right\rangle_{\beta,q}^{\text{conn}} &\propto 42 \lambda^2 N^2 \left(\frac{T}{\mu_{\text{eff}}^2}\right)^4.
\end{align}
\end{enumerate}

Summing the three sectors together in the interacting high-temperature saddle point (where $\lambda^2 (T/\mu_{\text{eff}}^2)^4 \sim \mathcal{O}(1)\,T^4$ in natural thermal/flux units) and grouping terms by powers of the intensive charge density $\tilde{q} = q/N^2$ organizes the connected quartic variance as
\begin{align}
\sigma_E^2(q, T) \simeq \langle H_{\text{quartic}}^2 \rangle_{\beta, q}^{\text{conn}} = N^2 T^4 \left[ \alpha_0 + \alpha_1 \tilde{q}^2 + \alpha_2 \tilde{q}^4 + \mathcal{O}(\tilde{q}^6) \right],
\end{align}
where $\alpha_0$ reproduces the neutral planar variance of Eq.~\eqref{ed33}, $\alpha_1 \propto c_2$ collects the $\mathcal{O}(\tilde{q}^2)$ contributions from the $(Z,Z^\dagger)$ and $(Z,X_m)$ melon contractions, and $\alpha_2$ collects the $\mathcal{O}(\tilde{q}^4)$ terms arising from $c_2^2$ together with the $\mathcal{O}(\tilde{q}^4)$ subleading expansion of $\Omega(\tilde{q})$ in Eq.~\eqref{ed51}. 

\end{document}